\documentclass{llncs}
\usepackage{epsfig}
\usepackage{latexsym} 
\begin{document}
\title{The bisimulation problem for equational graphs of finite out-degree.}
\author{G\'eraud S\'enizergues}
\institute
{LaBRI\\
Universit\'e de Bordeaux I\\
351, Cours de la Lib\'eration 33405 Talence, France
\footnote{
mailing adress:LaBRI and UFR Math-info, Universit\'e Bordeaux1\\
               351 Cours de la lib\'eration -33405- Talence Cedex.\\
email:ges@labri.u-bordeaux.fr\\
fax: 05-56-84-66-69
}
}
\newtheorem{theo}{Theorem}[section]
\newtheorem{prop}[theo]{Proposition}
\newtheorem{lem}[theo]{Lemma}
\newtheorem{cor}[theo]{Corollary}
\newtheorem{fact}[theo]{Fact}
\newtheorem{exle}[theo]{Example}
\newtheorem{rem}[theo]{Remark}
\newtheorem{defi}[theo]{Definition}
\newenvironment{prooof}[0]{\noindent{\bf Proof}:}{$\Box$\\}
\newenvironment{sketch}[0]{{\bf Sketch of proof}:}{$\Box$\\}

\newcommand{\lp}{\langle~\langle}
\newcommand{\rp}{\rangle~\rangle}
\newcommand{\pa}[1]{\langle\langle~#1~\rangle\rangle}
\newcommand{\dd}{{\rm d}}
\newcommand{\CO}{{\rm Cong}}
\newcommand{\NN}[1]{\mid \parallel#1\mid \parallel}
\newcommand{\N}{{\rm N}}
\newcommand{\Str}{{\cal S}}
\newcommand{\Sys}{{\cal S}}
\newcommand{\topo}{{\cal T}}
\newcommand{\Language}{{\rm L}}
\newcommand{\transpose}[1]{\;\stackrel{t}{\,}\!\!#1}
\newcommand{\supp}{{\rm supp}}
\newcommand{\ptcup}{\stackrel{.}{\cup}}
\newcommand{\bbbB}{{\sf B}}
\newcommand{\bbbS}{{\sf S}}
\newcommand{\bbbD}{{\sf D}}
\newcommand{\bbbM}{{\sf M}}
\newcommand{\bbbR}{{\sf R}}
\newcommand{\bbbQ}{{\sf Q}}
\newcommand{\bbbV}{{\sf V}}
\newcommand{\Bin}{{\bf \cal B}}
\newcommand{\card}{{\rm Card}}
\newcommand{\mi}{{\rm min}}
\newcommand{\borninf}{{\rm inf}}
\newcommand{\ma}{{\rm max}}
\newcommand{\rd}{{\rm rd}}
\newcommand{\di}{{\rm dim}}
\newcommand{\dom}{{\rm dom}}
\newcommand{\im}{{\rm im}}
\newcommand{\leaves}{{\cal L}}
\newcommand{\internal}{{\cal I}}
\newcommand{\support}{{\rm supp}}
\newcommand{\Div}{{\rm Div}}
\newcommand{\INV}{{\rm INV}^{(\cal O)}}
\newcommand{\W}{{\rm W}^{(\cal O)}}
\newcommand{\D}{{\rm D}^{(\cal O)}}
\newcommand{\OR}{{\cal O}}
\newcommand{\deriv}[1]{\stackrel{#1}{\longrightarrow}}
\newcommand{\deducedir}{\:|\!\!\!-\!\!-\:}

\newcommand{\ededucedir}{\:|\!\deducedir}

\newcommand{\antiededucedir}{-\!\!-\!\!\!||}

\newcommand{\deduce}[1]{\stackrel{<#1>}{\deducedir}}

\newcommand{\ededuce}[1]{\stackrel{<#1>}{\ededucedir}}

\newcommand{\antiededuce}[1]{\stackrel{<#1>}{\antiededucedir}}

\newcommand{\deducesq}[1]{\stackrel{[#1]}{\deducedir}}

\newcommand{\ededucesq}[1]{\stackrel{[#1]}{\ededucedir}}

\newcommand{\computto}[2]{\stackrel{#1}{\longmapsto_{#2}}}
\newcommand{\computdir}[1]{\longmapsto_{#1}}
\newcommand{\reduito}[2]{\stackrel{#1}{\longrightarrow_{#2}}}
\newcommand{\reduitdir}[1]{\longrightarrow_{#1}}

\newcommand{\derive}[1]{\raisebox{-0.2ex}
  {$\begin{array}{c}\ast \\[-1.6ex]
    \longleftarrow \\[-1.0ex] {\it #1}
  \end{array}$}}

\newcommand{\reduit}[1]{\raisebox{-0.2ex}
  {$\begin{array}{c}\ast \\[-1.6ex]
    \longrightarrow \\[-1.0ex] {\it #1}
  \end{array}$}}

\newcommand{\derivedir}[1]{\raisebox{-0.8ex}
  {$\begin{array}{c}\longleftarrow \\[-1.0ex] {\it #1}
  \end{array}$}}

\newcommand{\congru}[1]{\raisebox{-0.2ex}
  {$\begin{array}{c}\ast \\[-1.6ex]
    \longleftrightarrow \\[-1.0ex] {\it #1}
  \end{array}$}}

\newcommand{\congrudir}[1]{\raisebox{-0.8ex}
  {$\begin{array}{c}\longleftrightarrow \\[-1.0ex] {\it #1}
  \end{array}$}}

\newcommand{\derivelg}[1]{\raisebox{-0.2ex}
  {$\begin{array}{c}\ast \\[-1.6ex]
    \longleftarrow \\[-1.0ex] {\it #1}{\;}^{lg}
  \end{array}$}}

\newcommand{\reduitlg}[1]{\raisebox{-0.2ex}
  {$\begin{array}{c}\ast \\[-1.6ex]
    \longrightarrow \\[-1.0ex]{\!}^{lg} {\it #1}
  \end{array}$}}

\newcommand{\congrulg}[1]{\raisebox{-0.2ex}
  {$\begin{array}{c}\ast \\[-1.6ex]
    \longleftrightarrow \\[-1.0ex]{\!}^{lg} {\it #1}
  \end{array}$}}

\newcommand{\derlgdir}[1]{\raisebox{-0.8ex}
  {$\begin{array}{c}\longleftarrow \\[-1.0ex]{\it  #1}{\;}^{lg}
  \end{array}$}}

\newcommand{\redlgdir}[1]{\raisebox{-0.8ex}
  {$\begin{array}{c}\longrightarrow \\[-1.0ex]{\!}^{lg} {\it  #1}
  \end{array}$}}

\maketitle


\begin{abstract}
The {\em bisimulation} problem for equational graphs of finite out-degree is shown to be decidable.
We reduce this problem to the $\eta$-bisimulation problem for deterministic  rational (vectors of)
boolean series on the alphabet of a dpda $\cal M$. 
We then exhibit a {\em complete formal system} for deducing equivalent pairs of
such vectors.
\end{abstract}

\pagenumbering{arabic}
\section{Introduction}
\label{introduction}
\subsection{Motivations}
\label{Motivations}
\subsubsection{Processes}
In the context of concurrency theory , several notions of ``behaviour of a process'' and 
`` behavioural equivalence between processes''
have been proposed. Among them, the notion of {\em bisimulation} equivalence  seems to play a 
prominent role ( see \cite{Mil89}). The question of whether this equivalence
is {\em decidable} or not for various classes of infinite processes has been the subject of many
works in the last ten years (see for example \cite{Bae-Ber-Klo87,Cau90canon,Sti91,Chr-Hir-Mol93,Gro-Hut94},\\
\cite{Hir-Jer-Mol94,Chr-Hut-Sti95,Cau95,Sti96,Jan97,Sen98_focs}).
 
The aim of this work is to show decidability of the bisimulation 
equivalence for the class of all processes defined by pushdown automata whose 
$\epsilon$-transitions are deterministic and decreasing
(of course, we assume that $\epsilon$-transitions are {\em not} visible,
which implies that the graphs of the processes considered here, might have infinite in-degree). 
This problem was raised in \cite{Cau95} ( see Problem 6.2 of this reference ) 
and is a significant subcase of the problem raised in
\cite{Sti96} (as the bisimulation-problem for processes `` of type -1'').
\subsubsection{Infinite graphs}
A wide class of graphs enjoying interesting decidability properties has been defined in 
\cite{Cou89,Bau91,Bau92} (see \cite{Cou90b} for a survey). In particular it is known that the
problem
$$ 
`` \mbox{ are } \Gamma,\Gamma' \mbox{ isomorphic? }''
$$
 is decidable for pairs $\Gamma,\Gamma'$ of
equational graphs. It seems quite natural to investigate whether the problem
$$ 
`` \mbox{ are } \Gamma,\Gamma' \mbox{ bisimilar? }''
$$
 is decidable for pairs $\Gamma,\Gamma'$ of equational graphs.
We show here that this problem is decidable for equational graphs of finite out-degree.
\subsubsection{Formal languages}
Another classical equivalence relation between processes is the notion of {\em language} 
equivalence . The decidability of language equivalence for {\em deterministic } pushdown automata
has been recently established in \cite{Sen97} ( see also in \cite{Sen97ic,Sen97fty} shorter 
expositions of this result). It was first noticed in \cite{Bae-Ber-Klo87} that , in the case of
deterministic processes, language equivalence and bisimulation equivalence are identical.
Moreover deterministic pushdown automata can always be normalized ( with preservation of the
language) in such a way that $\epsilon$-transitions are all decreasing. Hence the main result of 
this work is a generalisation of the decidability of the equivalence problem for dpda's. 
\pagebreak[4]
\subsubsection{Mathematical generality}
More precisely, the present work
{\em extends the notions} developped in \cite{Sen97} so as to obtain a more general result.\\
As a by-product of this extension, we obtain a deduction system which, in the deterministic 
case, seems {\em simpler} than the one presented in \cite{Sen97} ( see system ${\cal B}_3$ in 
\S \ref{elimination}).\\
The present work can also be seen as a common generalization of 3 different results:
the results of \cite{Sti96,Jan97} establishing decidability of the bisimulation equivalence in 
two non-deterministic sub-classes of the class considered here, and the result of \cite{Sen97}
dealing only with deterministic pda's (or processes).  
\subsubsection{Logics}
Our solution consists in constructing a {\em complete} formal system , in the general 
sense taken by this word in mathematical logics i.e.: it consists of
a set of well-formed assertions, a subset of basic assertions, the axioms, 
and a set of deduction rules allowing to derive new assertions from assertions which are already generated. 
The well-formed assertions we are considering are pairs $(S,T)$ of rational 
boolean
series over the non-terminal alphabet $V$ of some strict-deterministic 
grammar $G=<X,V,P>$.
Such an assertion is true when the two series $S,T$ are bisimilar.\\ 
Several simple formal systems generating all the 
identities between boolean rational expressions have been the subject of
many works (\cite{Sal66,Bof90,Kro91}); the case of bisimilar 
rational expressions has been also adressed in \cite{Mil84,Koz91}.\\
A tableau proof-system generating all the bisimilar pairs of words with 
respect to
a given context-free grammar in Greibach normal form was also given in 
\cite{Sti91}.\\
Our complete formal systems can be seen as participating in this
general research stream (see in \cite{Sen00} an overview of this subject,
in the context of equivalence problems for pushdown automata).
\subsection{Results}
\label{Results}
The main results of this work are the following theorems.\\
{\bf Theorem} \ref{decidable}:\\
{\em The bisimulation  problem for rooted equational 1-graphs of finite out-degree is decidable.}\\
{\bf Theorem} \ref{B3_complete}:\\
{\em ${\cal B}_3$ is a complete deduction system}.\\
where ${\cal B}_3$ is a formal system whose elementary rules just express
the basic algebraic properties of bisimulation: the fact that it is
 an equivalence relation, that it is compatible with right and left (matricial) product, that Arden's lemma 
remains true modulo bisimulation and at last,
its link with one-step derivation (rule R34). Completeness means here that
{\em all} pairs of bisimilar rational ``deterministic'' boolean series are generated by this formal system.
\subsection{Main tools}
\label{Tools}
We re-use here the notions developed in \cite{Sen97} (1-4) and introduce new ideas (5-7):
\begin{enumerate}
\item the {\em deduction systems} ( which were in turn inspired by \cite{Cou83}).
\item the {\em deterministic boolean series}  ( which were in turn inspired by \cite{Har79}). 
\item the {\em deterministic spaces}  ( which were elaborated around Meitus notion of linear independence (\cite{Mei89,Mei92})).
\item the {\em analysis} of the proof-trees generated by a suitable strategy ( which was somehow similar
with the analysis of the parallel computations , interspersed with  replacement-moves, done in
\cite{Val74,Rom85,Oya87}).
\item the notion of { \em $\eta$-bisimulation} over deterministic row-vectors of boolean series
( which , in some sense,  translates the usual notion of bisimulation to the d-space of
row-vectors of series).
\item the notion of {\em oracle}, which is a choice of bisimulation for every pair of bisimilar 
vectors;
the notion of triangulation of systems of linear equations is now ``parametrized'' by such 
an oracle ${\cal O}$ ( see \S \ref{triangulations}); as well, the strategies are now 
parametrized by an oracle too.
\item an {\em elimination} argument: roughly speaking, this argument shows that, in a proof-tree $t$, if we take into 
account not only the {\em branch} ending at a node $x$, but also the {\em whole} proof-tree, then the 
meta-rule $R5$
$$
\; \; \{(p, S, S')\} {\ededucedir} (p+ 2, S \odot x, S' \odot x')
$$
is not needed to show that $\im(t) \deducedir \{t(x)\} $. A nice (and unexpected) by-product 
of this elimination is that the {\em weights} can
be removed from the equations ( see systems ${\cal B}_2,{\cal B}_3$ in \S \ref{elimination}).
\end{enumerate}
The proof exposed here is an updated version of the full proof given in \cite{Sen97gamma} and
exposed in a consise way in \cite{Sen98_focs}. 
Some simplifications of \cite{Sen97} found by C. Stirling (\cite{Sti99}) 
were taken in account in this proof too:\\
-the technical notion of ``N-stacking sequence'' is replaced by the slightly
simpler notion of ``B-stacking sequence''\\ 
-the analysis of section \ref{trees} uses a choice of ``generating set'' 
which is simpler 
than the choice given in \cite{Sen97gamma,Sen98_focs}.\\ 
-a main simplification linked with this more clever choice,
is that we can restrict ourselves to the case of a {\em proper, reduced} strict-deterministic grammar (as is done in \cite{Sti99}), 
while in \cite{Sen97gamma,Sen98_focs} we could not assume this restriction.

\tableofcontents
\section{Preliminaries}
\subsection{Graphs}
\label{graphs}
Let $X$ be a finite alphabet.
We call {\em graph over $X$} any pair $\Gamma = (V_{\Gamma}, E_{\Gamma})$ where $V_{\Gamma}$ is
a set and $E_{\Gamma}$ is a subset of $V_{\Gamma} \times X  \times V_{\Gamma}$. 
For every integer $n \in \bbbn$, we call an $n$-graph every $n+2$-tuple $\Gamma = (V_{\Gamma}, E_{\Gamma},v_1,\ldots,v_n)$ where $(V_{\Gamma}, E_{\Gamma})$ is a graph and $(v_1,\ldots,v_n)$ is a
sequence of distinguished vertices:  they are called the {\em sources } of $\Gamma$.\\
A 1-graph $(V, E,v_1)$ is said to be {\em rooted} iff $v_1$ is a root of $(V, E)$ and $V \neq \{v_1\}$.
A 2-graph  $(V, E,v_1,v_2)$ is said {\em bi-rooted} iff  $v_1$ is a root, $v_2$ is a co-root of 
$(V, E)$, $v_1 \neq v_2$  and there is no edge going out of $v_2$ ( this last technical condition will
be useful for reducing the bisimilarity notion for graphs to an analogous notion on series, see \S 
\ref{gamma_bisimulation}, \S  \ref{nondet_versus_det} and \S \ref{sigma_bisimulation}).\\
The {\em equational} graphs are the least solutions ( in a suitable sense) of the systems
of (hyperedge) graph-equations ( see in \cite{Cou90b} precise definitions). Let us mention
that the equational graphs of {\em finite degree} are exactly the {\em context-free} graphs 
defined in \cite{Mul-Sch85}.
\subsubsection{Bisimulations}$\;\;$\\
\label{gamma_bisimulation}
\begin{defi}
Let $\Gamma=(V_{\Gamma}, E_{\Gamma}, v_1,\ldots,v_n), \Gamma'=(V_{\Gamma'}, E_{\Gamma'},v'_1,\ldots,v'_n)$ be two n-graphs over an alphabet $X$.
Let ${\cal R}$ be some binary relation ${\cal R} \subseteq V_{\Gamma} \times V_{\Gamma'}$.\\ 
${\cal R}$ is a {\em simulation} from $\Gamma$ to $\Gamma'$ iff 
\begin{enumerate}
\item $\dom({\cal R})= V_{\Gamma}$,
\item $\forall i \in [1,n], (v_i,v'_i) \in {\cal R}$,
\item $\forall v \in V_{\Gamma}, w \in V_{\Gamma}, v' \in V_{\Gamma'}, x \in X,
\mbox{ such that } (v,x,w) \in E_{\Gamma} \mbox{ and } v {\cal R} v',$\\
$ \mbox{ there exists }
w' \in V_{\Gamma'} \mbox{ such that } (v',x,w') \in E_{\Gamma'} \mbox{ and } w {\cal R} w'.$ 
\end{enumerate}
${\cal R}$ is a {\em bisimulation} from $\Gamma$ to $\Gamma'$ iff ${\cal R}$ is a simulation 
from $\Gamma$ to $\Gamma'$ and ${\cal R}^{-1}$ is a simulation from $\Gamma'$ to $\Gamma$.
\label{def_gamma_bisimulation}
\end{defi}
This definition corresponds to the standard one (\cite{Par81,Mil89,Cau95}) in the case where 
$n = 0$.
The n-graphs $\Gamma, \Gamma'$ are said {\em bisimilar} , which we denote 
by $\Gamma \sim \Gamma'$,
iff there exists a bisimulation ${\cal R}$ from $\Gamma$ to $\Gamma'$.

Let us extend now this definition by means of a relational morphism between free monoids. 

\begin{defi}
Let $X,X'$ be two alphabets. A binary relation $\eta \subseteq X^* \times X'^*$ is called
a strong relational morphism from $X^*$ to $X'^*$ iff
\begin{enumerate}
\item $\eta$ is a submonoid of $X^* \times X'^*$
\item $\dom(\eta) = X^*, \im(\eta) = X'^*$
\item $\eta$ is generated (as a submonoid) by the subset $\eta \cap (X \times X')$.
\end{enumerate}
\label{sr_morphism}
\end{defi}
One can easily check that s.r. morphisms are preserved by inversion, composition and
that any surjective map $ \eta : X \rightarrow X'$ induces a s.r. morphism from $X^*$ to $X'^*$.
Let $\Gamma=(V_{\Gamma}, E_{\Gamma}, v_1,\ldots,v_n)$ be an n-graph over the alphabet $X$,
 $\Gamma'=(V_{\Gamma'}, E_{\Gamma'},v'_1,\ldots,v'_n)$ be an n-graph over the alphabet $X'$.
Let $\eta\subseteq X^* \times X'^*$ be some s.r. morphism, and let
${\cal R}$ be some binary relation ${\cal R} \subseteq V_{\Gamma} \times V_{\Gamma'}$.\\
\begin{defi} 
${\cal R}$ is a {\em $\eta$-simulation} from $\Gamma$ to $\Gamma'$ iff
\begin{enumerate}
\item  $\dom({\cal R}) = V_{\Gamma}$,
\item  $\forall i \in [1,n], (v_i,v'_i) \in {\cal R}$,
\item  
$ \forall  v, w \in V_{\Gamma}, v' \in V_{\Gamma'}, x \in X,
\mbox{ such that } (v,x,w) \in E_{\Gamma} \mbox{ and } v {\cal R} v',$\\
$$ \exists 
w' \in V_{\Gamma'}, x' \in \eta(x) \mbox{ such that } 
(v',x',w') \in E_{\Gamma'} \mbox{ and } w {\cal R} w'.$$
\end{enumerate}
${\cal R}$ is a {\em $\eta$-bisimulation} iff ${\cal R}$ is a $\eta$-simulation and 
${\cal R}^{-1}$ is a $\eta^{-1}$- simulation.
\label{def_eta_bisimulation}
\end{defi}
For every $v \in V_{\Gamma}, v' \in V_{\Gamma'}$, we denote by $v \sim v'$ the fact that there 
exists some $\eta$-bisimulation $\cal R$ from $\Gamma$ to $\Gamma'$ such that $(v,v')\in {\cal R}$.
In all this work, the composition of binary relations is denoted by $\circ$ and defined by:
if ${\cal R}_1 \subseteq E \times F$ and ${\cal R}_2 \subseteq F \times G$ then,
\begin{equation} 
{\cal R}_1 \circ {\cal R}_2= \{ (x,z) \in E \times G| \exists y \in F, (x,y) \in  {\cal R}_1,
(y,z) \in  {\cal R}_2 \}.
\label{composition_notation}
\end{equation}
\begin{fact}$\;\;$\\
\begin{enumerate}
\item if ${\cal R}$ is a $\eta$-bisimulation, then ${\cal R}^{-1}$ is a $\eta^{-1}$-bisimulation
\item if ${\cal R}_1$ is a $\eta_1$-bisimulation and ${\cal R}_2$ is a $\eta_2$-bisimulation, then
${\cal R}_1 \circ {\cal R}_2$ is a $\eta_1 \circ \eta_2$-bisimulation
\item if for every $i \in I, {\cal R}_i$ is a $\eta$-bisimulation, then 
$\bigcup_{i \in I} {\cal R}_i$ is a $\eta$-bisimulation.
\end{enumerate}
\end{fact}

\subsection{Pushdown automata}
\label{pda}
A {\em pushdown automaton} on the alphabet $X$ is a 7-tuple ${\cal M}$ =
 $<X, Z, Q,\delta, q_0, z_0,F>$ where $Z$ is the finite stack-alphabet, $Q$ is the finite set of 
states, $q_0 \in Q$ is the initial state,
$z_0$ is the initial stack-symbol, $F$ is a finite subset of $QZ^*$, the set of {\em final} 
configurations, and $\delta$, the transition function, is a mapping
 $\delta : QZ \times(X \cup \{\epsilon\}) \rightarrow {\cal P}_f(QZ^*)$. 

Let $q, q' \in Q, \omega, \omega' \in
Z^*, z \in Z, f \in X^*$ and $a \in X \cup \{\epsilon\}$ ;
 we note $(q z \omega ,af)\computdir{\cal M} (q' \omega' \omega, f)$ if
 $ q'\omega' \in \delta(qz,a)$.
$\computto{*}{\cal M}$  is the reflexive and transitive closure of
$\computdir{\cal M}$ .\\  

For every $q \omega , q' \omega'\in QZ^*$ and $f \in X^*$, we note $q \omega
\reduito{f}{\cal M}q' \omega'$ iff
$(q \omega , f)\computto{*}{\cal M}  (q' \omega',\epsilon)$.

${\cal M}$ is said {\em deterministic} iff, for every $z \in Z, q \in Q, x \in X$ :
\begin{equation}
\card (\delta (qz, \epsilon)) \in \{0,1\} 
\label{determinism0}
\end{equation}
\begin{equation}
\card (\delta (qz, \epsilon)) = 1 \Rightarrow
 \card (\delta(qz, x)) = 0,
\label{determinism1}
\end{equation}
\begin{equation}
\card(\delta (qz,\epsilon)) = 0 \Rightarrow
 \card(\delta (qz,x)) \leq 1.
\label{determinism2}
\end{equation}
 ${\cal M}$ is said {\em real-time} iff, for every $q \in Q, z \in Z$,
 Card$(\delta(qz, \epsilon)) = 0$.\\
A configuration $q\omega$ of ${\cal M}$ is said $\epsilon$-bound iff there exists a configuration
 $q' \omega'$ such that $(q \omega , \epsilon) \computdir{\cal M} (q' \omega' , \epsilon)$;
$q\omega$ is said $\epsilon$-free iff it is not $\epsilon$-bound.\\
A pda ${\cal M}$ is said {\em normalized} iff, it fufills conditions (\ref{determinism0}),
(\ref{determinism1}) (see above) and (\ref{normalization0}),(\ref{normalization1}),(\ref{normalization2}):
\begin{equation}
q_0 z_0 \mbox{ is } \epsilon-\mbox{free}
\label{normalization0}
\end{equation} 
 and for every $q\in Q, z \in Z, x \in X$: 
\begin{equation}
q'\omega'\in \delta(qz,x)  \Rightarrow \mid \omega' \mid \leq 2,
\label{normalization1}
\end{equation}
\begin{equation}
q'\omega'\in \delta(qz,\epsilon)  \Rightarrow \mid \omega' \mid = 0
\label{normalization2}
\end{equation}
All the pda considered here are assumed to fulfill condition (\ref{normalization0}).
A pda ${\cal M}$ will be said {\em bi-rooted} iff it fulfills (\ref{one_final state}) and
(\ref{birooted}):
\begin{equation}
\exists \bar{q} \in Q,  F = \{ \bar{q} \}\mbox{ and }
\label{one_final state}
\end{equation}
\begin{equation} 
\forall q \in Q, \omega \in Z^*, f \in X^*, 
q_0 z_0 \reduito{f}{\cal M} q \omega \Rightarrow \exists g \in X^*, 
q \omega \reduito{g}{\cal M} \bar{q}.
\label{birooted}
\end{equation}
The {\em language recognized} by ${\cal M}$ is
$$\Language({\cal M})=\{w \in X^* \mid \exists c\in F, 
q_0z_0 \stackrel{w}{\longrightarrow_{\cal M}}c\}.$$ 
It is a ``folklore'' result that , given a deterministic pda $\cal M$, one can effectively 
compute another dpda $\cal M'$ which is normalized and fulfills: 
$$\Language({\cal M}) = \Language({\cal M}')-\{\varepsilon\}.$$

\subsection{Graphs and pushdown automata}
\subsubsection{Equational graphs and pushdown automata}

We call {\em transition-graph} of a pda ${\cal M}$, denoted ${\cal T}({\cal M})$, the $0$-graph:\\ 
${\cal T}({\cal M}) = (V_{{\cal T}({\cal M})},E_{{\cal T}({\cal M})}) $ where
$ V_{{\cal T}({\cal M})} = \{ q \omega \mid q \in Q, \omega \in Z^*, q \omega \mbox { is }
\epsilon-\mbox{free}\}$ and 
\begin{equation}
 E_{{\cal T}({\cal A})} = \{ (c,x,c') \in V_{{\cal T}({\cal M})}\times V_{{\cal T}({\cal M})} \mid
c \reduito{x}{\cal M} c'\}.
\end{equation}
We call {\em computation 1-graph} of the pda ${\cal M}$, denoted $({\cal C}({\cal M}), 
v_{{\cal M}})$, the subgraph of ${\cal T}({\cal M})$ induced by the set of vertices which 
are accessible from the vertex $q_0 z_0$, together with the source $v_{{\cal M}}= q_0 z_0$.
In the case where ${\cal M}$ is bi-rooted, we call {\em computation 2-graph} of the pda ${\cal M}$, denoted $({\cal C}({\cal M}), 
v_{{\cal M}}, \bar{v}_{\cal M})$, the graph ${\cal C}({\cal M})$ defined just above, 
together with the sources $v_{{\cal M}}= q_0 z_0, \bar{v}_{\cal M}= \bar{q}$.
\begin{theo} Let $\Gamma= (\Gamma_0,v_0)$ be a rooted 1-graph over $X$. The following conditions are 
equivalent:
\begin{enumerate}
\item $\Gamma$ is equational and has finite out-degree.
\item $\Gamma$ is isomorphic to the computation 1-graph $({\cal C}({\cal M}), v_{{\cal M}})$ of some normalized
pushdown automaton ${\cal M}$.
\end{enumerate}
\label{1graph_and_pda}
\end{theo}
The formal proof of this theorem is quite technical and is omitted here.
(See the annex for a sketch of proof). 
\begin{cor} Let $\Gamma=(\Gamma_0,v_0,\bar{v})$ be a bi-rooted 2-graph over $X$. The following 
conditions are equivalent:
\begin{enumerate}
\item $\Gamma$ is equational and has finite out-degree.
\item $\Gamma$ is isomorphic to the computation 2-graph $({\cal C}({\cal M}), v_{{\cal M}},\bar{v}_{\cal M})$ 
of some bi-rooted normalized pushdown automaton ${\cal M}$.
\end{enumerate}
\label{2graph_and_pda}
\end{cor}
\subsubsection{Bisimulation for non-deterministic (versus deterministic) graphs}$\;\;$\\
\label{nondet_versus_det}
In this paragraph, we reduce the classical notion of {\em bisimulation} for equational graphs
to the notion of $\eta$-bisimulation  for {\em deterministic} equational graphs, where $\eta$ has 
been suitably choosen ( see definition \ref{def_eta_bisimulation}).\\
\begin{lem}
\label{reduction1}
Let $\Gamma_1$ be some rooted equational 1-graph over a finite alphabet $Y_1$ and let $\#$ be
a new letter $\# \notin Y_1$. Then one can construct an equational bi-rooted 2-graph
$\Gamma$ over the alphabet $Y = Y_1 \cup \{\#\}$ such that, 
\begin{enumerate}
\item $V_{\Gamma_1} \subseteq V_{\Gamma}$,
\item for every $v, v' \in  V_{\Gamma_1}$, ($v,v'$ are bisimilar in $\Gamma_1$) iff
($v,v'$ are bisimilar in $\Gamma$),
\item $\Gamma_1$ has finite out-degree iff  $\Gamma$ has finite out-degree.
\end{enumerate}
\end{lem}
\begin{sketch}
Let us define $\Gamma$ from $\Gamma_1$ by:
$$ V_{\Gamma} = V_{\Gamma_1} \cup \{\bar{v}\}, \;\;
E_{\Gamma} = E_{\Gamma_1} \cup \{(w,\#, \bar{v}) \mid w \in V_{\Gamma_1}\},\;\;
\Gamma = (\Gamma_1,\bar{v}),$$
where $\bar{v}$ is a new vertex $\bar{v} \notin V_{\Gamma_1}$.
One can easily check that $\Gamma$ is equational iff 
$\Gamma_1$ is equational and that , provided $\Gamma_1$ is rooted, $\Gamma$ is 
bi-rooted.  Points (1) and (3) of the lemma are clear.
One can check that the mapping ${\cal R} \mapsto {\cal R} \cup \{(\bar{v},\bar{v})\}$ is
a bijection from the set of all the bisimulations over $\Gamma_1$ ( i.e. from $\Gamma_1$ to
$\Gamma_1$) to the set of all the bisimulations over $\Gamma$. Hence point (2) is true.
\end{sketch}

Let us consider finite alphabets $X,Y$ , a length-preserving homomorphism 
$\psi: X^* \rightarrow Y^*$ and the s.r. morphism $\bar{\psi}= \psi \circ \psi^{-1} \subseteq 
X^* \times X^*$. A $n$-graph $\Gamma$ over $X$ will be said {\em $\bar{\psi}$-saturated} iff,
for every $v \in V_{\Gamma}$, for every $(x, x') \in \bar{\psi}$,
 $$ (\exists v_1 \in V_{\Gamma} , (v,x,v_1)\in E_{\Gamma})
\Leftrightarrow
(\exists v'_1 \in V_{\Gamma} , (v,x',v'_1)\in E_{\Gamma}).$$
\begin{lem}
Let $\Gamma_1$ be an equational bi-rooted 2-graph of finite out-degree over an  alphabet $Y$. 
One can construct a finite alphabet $X$, a surjective length-preserving homomorphism 
$\psi: X^* \rightarrow Y^*$ and
an equational, bi-rooted 2-graph  $\Gamma$ over the alphabet $X$, such that
\begin{enumerate}
\item $\Gamma$ is deterministic,
\item $\Gamma$ is $\bar{\psi}$-saturated,
\item $V_{\Gamma_1} = V_{\Gamma}$, 
\item ${\rm Id}: V_{\Gamma} \rightarrow V_{\Gamma_1}$ is a 
$\psi$-bisimulation from $\Gamma$ to $\Gamma_1$.
\end{enumerate}
\label{reduction2}
\end{lem}
\begin{sketch}
By lemma \ref{2graph_and_pda}, we can suppose that $\Gamma_1$ is the computation 2-graph 
$({\cal C}({\cal M}_1), v_{{\cal M}_1},\bar{v}_{{\cal M}_1})$ 
of some bi-rooted normalized pushdown automaton ${\cal M}_1=<Y, Z, Q,\delta_1, q_0, z_0,
\{\bar{q}\}>$.
Let us consider the following integers:
$$ \forall q \in Q, \forall z \in Z, \forall y \in Y,
t_1 (qz,y)= \card (\delta_1(qz,y)),\;\;
\bar{t_1} = \max\{ t_1 (qz,y) \mid q \in Q,z \in Z, y \in Y\}.$$
Let $X = Y \times [1,\bar{t_1}]$ and let $\psi: X \rightarrow Y$ be the first projection.\\
Let $\rho: QZ \times Y \times \bbbn \rightarrow QZ^*$ 
such that
$\dom(\rho) = \bigcup_{q \in Q, z \in Z, y \in Y} \{qz\}\times \{y\} \times [1,t_1 (qz,y)]$ and
$$\rho ( qz,y,\star): \{qz\}\times \{y\} \times [1,t_1 (qz,y)] \rightarrow \delta_1(qz,y)$$
is a bijection ( for every $q,z,y)$).
We then define  ${\cal M}=<X, Z, Q,\delta, q_0, z_0,\{\bar{q}\}>$ by:
for every $q \in Q, z \in Z, y \in Y, i \in [1,\bar{t_1}]$ 
$$ \delta(qz,\epsilon) = \delta_1(qz,\epsilon) \mbox{ if } qz \mbox{ is } \epsilon-\mbox{bound}.$$
$$ \delta(qz,(y,i)) = \{q' \omega'\} \mbox{ if } \rho(qz,y,i) = q' \omega' \mbox{ or }
(1 \leq t_1(qz,y) < i \leq \bar{t_1} \mbox {and } \rho(qz,y,1) = q' \omega').$$
The 2-graph  $\Gamma=({\cal C}({\cal M}), v_{{\cal M}},\bar{v}_{\cal M})$ fulfills the required 
properties. 
\end{sketch}
Let us remark that, by point (4) and by composition of $\eta$-bisimulations, for every
$v,v' \in V_{\Gamma}$, $v,v'$
are $\bar{\psi}$-bisimilar (w.r.t. $\Gamma$) iff $v,v'$ are bisimilar (w.r.t. $\Gamma_1$).
\subsection{Deterministic context-free grammars}
\label{grammars}
Let $\cal M$ be some deterministic pushdown automaton ( we suppose
here that $\cal M$ is normalized). 
The {\em variable} alphabet $V_{\cal M}$ associated to $\cal M$ is defined as:
$$V_{\cal M} = \{ [p,z,q] \mid p,q \in Q, z \in Z \}.$$ 
The {\em context-free} grammar $G_{\cal M} $ associated to $\cal M$ is then 
$$G_{\cal M} = < X,V_{\cal M},P_{\cal M} >$$
where\\
\noindent $P_{\cal M}$ is the set of all the pairs of one of the following forms:
\begin{equation}
([p,z,q], x [p', z_1,p''] [p'', z_2, q])
\end{equation} where  
$p,q,p',p'' \in Q, x \in X,  p'z_1z_2 \in \delta(pz, x)$
\begin{equation}
([p,z,q], x [p', z',q] )
\end{equation} where  
$p,q,p'\in Q, x \in X, p'z' \in \delta(pz, x)$
\begin{equation}
([p,z,q], a)
\end{equation}  where 
$p,q,\in Q, a \in X \cup \{ \epsilon\}, q \in \delta(pz,a)$.
$G_{\cal M}$ is a {\em strict-deterministic} grammar 
(see definition \ref{determinsitic_grammar} below) . 
A general theory of this class of  grammars is exposed in \cite{Har78} and used in \cite{Har79}.

\subsection{Free monoids acting on semi-rings}
\label{W*_action}
\subsubsection{Semi-ring $\bbbB\pa{W}$}
Let $({\sf B},+,\cdot,0,1)$ where ${\sf B} = \{0,1\}$ denote the semi-ring of ``booleans''. Let
$W$ be some alphabet.
By  $(\bbbB\pa{W},+,\cdot,\emptyset,\epsilon)$  we denote the semi-ring of 
{\em boolean series} over $W$:\\
the set $\bbbB\pa{W}$ is defined as $\bbbB^{W^*}$; the sum and product are defined as usual; 
each word $w \in W^*$ can be identified with the element of $\bbbB^{W^*}$ mapping the word $w$ 
on $1$ and every other word $w'\neq w$ on $0$; every boolean series $S \in \bbbB\pa{W}$ 
can then be written in a unique way as:
$$S = \Sigma_{w \in W^*} S_w \cdot w, $$
where, for every $w \in W^*$, $S_w \in \bbbB$.\\
The {\em support} of $S$ is the language
$$ \support(S)= \{w \in W^* \mid S_w \neq 0 \}.$$
In the particular case where the semi-ring of coefficients is $\bbbB$ ( which is the only 
case considered in this article) we sometimes identify the series $S$ with its support.
A series $S \in \bbbB\pa{W}$ is called a boolean {\em polynomial} over $W$ if and only if 
its support is {\em finite}. The set of all boolean polynomials over $W$ is denoted by 
$\bbbB \langle W \rangle$.\\
The usual ordering $\leq$ on $\bbbB$ extends to  $\bbbB\pa{W}$ by: 
$$S \leq S' \mbox {  iff  } \forall w \in W^*, S_w \leq S_w'.$$
We recall that for every $S \in \bbbB\pa{W}$, $S^*$ is the series defined by:
\begin{equation}
S^* = \sum_{0 \leq n}S^n.
\end{equation}

Given two alphabets $W,W'$,
a map $\psi: \bbbB\pa{W}\; \rightarrow \bbbB\pa{W'}$ is said {\em $\sigma$-additive} iff it fulfills: 
for every denumerable family $(S_i)_{i \in \bbbn}$ of elements of $\bbbB\pa{W}$,
\begin{equation}
\psi(\sum_{i \in \bbbn} S_i)= \sum_{i \in \bbbn}\psi(S_i).
\label{psi_sigma_additive}
\end{equation}
A map $\psi: \bbbB\pa{W}\; \rightarrow \bbbB\pa{W'}$ which is both a semi-ring homomorphism
and a $\sigma$-additive map  is usually called a {\em substitution}.

\subsubsection{Actions of monoids}
Given a semi-ring $(\bbbS,+,\cdot,0,1)$ and a monoid $(\bbbM,\cdot,1_M)$, a map
$\circ: \bbbS \times \bbbM \rightarrow \bbbS$ is called a {\em right-action} of the 
monoid $\bbbM$ 
over the semi-ring $\bbbS$ iff, for every $ S,T \in \bbbS, m,m' \in \bbbM$:
\begin{equation}
0 \circ m = 0, \;\;
S \circ 1_M = S,\;\;
(S + T) \circ m = (S \circ m) + (T \circ m)\;\;\mbox{ and }
S \circ ( m \cdot m') = (S \circ m ) \circ m'.
\label{monoid_action}
\end{equation}
In the particular case where $\bbbS=\bbbB\pa{W}$, $\circ$ is said to be a $\sigma$-right-action if it 
fulfills the additional property that, 
for every denumerable family $(S_i)_{i \in \bbbn}$ of elements of $\bbbS$ and $m \in \bbbM$:
\begin{equation}
(\sum_{i \in \bbbn} S_i) \circ m= \sum_{i \in \bbbn}(S_i \circ m).
\label{sigma_action}
\end{equation}

\subsubsection{The action of $W^*$ on $\bbbB\pa{W}$}
We recall the following classical $\sigma$-right-action  $\bullet$ of the monoid $W^*$ over 
the semi-ring $\bbbB\pa{W}$ : for all $S, S' \in \bbbB\pa{W}, u \in W^*$
$$ S \bullet u= S' \Leftrightarrow \forall w \in W^*, ( S'_w = S_{u\cdot w}),$$
(i.e. $ S \bullet u$ is the {\em left-quotient} of $S$ by $u$ , or the {\em residual} of $S$ 
by $u$ ).

For every $S \in \bbbB\pa{W}$ we denote by $\bbbQ(S)$ the set of residuals of $S$:
$$\bbbQ(S)= \{ S \bullet u \mid u \in W^* \} .$$

We recall that $S$ is said {\em rational} iff the set $\bbbQ(S)$ is {\em finite}.
We define the {\em norm} of a series $S \in \bbbB\pa{W}$, denoted $\|S\|$ by:
$$\|S\| = \card(\bbbQ(S)) \in \bbbn \cup \{\infty\}.$$
\subsubsection{The reduced grammar $G$}
\label{reduced_grammar}
The classical reduced and $\epsilon$-free  grammar associated with $G_{\cal M}$ is 
$G_0= < X,V_0,P_0 >$
where:
\begin{equation}
V_0= \{ v \in V_{\cal M} \mid \exists w \in X^+, v \reduito{*}{P_{\cal M}} w\},
\label{proper_alphabet}
\end{equation}
$$ \varphi_0: \bbbB\pa{V} \rightarrow \bbbB\pa{V_0}$$ is the
unique substitution such that, for every $v \in V$: \\
$$ \varphi_0 (v) = v\; (\mbox{ if }v \in V_0),\;\;
\varphi_0 (v) = \epsilon \;(\mbox{ if }v \reduito{*}{P_{\cal M}} \epsilon) ,\;\;
\varphi_0 (v) = \emptyset\; (\mbox{ otherwise }),$$
\begin{equation}
P_0= \{ (v,w') \in V_0 \times (X \cup V_0)^+ \mid v \in V_0, \exists w \in (X \cup V_{\cal M})^*,
(v,w) \in P_{\cal M},w' =\varphi_0(w)\}.
\label{proper_productions}
\end{equation}
$G_0$ is the {\em reduced } and {\em $\epsilon$-free }form of $G_{\cal M}$. It is well-known that, for all $v \in V_0$:
$$\exists w \in X^+, v \reduito{*}{P_0} w \mbox{ and }$$
$$\{w \in X^*, v \reduito{*}{P_{\cal M}} w \} =
\{w \in X^*, v \reduito{*}{P_0} w \}.$$
For technical reasons ( which will be made clear in section \ref{strategies_B0}),
we introduce an  alphabet of ``marked variables'' $\bar{V}_0$ together with a fixed bijection:
$v \mapsto \bar{v}$ from $V_0$ to $\bar{V}_0$. Let $V = V_0 \cup \bar{V}_0$.
We denote by $\rho_e$ (letter $e$
stands here for ``erasing the marks'') the litteral morphism $V^* \rightarrow V_0^*$ defined by:
for every $v \in V_0$,
$$\rho_e (v) = v,\;\;\rho_e (\bar{v}) = v.$$
Similarly, $\bar{\rho}_e$ is the litteral morphism $V^* \rightarrow \bar{V}_0^*$ defined by:
for every $v \in V_0$,
$$\bar{\rho}_e (v) = \bar{v},\;\;\bar{\rho}_e (\bar{v}) = \bar{v}.$$
We denote also by $\rho_e,\bar{\rho}_e$ the unique substitutions extending these monoid
homomorphisms.\\
At last, the grammar $G$ is defined by, 
$G= < X,V,P >$  where 
$$P = P_0 \cup \{ (\bar{\rho}_e (v),\bar{\rho}_e (w)\mid (v,w) \in P_0\}.$$
\label{marked_variables}
In other words, the rules of $G$ consist of the rules of the usual proper and reduced grammar
associated with ${\cal M}$ to-gether with their marked copies.\\

\subsubsection{The action of $X^*$ on $\bbbB\pa{V}$}
Let us fix now a deterministic (normalized) pda ${\cal M}$ and consider the associated grammar $G$. We define a  
$\sigma$-right-action $\odot$ of the monoid $X^*$ over the semi-ring $\bbbB\pa{V}$ by:
for every $v \in V,\beta \in V^*, x \in X$
\begin{equation}
 (v \cdot \beta) \odot x = (\sum_{(v, h) \in P} h \bullet x) \cdot \beta,\;\;
\label{odot1}
\end{equation}
\begin{equation}
\epsilon \odot x = \emptyset.\;\;  
\label{odot2}  
\end{equation}

Let us consider the unique substitution 
$\varphi: \bbbB\pa{V}\; \rightarrow \bbbB\pa{X}$ fulfilling: for every $v \in V$,
$$\varphi(v)= \{ u \in X^* \mid v \reduito{*}{P} u \}, $$
(in other words, $\varphi$ maps every subset $ L \subseteq V^* $ on the language generated by the 
grammar $G$ from the set of axioms $L$).
\begin{lem}
For every $S \in \bbbB\pa{V}, u \in X^*$,
$\varphi(S \odot u)= \varphi(S) \bullet u$ ( i.e. $\varphi$ is a morphism of right-actions).
\label{varphi_morphic}
\end{lem}
\begin{prooof}
Let $v \in V, \beta \in V^*, x \in X$.
Recall that $G$ is in Greibach normal form (i.e. $P \subseteq V \times X\cdot V^*$).
One can then check on formulas (\ref{odot1,odot2}) that:
$$ \varphi(\epsilon \odot x)= \varphi(\epsilon) \bullet x\mbox{ and }
\varphi((v \cdot \beta) \odot x) = \varphi(v \cdot \beta)\bullet x. $$
By induction on $|w|$, it follows that, $\forall w \in V^*$,
$$ \varphi(w \odot x)= \varphi(w) \bullet x.$$
By $\sigma$-additivity of $\varphi$, it follows that,$\forall S \in \bbbB\pa{V}$,
$$ \varphi(S \odot x)= \varphi(S) \bullet x.$$
By induction on $u$, it follows that,$\forall u \in X^*$,
$$ \varphi(S \odot u)= \varphi(S) \bullet u.$$
\end{prooof}

We denote by $\equiv$ the kernel of $\varphi$ i.e.: for every $S,T \in \bbbB\pa{V}$,\\
$$ S \equiv T \Leftrightarrow \varphi(S) = \varphi(T).$$


\section{Series and matrices}
\label{algebra}
\subsection{Deterministic series, vectors and matrices}
\label{series}
We introduce here a notion of {\em deterministic} series which, in the case of the alphabet 
$V$ associated to a dpda $\cal M$, generalizes the classical notion of {\em configuration} of
$\cal M$. The main advantage of this notion is that, unlike for configurations, we shall 
be able to define {\em nice algebraic operations} on these series (see, in particular, \S \ref{spaces}).
Let us consider a pair $(W,\smile)$ where $W$ is an alphabet and $\smile$ is an equivalence 
relation over $W$. We call $(W,\smile)$ a {\em structured} alphabet. The two examples we have 
in mind are:
\begin{itemize}
\item the case where $W=V_{\cal M}$, the variable alphabet associated to $\cal M$ and $[p,z,q] \smile 
 [p',z',q']$ iff $p=p'$ and $z=z'$  (see \cite{Har78})
\item the case where $W=X$, the terminal alphabet of $\cal M$ and $x \smile y$ holds for every 
$x,y \in X$  (see \cite{Har78}).
\end{itemize}
\subsubsection{Definitions}
\begin{defi}
Let $S \in \bbbB\pa{W}$. $S$ is said {\em left-deterministic} iff either
\begin{description}
\item{(1)} $S= \emptyset$ or  
\item{(2)} $S = \epsilon$ or
\item{(3)} $\exists i_0 \in [1,m], S_{i_0} \neq \emptyset$ and $\forall w, w' \in W^*$,
$$ S_w = S_{w'} = 1 \Rightarrow [ \exists A,A' \in W, w_1, w_1' \in W^*,
 A \smile A', w=A\cdot w_1 \mbox{ and } w'= A'\cdot w_1'].$$
\end{description}
\end{defi}
A left-deterministic series $S$ is said to have the type  $\emptyset$ 
(resp. $\epsilon$, $[A]_{\smile}$) if case (1) (resp. (2), (3)) occurs.
\begin{defi}
Let $S \in \bbbB\pa{W}$. $S$ is said {\em deterministic} iff, for every $u \in W^*$,
$S \bullet u$ is left-deterministic.
\end{defi}
This notion is the straighforward extension to the infinite case of
the notion of (finite) {\em set of associates} defined in \cite[definition 3.2 p. 188]{Har79}.\\
We denote by $\bbbD \bbbB\pa{W}$ the subset of deterministic boolean series over $W$.
Let us denote by $\bbbB_{n,m} \pa{W}$ the set of $(n,m)$-matrices with entries in the semi-ring
$\bbbB\pa{W}$. 
\begin{defi}
Let $m \in \bbbn, S \in \bbbB_{1,m}\pa{W}: S= ( S_1, \cdots,S_m)$. $S$ is said 
{\em left-deterministic} iff either
\begin{description} 
\item{(1)} $\forall i \in [1,m] , S_i= \emptyset$ or
\item{(2)}  $\exists i_0 \in [1,m], S_{i_0} = \epsilon \mbox{ and  } 
\forall i \neq i_0,  S_{i} = \emptyset$ or
\item{(3)} $\forall w, w' \in W^*, \forall i,j \in [1,m],
{(S_i)}_w = {(S_j)}_{w'} = 1 \Rightarrow [ \exists A,A' \in W, w_1, w_1' \in V^*,
A \smile A', w= A \cdot w_1 \mbox{ and } w'= A' \cdot w_1'].$
\end{description}
\end{defi}
A left-deterministic row-vector $S$ is said to have the type  $\emptyset$ 
(resp. $(\epsilon, i_0)$, $[A]_{\smile}$) if case (1) (resp. (2), (3)) occurs.

The right-action $\bullet$ on $\bbbB\pa{W}$ is extended componentwise to $\bbbB_{n,m}\pa{W}$:
for every  $S= (s_{i,j})$, $u \in W^*$, the matrix $T= S \bullet u$ is defined by
$$ t_{i,j} = s_{i,j} \bullet u.$$
The ordering $\leq$ on $\bbbB$ is also extended componentwise to $\bbbB_{n,m}\pa{W}$.

\begin{defi}
Let $S \in \bbbB_{1,m}\pa{W}$. $S$ is said {\em deterministic} iff, for every $u \in W^*$,
$S \bullet u$ is left-deterministic.
\end{defi}

We denote by $\bbbD \bbbB_{1,m}\pa{W}$ the subset of deterministic row-vectors of dimension $m$ 
over $\bbbB\pa{W}$.

\begin{defi}
Let $S \in \bbbB_{n,m}\pa{W}$. $S$ is said {\em deterministic} iff, for every $i \in [1,n]$,
$S _{i,.}$ is a deterministic row-vector.
\end{defi}

Let us notice first some easy facts about deterministic matrices.
\begin{fact}
Let $S \in \bbbD \bbbB\pa{W}$. For every $T \in \bbbB\pa{W}, u \in W^*$
\begin{description}
\item{(1)} $T \leq S \Rightarrow T \in \bbbD \bbbB\pa{W}$
\item{(2)} $T =S \bullet u \Rightarrow T \in \bbbD \bbbB\pa{W}$
\end{description}
\end{fact}
\subsubsection{Norm}
Let us generalize the classical definition of {\em rationality} of series in 
$\bbbB\pa{W}$ to matrices.
Given $M \in \bbbB_{n,m}\pa{W}$ we denote by $\bbbQ(M)$ the set of {\em residuals} of $M$:
$$\bbbQ(M)= \{ M \bullet u \mid u \in W^* \} .$$
Similarly, we denote by $\bbbQ_r(M)$ the set of {\em row-residuals} of $M$:
$$\bbbQ_r(M)= \bigcup_{1 \leq i \leq n} \bbbQ( M_{i,*}).$$
$M$ is said {\em rational} iff the set $\bbbQ(M)$ is finite. One can check that it is equivalent 
to the property that every coefficient $M_{i,j}$ is rational, or to the property that 
$\bbbQ_r(M)$ is finite. We denote by $\bbbR \bbbB_{n,m}\pa{W}$ 
(resp. $\bbbD \bbbR \bbbB_{n,m}\pa{W}$) the set of rational ( resp. deterministic, rational) 
matrices over  $\bbbB\pa{W}$. For every $M \in \bbbR \bbbB_{n,m}\pa{W}$, we define the norm of $M$ as:
$$ \|M\| = \card(\bbbQ_r(M)).$$
\subsubsection{Grammars}

\begin{defi}
\label{deterministic_grammar}
Let $G = < X,V,P>$ be a context-free grammar in Greibach normal form. 
$G$ is said {\em strict-deterministic}
iff there exists an equivalence relation $\smile$ over $V$ fulfilling the following condition:
for every $E \in V, x \in X$, if $(E_k)_{1 \leq k \leq m}$ is a bijection 
$[1,m] \rightarrow [E]_\smile$, and $H_k = \sum_{(E_k, h) \in P} h \bullet x$, then
$$ (H_1,H_2, \ldots, H_m) \mbox{ is a deterministic vector}.$$
Any equivalence $\smile$ satisfying the above condition is said to be a {\em strict equivalence}
for the grammar $G$.
\end{defi}
This definition is a reformulation of \cite[Definition 11.4.1 p.347]{Har78}
adapted to the case of a Greibach normal-form.
\begin{theo}
Let $G_1 = < X,V_1,P_1>$ be a strict-deterministic grammar.
Then its reduced form $G_0 = < X,V_0,P_0>$, as defined in 
formulas (\ref{proper_alphabet},
\ref{proper_productions}), is strict-deterministic too. Moreover, if 
$\smile$ is a strict equivalence for $G_1$, its restriction over $V_0$ is a strict 
equivalence for $G_0$.
\label{proper_grammar}
\end{theo}
The proof would consist in slightly extending the proof of \cite[Theorem 11.4.1 p.350]{Har78}.\\
It is known that, given a dpda ${\cal M}$, its associated grammar $G_{\cal M}$ is 
strict-deterministic. By theorem \ref{proper_grammar} $G_0$ is strict-deterministic too.
Let us consider the minimal strict equivalence ${\smile}$ for $G_0$ and extend
it to $V$ by, $\forall v,v' \in V_0$:
$$
\bar{v} \smile \bar{v}' \Leftrightarrow v \smile v';\;\;\;\;\bar{v} \not\smile v'.
$$
Then $\smile$ is a strict equivalence for $G$ (the grammar $G$ is defined in \S \ref{marked_variables}). 
This ensures that $G$ is strict-deterministic.
\subsubsection{Residuals}
\begin{lem}
Let $S \in \bbbD \bbbB\pa{W}, T \in \bbbB\pa{W}, u \in W^*$.
If $S \bullet u \neq \emptyset$ then $ (S \cdot T)\bullet u = (S \bullet u ) \cdot T$.
\label{residual_case1}
\end{lem}

\begin{prooof}
Let $S \in \bbbD \bbbB\pa{W}, T \in \bbbB\pa{W}, u \in W^*$, such that $S \bullet u \neq \emptyset$.
Let $u',u'' \in W^*$ such that $u = u' \cdot u'', u'' \neq \epsilon$ and let $w \in \supp(S)$.
If $w \bullet u' = \epsilon$ then $S \bullet u' = \epsilon$ ( because $S \bullet u'$ is 
left-deterministic), hence $S \bullet u = \epsilon \bullet u''=\emptyset$, which would
contradict the hypothesis. It follows that
$$\forall u' \prec u, \forall w \in \supp(S), w \bullet u' \neq \epsilon.$$
Hence 
$$\forall w_1 \in \supp(S), \forall w_2 \in \supp(T),
 (w_1 \cdot w_2) \bullet u = (w_1 \bullet u) \cdot w_2.$$
This proves that $ (S \cdot T)\bullet u = (S \bullet u ) \cdot T$.
\end{prooof}

\begin{lem}
Let $S \in \bbbD \bbbB\pa{W}, T \in \bbbB\pa{W}, u \in W^*$ and $U = S \cdot T$. Exactly one of 
the following cases is true:
\begin{description}
\item{(1)} $S \bullet u \neq \emptyset;\\ 
\mbox{ in this case } U  \bullet u = (S \bullet u ) \cdot T$.
\item{(2)} $S \bullet u = \emptyset, \exists u', u'',  u=u' \cdot u'',
S \bullet u' = \epsilon;\\ 
\mbox{ in this case } U \bullet u = T \bullet u'' $.
\item{(3)} $S \bullet u = \emptyset, \forall u'\preceq u,
S \bullet u' \neq \epsilon ;\\
\mbox{ in this case } U \bullet u = \emptyset = (S \bullet u ) \cdot T$.
\label{residual_product}
\end{description}
\end{lem}
\begin{prooof}
Clearly, one of the hypotheses (1-3) must occur. Let us examine each one of these cases.\\
In case (1), by lemma \ref{residual_case1}, $U  \bullet u = (S \bullet u ) \cdot T$.\\
In case (2), $U \bullet u= (U \bullet u') \bullet u''$ and by case (1), $U \bullet u'=
(S \bullet u' ) \cdot T$. It follows that
$U \bullet u= T \bullet u''$.\\
In case (3), if $S=\emptyset$, the conclusion of the lemma is clearly true. Let us suppose
now that $S \neq \emptyset$ and let $u' \prec u$ be the maximum prefix of $u$ such that
$S \bullet u' \neq \emptyset$. Then, there exist some $A \in W, u'' \in W^*$such that  
$ u = u'\cdot A \cdot u''$ 
and there exist some $B_1,\cdots, B_q  \in W, S_1, \cdots,S_q \in \bbbB\pa{W}- \{\emptyset\}$ 
such that
$S \bullet u' = \sum_{1 \leq i \leq q} B_q \cdot S_q$ and  
$B_1 \smile \cdots \smile B_i \smile \cdots \smile B_q$( because $S \bullet u'$ is left-deterministic).
By maximality of $u'$, $A$ does not belong to $\{ B_1,\cdots, B_q\}$, hence
$$U \bullet u = ((\sum_{1 \leq i \leq q} B_i \cdot S_i \cdot T) \bullet A) \bullet u'' =
\emptyset \bullet u'' = \emptyset.$$
\end{prooof}

\begin{lem}
Let $S \in \bbbD \bbbB_{1,m}\pa{W}, T \in \bbbB_{m,1}\pa{W}, u \in W^*$ and $U = S \cdot T$. Exactly one of the following cases is true:
\begin{description}
\item{(1)} $\exists j, S_j \bullet u \not\in \{\emptyset,\epsilon\};\\ 
\mbox{ in this case  } U  \bullet u = (S \bullet u ) \cdot T$.
\item{(2)} $\exists j_0, \exists u', u'',  u=u' \cdot u'',
S_{j_0} \bullet u' = \epsilon;\\
 \mbox{ in this case } U \bullet u = T_{j_0} \bullet u'' $.
\item{(3)} $\forall j, S_j \bullet u = \emptyset, \forall u' \preceq u, S_j \bullet u' \neq 
\epsilon;\\
\mbox{ in this case } U \bullet u = \emptyset = (S \bullet u ) \cdot T$.
\label{residual_linear_combination}
\end{description}
\end{lem}
\begin{prooof}
Let us note $S= (S_j)_{1 \leq j \leq m}, T= (T_j)_{1 \leq j \leq m}$. 
Clearly, one of the hypotheses (1-3) must occur. Let us examine each one of these cases.\\
In case (1), every 3-tuple $(S_j, T_j, u)$ fulfills case (1) or (3) of lemma 
\ref{residual_product}, hence $ (S_j \cdot T_j) \bullet u = (S_j \bullet u) \cdot T_j$.
Hence
$$ U  \bullet u = \sum_{1 \leq j \leq m} (S_j \cdot T_j) \bullet u
= \sum_{1 \leq j \leq m}(S_j \bullet u) \cdot T_j
= (S \bullet u) \cdot T.$$
In case (2), $ S \bullet u'$ must be left-deterministic of type $(\epsilon, j_0)$ , hence
$ \forall j \neq j_0, S_j \bullet u'= \emptyset$. It follows that
$$ U \bullet u = T_{j_0} \bullet u''.$$
In case (3), every 3-tuple $(S_j, T_j, u)$ fulfills case (3) of lemma 
\ref{residual_product}, hence $ (S_j \cdot T_j) \bullet u = \emptyset = 
(S_j \bullet u) \cdot T_j$. It follows that
$$U \bullet u = \emptyset = (S \bullet u ) \cdot T.$$
\end{prooof}
\begin{lem}
Let $S \in \bbbD \bbbB_{1,m}\pa{W}, T \in \bbbB_{m,s}\pa{W}, u \in W^*$ and $U = S \cdot T$. Exactly 
one of the following cases is true:
\begin{description}
\item{(1)} $\exists j, S_j \bullet u \not\in \{\emptyset,\epsilon\}\\ 
\mbox{ in this case  } U  \bullet u = (S \bullet u ) \cdot T$.
\item{(2)} $\exists j_0, \exists u', u'',  u=u' \cdot u'',
S_{j_0} \bullet u' = \epsilon;\\
 \mbox{ in this case } U \bullet u = T_{j_0} \bullet u'' $.
\item{(3)} $\forall j, \forall u' \preceq u, S_j \bullet u = \emptyset, S_j \bullet u' \neq 
\epsilon;\\
\mbox{ in this case } U \bullet u = \emptyset = (S \bullet u ) \cdot T$.
\label{residual_linear_combination_vectors}
\end{description}
\end{lem}
\begin{prooof}
Let us notice that for every $k \in [1,s]$:
\begin{equation}
U_k = S \cdot T_{\ast,k},
\end{equation}
and that the hypothesis of the 3 cases considered in lemma \ref{residual_linear_combination} 
depend on the vector $S$ and the word $u$ only ( but not on the integer $k \in [1,s])$.
In case (1), by lemma \ref{residual_linear_combination},$\forall k \in [1,s]$
$$U_k \bullet u = (S \bullet u) \cdot T_{\ast,k},$$
hence $U \bullet u =  (S \bullet u) \cdot T.$
Cases 2,3 can be treated in the same way.
\end{prooof}
\begin{lem}
For every $S \in \bbbB_{n,m}\pa{W}, T \in \bbbB_{m,s}\pa{W}$, if $S$ and $T$ are both
left-deterministic, then $S \cdot T$ is left-deterministic.
\label{product_is_left_deterministic} 
\end{lem}
\begin{lem}
For every $S \in \bbbD \bbbB_{n,m}\pa{W}, T \in \bbbD \bbbB_{m,s}\pa{W}, 
S \cdot T \in \bbbD \bbbB_{n,s}\pa{W}$.
\label{product_is_deterministic} 
\end{lem}
\begin{prooof}
As the notion of deterministic matrix is defined row by row, it is sufficient to prove
this lemma in the particular case where $n=1$. Let us note $U = S \cdot T$. 
Let $u \in W^*$. Let us show that $U \bullet u$ is left-deterministic.
Let us consider every one of the 3 cases considered in 
lemma \ref{residual_linear_combination_vectors} . 
In case (1) or (3), 
$$U \bullet u = (S \bullet u ) \cdot T,$$
and in case (2),
$$U \bullet u = T \bullet u''.$$
In both cases, by lemma \ref{product_is_left_deterministic}, $U \bullet u$ is left-deterministic.
\end{prooof}
 
\begin{lem}
Let $A \in \bbbD \bbbB_{n,m}\pa{W}, B \in \bbbB_{m,s}\pa{W}$. Then
$\|A \cdot B \| \leq \|A\| + \|B\|$.
\label{norm_of_product}
\end{lem}
\begin{prooof}
Let $A = (a_{i,k}), B=(b_{k,j}), C=A \cdot B, C=(c_{i,j}).$ 
Let $1 \leq i \leq n, H \in \bbbQ(C_{i,*})$. Let $u \in W^*$ such that
$$ H = C_{i,\ast} \bullet u = (A_{i,\ast}\cdot B) \bullet u.$$
We apply lemma \ref{residual_linear_combination_vectors} to $S= A_{i,*}$ and $T= B$. 
If case (1) or (3) of lemma \ref{residual_linear_combination_vectors} is realized then
$$
H = (A_{i,*}\bullet u) \cdot B.
$$
If case (2) of lemma \ref{residual_linear_combination_vectors} is realized then
$$ H = B_{k_0,*} \bullet u''.$$
The number of residuals $H$ obtained by case (1) is less or equal than $\|A\|$ and
the number obtained by case (2) is less or equal than $\|B\|$. This proves the inequality.
\end{prooof}

\subsubsection{W=V}
Let $(W,\smile)$ be the structured alphabet $(V, \smile)$ associated with 
$\cal M$ and let us consider a bijective numbering of the elements of  $Q$:
$(q_1,q_2,\ldots,q_{n_Q})$.
Let us define here handful notations for some particular vectors or matrices.
Let us use the {\em Kronecker symbol} $\delta_{i,j}$ meaning $\epsilon $ if $i=j$ and $\emptyset$
if $i \neq j$.
For every $ 1 \leq n, 1 \leq i \leq n$, we define the row-vector $\epsilon_i^n$ as:
$$\epsilon_i^n = (\epsilon_{i,j}^n)_{1 \leq j \leq n} \mbox { where } 
\forall j , \epsilon_{i,j}^n = \delta_{i,j}.
$$
We call {\em unit row-vector} any vector of the form $\epsilon_i^n$.\\
For every $1 \leq n$, we denote by $\emptyset^n \in \bbbD \bbbB_{1,n}\pa{V}$ the row-vector:
$$\emptyset^n=(\emptyset, \ldots,\emptyset).$$

For every $\omega \in Z^*, p, q \in Q$, $[p \omega q]$ is the deterministic series
defined inductively by:
$$[p \epsilon q] = \emptyset \mbox{ if } p \neq q , [p \epsilon q] = \epsilon \mbox{ if } p= q,$$
$$[p \omega q] = \sum_{r \in Q}[p,z,r]\cdot [r \omega' q]  \mbox{ if } \omega = z \cdot \omega'
\mbox{ for some } z \in Z, \omega' \in Z^*.$$
Let us define
\begin{eqnarray}
K_0 &=& \ma\{ \| (E_1,E_2, \ldots,E_n) \odot x\|\mid (E_i)_{1 \leq i \leq n} 
\mbox{ is a bijective numbering } \nonumber\\
&& \mbox{ of some class in } V/\smile, x \in X\}.
\label{K0}
\end{eqnarray} 
\begin{lem}
For every $S \in \bbbD \bbbB_{1,\lambda}\pa{V}, u \in X^*$,
\begin{description}
\item{(1)} $S \odot u \in \bbbD \bbbB_{1,\lambda}\pa{V}$
\item{(2)} $\|S \odot u\| \leq \|S\| + K_0 \cdot |u|.$
\end{description}
\label{norm_of_action}
\end{lem}
\begin{prooof}
We treat first the case where $u$ is just a letter.\\
Let $S \in \bbbD \bbbB_{1,\lambda}\pa{V}$ and $x \in X$.
If $S = \emptyset^\lambda$ or $ S= \varepsilon ^\lambda_i$ ( for some $i \in [1,\lambda]$),
then $S \odot x = \emptyset^\lambda$  and points (1)(2) are both true.\\
Otherwise
$$S = \sum_{k=1}^q E_k \cdot \Phi_k$$
for some $q \in \bbbn, \Phi_k \in \bbbD \bbbB_{1,\lambda}\pa{V}, (E_k)_{1 \leq k \leq q} $ bijective
numbering of some class of $V/\smile$.\\
By equation (\ref{odot1}) , which defines the right-action $\odot$,
$$S \odot x = \sum_{k=1}^q (E_k \odot x) \cdot \Phi_k,$$
hence $S \odot x$ has the form $H \cdot \Phi$ where $H \in \bbbD \bbbB_{1,q}\pa{V}$ 
(see definition \ref{deterministic_grammar}),
$\|H\| \leq K_0$ (see inequation (\ref{K0})) and $\Phi \in \bbbD \bbbB_{q,\lambda}\pa{V}$.\\
By lemma \ref{product_is_deterministic}, $H \cdot \Phi$ is deterministic and by lemma \ref{norm_of_product} 
$\| H \cdot \Phi\| \leq \| \Phi\| + K_0 $ .
As every $\Phi_k \in {\sf Q}_r(S)$ we obtain:
$$\| H \cdot \Phi\| \leq \| \Phi\| + K_0 \leq \|S\| + K_0.$$
Both points (1)(2) are proved.\\
The general case where $u$ is any word of $X^*$ can be deduced by indiuction on $|u|$ 
from this particular case.
\end{prooof}
\begin{lem}
Let $ \lambda \in \bbbn-\{0\},S \in \bbbD \bbbR \bbbB_{1,\lambda}\pa{V}, u \in X^*$. One of 
the three following cases must occur:
\begin{description}
\item{(1)} $S \odot u = \emptyset^\lambda,$
\item{(2)} $S \odot u = \epsilon^\lambda_j$ for some $j \in [1,\lambda]$ ,
\item{(3)} $\exists u_1, u_2 \in X^*, v_1 \in V^*, q \in \bbbn, E_1,\ldots, E_k,\ldots,E_q \in V, 
\Phi \in \bbbD \bbbR \bbbB_{q,\lambda}\pa{V}$ such that
$$ u = u_1 \cdot u_2, S \odot u_1 = S \bullet v_1 = \sum_{k=1}^q E_k \cdot \Phi_k,\;\;
 S \odot u = \sum_{k=1}^q (E_k \odot u_2) \cdot \Phi_k,\mbox{ and }
$$
$$\forall k \in [1,q], E_k \smile E_1, E_k \odot u_2 \notin \{ \epsilon, \emptyset\}.$$
\end{description}
\label{bullet_and_odot}
\end{lem}
\begin{prooof}
Let $u \in X^*$. Let us prove the lemma by induction on $|u|$.\\
$ u = \epsilon$:\\ 
if $S \in {\emptyset^\lambda} \cup \{\epsilon^\lambda_j| 1 \leq j \leq \lambda\}$
then clearly the conclusion of case (1) or (2) is realized.
Otherwise, as $S$ is left-deterministic, $S$ has a decomposition as 
$S = \sum_{k=1}^q E_k \cdot  \Phi_k$ 
such that the conclusion of case (3) is realized
with $ u_1 = u_2 = \epsilon$,$v_1 = \epsilon$, the given integer $q$  and the letters 
$E_1 \smile \ldots \smile E_q \in V$.\\
$ u = u_0 \cdot a, a \in X $:\\
Let us consider the $u_1, u_2 , v_1 , q , (E_k)_{1 \leq k \leq q} , (\Phi_k)_{1 \leq k \leq q}$ 
given by the induction hypothesis 
on $u_0$. 
$$ (S \odot u) \odot a = (\sum_{k=1}^q(E_k \odot u_2) \cdot \Phi_k) \odot a \mbox{  and }$$
$$ \forall k \in [1,q], \|E_k \odot u_2 \| \geq 3.$$
{\bf case 1}: $\forall k \in [1,q], \|E_k \odot u_2a \| \geq 3$.\\
Then $S \odot ua = \sum_{k=1}^q (E_k \odot u_2a) \cdot \Phi_k$. 
Hence conclusion (3) of the lemma is fulfilled by
$u'_1 = u_1,u'_2 = u_2 a, v'_1 = v_1, q' = q, E'_k = E_k, \Phi'_k = \Phi_k$. \\
{\bf case 2}: $\exists r \in [1,q], \|E_r \odot u_2a \| = 2 $.\\
In other words: there exists some $r \in [1,q]$ such that $E_r \odot u_2a= \epsilon$, hence
$$ S \odot ua = \Phi_r.$$
{\bf subcase 1}: $\Phi_{r} \in \{\emptyset^\lambda\} \cup 
\{\epsilon^\lambda_j | 1 \leq j \leq \lambda\}$.\\
Conclusion (1) or (2) of the lemma is then realized.\\
{\bf subcase 2}: $\Phi_{r}= \sum_{\ell=1}^{r'} F_\ell \cdot \Psi_\ell$ for some $r' \in \bbbn, 
 F_1 \smile \ldots F_{r'} \in V, \Psi \in \bbbD \bbbR \bbbB_{r',\lambda}\pa{V}$.\\
Then
$$S \odot ua = \sum_{\ell=1}^{r'} F_\ell \cdot \Psi_\ell ; \;\;
S \bullet(v_1 E_r)= \Phi_r= \sum_{\ell=1}^{r'} F_\ell \cdot \Psi_\ell.$$
Conclusion (3) of the lemma is then realized by $u'_1 = ua, u'_2 = \epsilon, 
v'_1 = v_1 E_r, q' = r', E'_k = F_k, \Phi' = \Psi$.\\
{\bf case 3}: $\forall k \in [1,q], \|E_k \odot u_2a \| = 1 $.\\
This means that $E \odot u_2a = \emptyset^q$, hence that case (1) is realized.
\end{prooof}

We give now an adaptation of lemma \ref{residual_linear_combination_vectors} to the 
action $\odot$ in place of $\bullet$.
\begin{lem}
Let $S \in \bbbD \bbbB_{1,m}\pa{V}, T \in \bbbB_{m,s}\pa{V}, u \in X^*$ and 
$U = S \cdot T$.Exactly one of the following cases is true:
\begin{description}
\item{(1)} $S \odot u \not\in \{\emptyset^m\} \cup \{\epsilon^m_j | 1 \leq j \leq m \}\\ 
\mbox{ in this case  } U  \odot u = (S \odot u ) \cdot T$.
\item{(2)} $\exists j_0, \exists u', u'',  u=u' \cdot u'',
S \odot u' = \epsilon^s_{j_0};\\
\mbox{ in this case } U \odot u = T_{j_0} \odot u'' $.
\item{(3)} $\forall j, \forall u' \preceq u, S \odot u = \emptyset^m$ and $S \odot u' \neq 
\epsilon^m_j;\\
\mbox{ in this case } U \odot u = \emptyset^s = (S \odot u ) \cdot T$.
\label{RA_odot}
\end{description}
\end{lem}
\begin{prooof}
The arguments used in the proofs of lemma \ref{residual_case1}, \ref{residual_product}, 
\ref{residual_linear_combination}, \ref{residual_linear_combination_vectors} can be adapted to 
$\odot$ in place of $\bullet$. The only non-trivial adaptation is that of lines 6-7
of the proof of lemma \ref{residual_case1}: let us suppose that $u \in X^*$ is such 
that
\begin{equation}
\forall u' \prec u, S \odot u' \neq \epsilon,
\label{hypo_neq_epsilon}
\end{equation}
and let us prove that
\begin{equation}
(S \cdot T) \odot u = (S \odot u) \cdot T.
\label{conlusion_neq_epsilon}
\end{equation}
We prove by induction on $|u|$ that (\ref{hypo_neq_epsilon}) implies (\ref{conlusion_neq_epsilon})
.\\
$|u| = 0$: by definition of a right-action, $\forall S' \in \bbbD \bbbB \pa{V}, S' \odot \epsilon = S'$.
Hence conclusion (\ref{conlusion_neq_epsilon}) is true.\\
$u = u_0 \cdot a$, where $ u_0 \in X^*, a \in X$:\\
Hypothesis (\ref{hypo_neq_epsilon}) is fulfilled by $u_0$ too, hence, by induction hypothesis,
\begin{equation}
(S \cdot T) \odot u_0 = (S \odot u_0)\cdot T. 
\label{induction_hypo_u0}
\end{equation}
If $S \odot u_0 = \emptyset$, then , by the above equality $(S \cdot T) \odot u_0 =
\emptyset$ too, hence 
$$(S \cdot T) \odot u_0a = \emptyset = (S \odot u_0a)\cdot T, $$
hence (\ref{conlusion_neq_epsilon}) is true.\\
Otherwise, by hypothesis (\ref{hypo_neq_epsilon}) $ S \odot u_0 \not\in \{\emptyset,\epsilon\}$, hence there exists
$q \in \bbbn, E_1 \smile \ldots \smile E_q \in V, \Phi \in \bbbD \bbbB_{m,s}\pa{V}$ such that
\begin{equation}
S \odot u_0 = \sum_{k=1}^q E_k \cdot \Phi_k.
\label{decomposition_Su0}
\end{equation}
By definition (\ref{odot1}) and the fact that $\odot$ is a $\sigma$-action:
$$ (E_k \cdot \Phi_k ) \odot a = (E_k \odot a) \cdot \Phi_k ,$$
hence, by $\sigma$-additivity ,  
$$(\sum _{k=1}^q E_k \cdot \Phi_k ) \odot a = \sum_{k=1}^q (E_k \odot a) \cdot \Phi_k $$
and by product by $T$:
\begin{equation}
(S \odot u_0a)\cdot T = \sum_{k=1}^q (E_k \odot a)  \cdot \Phi_k \cdot T.
\label{decomposition_S_u_T}
\end{equation}
Let us examine now $(ST) \odot u_0a$.
By (\ref{induction_hypo_u0}):
\begin{equation}
(S \cdot T) \odot u_0 = \sum_{k=1}^q E_k \cdot \Phi_k \cdot T.
\label{decomposition_ST_u0}
\end{equation}
By definition (\ref{odot1}) and the fact that $\odot$ is a $\sigma$-action:
$$ (E_k \cdot \Phi_k \cdot T) \odot a = (E_k \odot a) \cdot \Phi_k \cdot T,$$
hence, by $\sigma$-additivity ,  
$$(\sum _{k=1}^q E_k \cdot \Phi_k \cdot T) \odot a = 
\sum_{k=1}^q (E_k \odot a) \cdot \Phi_k \cdot T $$
Using (\ref{decomposition_ST_u0}) this last equality can be read:
\begin{equation}
(ST) \odot u_0a  = \sum_{k=1}^q (E_k \odot a)  \cdot \Phi_k \cdot T.
\label{decomposition_ST_u}
\end{equation}
As equalities (\ref{decomposition_ST_u}),(\ref{decomposition_S_u_T}) 
have the same righthand-side, we conclude
that (\ref{conlusion_neq_epsilon}) is true.
\end{prooof}

\subsubsection{Marks}
A word $w \in V^*$ is said {\em marked} iff $w \in V^* \cdot \bar{V}_0 \cdot V^*$; it is said
{\em fully marked} iff $w \in \bar{V}_0^*$.\\
A series $S \in \bbbB\pa{V}$ is said  {\em marked} iff $\exists w \in \supp(S)$, $w$ is marked; 
it is said {\em fully marked} iff $\forall w \in \supp(S)$, $w$ is fully marked. It is said {\em unmarked} iff it is {\em not} marked.
A matrix $S \in \bbbB_{m,n}\pa{V}$
is said marked (resp. fully marked, unmarked) iff, for every $i \in [1,m]$, 
the series $\sum_{j=1}^n S_{i,j}$ is marked 
(resp. fully marked, unmarked).\\
\begin{defi}
Let $d \in \bbbn$. A vector $S \in \bbbD \bbbB_{1,\lambda}\pa{V}$ is said $d$-marked iff
there exists $q \in \bbbn, \alpha \in \bbbD \bbbR \bbbB_{1,q}\langle V \rangle, \Phi \in
\bbbD \bbbR \bbbB_{q,\lambda}\pa{V}$ such that
$$S = \sum_{k=1}^q \alpha_k \cdot \Phi_k \mbox{ and }
\|\alpha\| \leq d,$$
and $\Phi$ is unmarked.
\end{defi}
\begin{lem}
For every $S \in \bbbD \bbbB_{1,\lambda}\pa{V}$ 
\begin{description}
\item{(1)} $\rho_e(S) \in \bbbD \bbbB_{1,\lambda}\pa{V}$
\item{(2)} $\|\rho_e(S)\| \leq \|S\|.$
\end{description}
\label{rhoe_is_nice}
\end{lem}
\begin{sketch}$\;\;\;$\\
(1)-Let us notice that the homomorphism 
$\rho_e: V^* \rightarrow V^*$ preserves the equivalence $\smile$:
for every $v,v' \in V$, if $v \smile v'$ then $\rho_e(v) \smile \rho_e(v')$. It follows that 
the corresponding substitution $\rho_e$ preserves determinism.\\
(2)-Let $S \in \bbbD \bbbB_{1,\lambda}\pa{V}$. For every $v \in V_0$
$$\rho_e(S) \bullet v = \rho_e(S \bullet v) \mbox{ or } \rho_e(S) \bullet v = \rho_e(S \bullet \bar{v})$$
according to the fact that the leftmost letters of the monomials of $S$ are in $[v]_\smile$ or in
$[\bar{v}]_\smile$; both formulas are true when $S$ is null or is a unit.\\
By induction on the length, it follows that, for every $w \in V_0^*$, there exists $w' \in V^*$ such that:
$$ \rho_e(w') = w \mbox{ and } \rho_e(S)\bullet w = \rho_e(S \bullet w').$$
Moreover, for every $w \in V^* \bar{V_0} V^*$,
$$\rho_e(S)\bullet w= \emptyset^\lambda ,$$
but in this case too, there exists some $w' \in V^*$ such that
$\rho_e(S)\bullet w = \rho_e(S \bullet w').$\\
The map $T \mapsto \rho_e(T)$ is then a surjective map from $\bbbQ(S)$ onto $\bbbQ(\rho_e(S))$,
which proves that $\|\rho_e(S)\|\leq \|S\|.$
\end{sketch}

\subsubsection{Operations on row-vectors}$\;\;$\\
Let us introduce two new operations on row-vectors and prove some technical lemmas about them.\\
Given $ A, B \in \bbbB_{1,m}\pa{W}$ and $1 \leq j_0 \leq m $ we define the vector
$C = A \nabla_{j_0} B$ as follows:\\ 
if $A= ( a_1, \ldots,a_j, \ldots, a_m), B= ( b_1, \ldots, b_j, \ldots, b_m)$ then 
$C= ( c_1, \ldots, c_j, \ldots, c_m)$ where
$$c_j = a_j + a_{j_0} \cdot b_j \mbox{ if }j \neq j_0 \;,\;\; c_j= \emptyset \mbox{ if }j = j_0.$$
\begin{lem}
Let $A, B \in \bbbB_{1,m}\pa{W}$ and $1 \leq j_0 \leq m $.
\begin{enumerate}
\item if $A,B$ are left-deterministic, then $A \nabla_{j_0} B$ is left-deterministic.
\item if $A,B$ are deterministic, then $A \nabla_{j_0} B$ is deterministic.
\item if $A,B$ are deterministic, then $\|A \nabla_{j_0} B\| \leq \|A\|+ \|B\|.$
\end{enumerate}
\label{Boxj}
\end{lem}
\begin{prooof} $\;$\\
Let $C= A \nabla_{j_0} B$.\\
\noindent{\bf 1} Let us prove first that if $A,B$ are both left-deterministic, then $C$ is 
left-deterministic too.\\
If $A$ is left-deterministic of type $[pz]$ , then $C$ is 
left-deterministic of the same type.\\
If $A$ is left-deterministic of type $(\epsilon, j_1)$ with $j_1 \neq j_0$, then $C = A$, hence
$C$ is left-deterministic.\\
If $A$ is left-deterministic of type $(\epsilon, j_0)$, then $C \leq B$, hence $C$ is 
left-deterministic.\\
If $A$ is left-deterministic of type $(\emptyset)$, then $C= \emptyset$, hence $C$ is 
left-deterministic.\\
\noindent{\bf 2} Let us suppose now that $A$ is deterministic and let us examine a residual 
$C\bullet u$ , for some $u \in W^*$.
Lemma \ref{residual_product} applies on $S= a_{j_0}$ and $T= b_j$ for every $j \neq j_0$.
But the case of the lemma fulfilled by $(S,T_j,u)$ depends on $(S,u)$ only.\\
Suppose $a_{j_0}\bullet u \neq \emptyset$ ( case 1); in this case 
\begin{equation} 
C \bullet u = (A \bullet u) \nabla_{j_0} B
\label{AuB}
\end{equation}
Suppose $a_{j_0} \bullet u = \emptyset, \exists u', u'',  u=u' \cdot u'',
a_{j_0} \bullet u' = \epsilon$ ( case 2); in this case
\begin{equation}
C \bullet u = <(B \bullet u'')| \emptyset_{j_0}^m> 
\label{Bu}
\end{equation}
where ${\emptyset}_{j_0}^m$ is the row vector ${\epsilon}_{j_0}^m$ in which $ \emptyset$ and
$\epsilon$ have been exchanged and $<*,*>$ is the ``scalar product'' defined by
$<S,T> = \sum^{m}_{j=1} S_j \cdot T_j$.\\
Suppose $a_{j_0} \bullet u = \emptyset, \forall u'\preceq u, a_{j_0} \bullet u' \neq \epsilon$ 
( case 3);
in this case, equation (\ref{AuB}) is true again. When equation (\ref{AuB}) is true, $C \bullet u$
is left-deterministic by part (1) of this proof, and when equation (\ref{Bu}) is true, 
$C \bullet u$ is left-deterministic because $B$ is assumed deterministic. We have proved that 
$C \in \bbbD \bbbB_{1,m}\pa{W}$.\\
\noindent{\bf 3}
The number of residuals of the form (\ref{AuB}) is bounded above by $\|A\|$ and 
the number of residuals of the form (\ref{Bu}) is bounded above by $\|B\|$. Hence
$\|C\| \leq \|A\| + \|B\|.$
\end{prooof}
Given $ A \in \bbbD \bbbB_{1,m}\pa{W}$ and $1 \leq j_0 \leq m $ we define the vector
$A' = \nabla_{j_0}^* (A)$ as follows:\\
if $A= ( a_1, \ldots, a_j, \ldots, a_m)$ then 
$A'= ( a'_1, \ldots, a'_j, \ldots, a'_m)$ where
$$a'_j = a_{j_0}^* \cdot a_j \mbox{ if }j \neq j_0 \;,\;\; a'_j= \emptyset \mbox{ if }j = j_0.$$
\begin{lem}
Let $A \in \bbbD \bbbB_{1,m}\pa{W}$ and $1 \leq j_0 \leq m $.\\
Then $\nabla_{j_0}^*(A) \in \bbbD \bbbB_{1,m}\pa{W}$ and $\|\nabla_{j_0}^* (A)\| \leq \|A\|.$
\label{Boxj*}
\end{lem}
\begin{prooof} Let us examine a residual $A' \bullet u$ , for some $u \in W^*$.
Let $u' = {\rm max} \{ v \preceq u \mid v \in a_{j_0}^* \}$. Let $u''\in W^*$ such that
$u = u' \cdot u''$. One can check that for every $S, T \in \bbbB\pa{W}$
$$ (S \cdot T) \bullet u = (S \bullet u) \cdot T + 
\sum_{\stackrel{u = u_1 \cdot u_2,}{ \epsilon \in S \bullet u_1}} a_j \bullet u_2.$$
Applying this formula to $S= a_{j_0}^*$ and $T= a_j $, with $j \neq j_0$ we obtain
\begin{equation}
a_j'\bullet u = (a_{j_0}^* \bullet u) \cdot a_j  + 
\sum_{\stackrel{u = u_1 \cdot u_2,}{\epsilon \in a_{j_0}^* \bullet u_1}} a_j \bullet u_2.
\label{residual_a'}
\end{equation}
As $a_{j_0}$ is deterministic, one can check that 
$$a_{j_0}^* \bullet u= (a_{j_0} \bullet u'') \cdot a_{j_0}^*.$$
As $A$ is deterministic, if $u_2$ has some prefix $u_2'$ in $a_{j_0}$, then 
$a_j \bullet u_2' = \emptyset$ so that $a_j \bullet u_2 = \emptyset$.
Hence 
$$\sum_{\stackrel{u = u_1 \cdot u_2,}{\epsilon \in a_{j_0}^* \bullet u_1}} a_j \bullet u_2 
= a_j \bullet u''.$$
Plugging the two last equations into (\ref{residual_a'}) we obtain
$$
a_j'\bullet u = (a_{j_0} \bullet u'') \cdot a_{j_0}^* \cdot a_j  + a_j \bullet u''\;
(\mbox{ for } j \neq j_0), \;\;
\mbox{ and }a_{j}'\bullet u = \emptyset\; (\mbox{ for } j = j_0)
\label{residual_a'_simplified}
$$
which can be rewritten as
\begin{equation}
A' \bullet u = ( A \bullet u'') \nabla_{j_0} A'
\label{residual_a'_simplified}
\end{equation}

\noindent Let us show that $A'$ is left-deterministic. 
If $A$ is left-deterministic of type $[pz]$ , then $A'$ is 
left-deterministic of the same type.\\
If $A$ is left-deterministic of type $(\epsilon, j_1)$ with $j_1 \neq j_0$, then $A' = A$
 ( notice that $\emptyset^* = \epsilon$), hence $A'$ is left-deterministic.\\ 
If $A$ is left-deterministic of type $(\epsilon, j_0)$ or $(\emptyset)$, then $A'= \emptyset$, 
hence $A'$ is left-deterministic.\\
By point (1) of lemma \ref{Boxj}, the fact that $A \bullet u''$ and $A'$ are both
left-deterministic implies that $( A \bullet u'') \nabla_{j_0} A'$ is left-deterministic too.
By formula (\ref{residual_a'_simplified}), $A' \bullet u$ is left-deterministic. We have 
proved that $A' \in \bbbD \bbbB_{1,m}\pa{W}$.\\ 
Moreover, by formula (\ref{residual_a'_simplified}), $\card(\bbbQ(A')) \leq \card(\bbbQ(A))$, i.e.
$\|A'\| \leq \|A\|$.
\end{prooof}

\subsection{Bisimulation of series}
\label{sigma_bisimulation}
Up to the end of this section, we consider the structured alphabet $V$ associated with a dpda 
$\cal M$ over $X$ . We suppose
a s.r. morphism $\eta \subseteq X^* \times X^*$ is given (see definition \ref{sr_morphism}).
\subsubsection{Series, words and graphs}
Let us give first a slight adaptation of definition \ref{def_gamma_bisimulation} to the $n$-graph 
$(\bbbD \bbbR \bbbB_{1,n}\pa{V},\odot, (\epsilon^n_i)_{1 \leq i \leq n})$. 
\begin{defi}
Let ${\cal R}$ be some binary relation ${\cal R} \subseteq \bbbD \bbbR \bbbB_{1,n}\pa{V} \times
\bbbD \bbbR \bbbB_{1,n}\pa{V}$. ${\cal R}$ is a $\sigma-\eta$-bisimulation iff
\begin{enumerate}
\item $\forall (S,S') \in {\cal R}, \forall x \in X$,
$$\exists x' \in \eta(x), (S \odot x , S' \odot x') \in {\cal R} \mbox { and }
  \exists x'' \in \eta^{-1}(x), (S \odot x'' , S' \odot x) \in {\cal R},$$
\item $\forall (S,S') \in {\cal R}, \forall i \in [1,n], (S = \epsilon^n_i
\Leftrightarrow S' = \epsilon^n_i).$
\end{enumerate}
\label{def_sigma_bisimulation}
\end{defi}
We denote by $S \sim S'$ the fact that there exists some $\sigma-\eta$-bisimulation $\cal R$
such that $(S,S') \in \cal R$. One can notice that $\sim$ is the greatest 
$\sigma-\eta$-bisimulation
( with respect to the inclusion ordering) over $\bbbD \bbbR \bbbB_{1,n}\pa{V}$.
The $\sigma$-bisimulation relations can be conveniently expressed in terms of 
{\em word}-bisimulations.
\begin{defi}
Let $S,S' \in \bbbD \bbbR \bbbB_{1,n}\pa{V}$ and ${\cal R} \subseteq X^* \times X^*$.
${\cal R}$ is a $w-\eta$-bisimulation with respect to $(S,S')$ iff ${\cal R} \subseteq \eta$ and
\begin{description}
\item[(1) totality: ] $\dom({\cal R}) = X^*, \im({\cal R}) = X^*,$
\item[(2) extension: ] $\forall (u,u') \in {\cal R}, \forall x \in X ,$ 
$$\exists x' \in \eta(x),
(u \cdot x, u' \cdot x') \in {\cal R} \mbox{ and }
\exists x'' \in \eta^{-1}(x),
(u \cdot x'', u' \cdot x) \in {\cal R}.$$
\item[(3) coherence: ] $\forall (u,u') \in {\cal R},\forall i \in [1,n], (S \odot u = \epsilon^n_i)
\Leftrightarrow (S' \odot u' = \epsilon^n_i),$
\item[(4) prefix: ] $\forall (u,u') \in X^* \times X^*, \forall (x,x') \in X \times X,
(u \cdot x, u' \cdot x') \in {\cal R} \Rightarrow (u , u') \in {\cal R}$.
\end{description}
\label{def_word_bisimulation}
\end{defi}
(Condition (1) can be equivalently replaced by ``$(\epsilon,\epsilon) \in {\cal R}$''.)
$\cal R$ is said to be a $w-\eta$-bisimulation of {\em order $m$}  with respect to $(S,S')$ iff it fulfills conditions
(3-4) above and the modified conditions\\
{\bf (1')}: $\dom({\cal R}) = X^{\leq m}, \im({\cal R}) = X^{\leq m},$\\
{\bf (2')}: $\forall (u,u') \in {\cal R} \cap (X^{\leq m-1} \times X^{\leq m-1}), \forall x \in X ,$ 
$$\exists x' \in \eta(x),
(u \cdot x, u' \cdot x') \in {\cal R} \mbox{ and }
\exists x'' \in \eta^{-1}(x),
(u \cdot x'', u' \cdot x) \in {\cal R}.$$
The $w-\eta$-bisimulations are also called $w-\eta$-bisimulations of {\em order $\infty$}.
The two next lemmas are relating the notions of $w-\eta$-bisimulation ( on words), 
$\sigma-\eta$-bisimulation ( on series), and  $\eta$-bisimulation (on the vertices of the computation 2-graph of ${\cal M}$).

\begin{lem}
Let $S,S' \in \bbbD \bbbR \bbbB_{1,n}\pa{V}$. The following properties are equivalent:
\begin{description}
\item[(i)] $S \sim S'$
\item[(ii)] there exists ${\cal R} \subseteq X^* \times X^*$ which is a $w-\eta$-bisimulation
w.r.t. $(S,S')$
\item[(iii)] $\forall m \in \bbbn$, there exists  ${\cal R}_m \subseteq 
X^{\leq m} \times X^{\leq m}$ which is a $w-\eta$-bisimulation of order $m$ w.r.t. $(S,S')$.
\end{description}
\label{sigma_reducesto_w}
\end{lem}
\begin{prooof}$\;\;$\\
{\bf (i) $\Rightarrow$ (iii):}
Suppose that ${\cal S}$ is a $\sigma$-$\eta$-bisimulation w.r.t. $(S,S')$.
Let us prove by induction on the integer $m$, the following property ${\cal P}(m)$:\\
$ \exists {\cal R}_m , w-\eta-\mbox{bisimulation of order } m 
\mbox{ w.r.t. } (S,S') \mbox{ such that } $
\begin{equation}
\forall (u,u') \in {\cal R}_m, 
(S \odot u , S' \odot u' ) \in {\cal S}.
\label{inclusion}
\end{equation}
{\bf m=0: } Let ${\cal R}_0 = \{(\epsilon,\epsilon)\}$. ${\cal R}_0$ clearly fulfills points
(1'),(2'),(4) of the above definition. Moreover, as $(S,S')\in {\cal S}$ where ${\cal S}$
fulfills condition (2) of definition \ref{def_sigma_bisimulation}, ${\cal R}_0$ fulfills
point (3) of definition \ref{def_word_bisimulation}.\\
{\bf m=m'+1: } Let ${\cal R}_{m'}$ be some $w-\eta$-bisimulation of order $m'$ w.r.t. $(S,S')$.
Let us define  ${\cal R}_{m} = {\cal R}_{m'} \cup \{ (u \cdot x, u' \cdot x') \mid
(u,u') \in {\cal R}_{m'}, (S \odot u x, S' \odot u' x') \in {\cal S} \mbox{ and }
(x,x') \in \eta \}$. Property (1) of ${\cal S}$ and property (1') of ${\cal R}_{m'}$ imply that
\begin{equation}
\dom({\cal R}_m) = X^{\leq m}, \im({\cal R}_m) = X^{\leq m}.
\label{is_total}
\end{equation}
Property (1) of ${\cal S}$ and property (2') of ${\cal R}_{m'}$ imply that 
$\forall (u,u') \in {\cal R}_m \cap (X^{\leq m-1} \times X^{\leq m-1}), \forall x \in X ,$
\begin{equation}
\exists x' \in \eta(x),
(u \cdot x, u' \cdot x') \in {\cal R}_m \mbox{ and }
\exists x'' \in \eta^{-1}(x),
(u \cdot x'', u' \cdot x) \in {\cal R}_m.
\label{does_extend}
\end{equation}
Property (2) of ${\cal S}$ and property (3) of ${\cal R}_{m'}$ imply that
\begin{equation}
\forall (u,u') \in {\cal R}_m,\forall i \in [1,n], (S \odot u = \epsilon^n_i)
\Leftrightarrow (S' \odot u' = \epsilon^n_i).
\label{is_coherent}
\end{equation}
Property (4) of ${\cal R}_{m'}$ and the definition of ${\cal R}_{m}$ imply that
\begin{equation}
\forall (u,u') \in X^* \times X^*, \forall (x,x') \in X \times X,
(u \cdot x, u' \cdot x') \in {\cal R}_m \Rightarrow (u , u') \in {\cal R}_m.
\label{is_prefix}
\end{equation}
Property (\ref{inclusion}) for ${\cal R}_{m'}$ and the definition of ${\cal R}_{m}$ imply that
(\ref{inclusion}) is fulfilled by ${\cal R}_{m}$ too.
Equations (\ref{is_total},\ref{does_extend},\ref{is_coherent},\ref{is_prefix}) prove that
${\cal R}_m$ is a $w$-$\eta$-bisimulation of order  $m$ w.r.t.  $(S,S')$, hence 
${\cal P}(m)$ is proved.\\
{\bf (iii) $\Rightarrow$ (ii):} Let us notice that, as the alphabet $X$ is finite,
for every $w$-$\eta$-bisimulation ${\cal R}$ of order  $m$ w.r.t.  $(S,S')$, 
$$\card\{ {\cal R}' \subseteq X^* \times X^* \mid {\cal R} \subseteq {\cal R}'\mbox{ and }
{\cal R}' \mbox{ is a } w-\eta-\mbox{bisimulation of order } m+1 \mbox{ w.r.t. }  (S,S')\}
< \infty.$$
Hence, by Koenig's lemma, if (iii) is true, then there exists an infinite sequence
$({\cal R}_m)_{m \in \bbbn}$ such that for every $m \in \bbbn$,
${\cal R}_m$ is a $w-\eta-\mbox{bisimulation of order } m \mbox{ w.r.t. } (S,S')$ and
${\cal R}_m \subseteq {\cal R}_{m+1}$. Let us define then
$$ {\cal R}= \bigcup_{m \geq 0} {\cal R}_m.$$
${\cal R}$ is a $w-\eta-\mbox{bisimulation of order } \infty$ w.r.t. $(S,S')$.\\
{\bf (ii) $\Rightarrow$ (i):} Let ${\cal R}$ be a $w-\eta-\mbox{bisimulation of order } \infty$
w.r.t. $(S,S')$. Let us define a relation ${\cal S}$ by: 
$${\cal S} = \{ (S \odot u, S' \odot u') \mid (u,u') \in {\cal R} \}.$$ 
The totality property of ${\cal R}$ implies that $(S,S') \in {\cal S}$.
The extension
property of ${\cal R}$ implies that ${\cal S}$ fulfills condition (1) of definition \ref{def_sigma_bisimulation} and the coherence property of ${\cal R}$ implies ${\cal S}$ fulfills condition (2).
\end{prooof}
Lemma \ref{sigma_reducesto_w} leads naturally to the following
\begin{defi}
Let $\lambda \in \bbbn-\{0\}, S,S' \in {\sf D R B}_{1,\lambda} \pa{V}$. We define the
{\em divergence} between $S$ and $S'$ as:
$$\Div(S, S') = \borninf \{ n \in \bbbn \mid {\cal B}_n(S,S') = \emptyset\}.$$
(It is understood that  $\borninf(\emptyset) = \infty$).
\label{divergence}
\end{defi}
Let us suppose that the dpda ${\cal M} = <X, Z, Q,\delta, q_0, z_0,\{\bar{q}\}>$ is 
normalized and bi-rooted. Let $\psi: X^* \rightarrow Y^*$ be a monoid homomorphism such
that $\psi(X) \subseteq Y$ and let $\bar{\psi} = \psi \circ \psi^{-1}$
( $\bar{\psi}$, the kernel of $\psi$, is a s.r. morphism which is also
 an equivalence relation; this additional property will be used in the sequel).
Let $\Gamma$ be the computation 2-graph of ${\cal M}$ and let us suppose $\Gamma$ is
$\bar{\psi}$-saturated.\\
Let $\theta: V_{\Gamma} \rightarrow \bbbD \bbbR \bbbB \pa{V}$ the mapping defined by:
$\forall q \in Q, \forall \omega \in Z^*$, such that $q \omega \in V_{\Gamma}$,
$$\theta (q \omega) = \varphi_0([q \omega \bar{q}]).$$

For every $q \omega \in V_{\Gamma}, S \in \bbbD \bbbR \bbbB \pa{V}$ we also define:
$$
\Language(q \omega)=\{ u \in X^*, q \omega \reduito{u}{\Gamma}\bar{q}, \};\;\;
\Language(S)=\{ u \in X^*, S \odot u= \epsilon\}.
$$
\begin{lem}
For every $q \omega \in V_{\Gamma},
\Language(q \omega)= \Language(\varphi_0([q \omega \bar{q}]))$.
\label{language_equivalence}
\end{lem}
This lemma follows from the classical result that the language recognized 
by ${\cal M}$ with starting congiguration $q \omega$ and final configuration
$\bar q$ is exactly the language generated by $G_{\cal M}$ from the polynomial $[q \omega \bar{q}]$ which, in turn, is 
equal to the language generated by $G_0$ 
from the polynomial $\varphi_0([q \omega \bar{q}])$. At last, $G$ and $G_0$ 
generate the same language from any given polynomial over $V_0$.
\begin{lem}
Let  
$v,v'$ be vertices of  $\Gamma$. Then
$v \sim v'$, in the sense of definition \ref{def_gamma_bisimulation}  iff  
$\theta (v) \sim \theta(v')$, in the sense of definition \ref{def_sigma_bisimulation}.
\label{gamma_reducesto_sigma}

\end{lem}
\begin{prooof}
In this proof we denote by $\odot_\Gamma$ the right-action of $X^*$ over
$V_\Gamma \cup \{ \bot \}$ defined by: for every $v,v \in V_\Gamma, u \in X^*$,
\begin{eqnarray*}
v \odot_\Gamma u &=& v' \mbox{ if } v \reduito{u}{\Gamma} v',\\
 v \odot_\Gamma u &=& \bot \mbox{ if } \mbox{ there is no } v' , 
\mbox{ such that } v \reduito{u}{\Gamma} v'\\
\bot \odot_\Gamma u&=& \bot.
\end{eqnarray*}
1-Let us suppose that $(v,v') \in {\cal R}$, where ${\cal R}$
is some $\bar{\psi}$-bisimulation over $\Gamma$.\\
Let ${\cal S}=\{ (\theta (v) \odot u, \theta (v') \odot u') \mid
(u,u') \in \bar{\psi}, (v \odot_\Gamma u, v' \odot_\Gamma u') \in {\cal R}\} 
\cup \{(\emptyset,\emptyset)\}$.
Let us show that ${\cal S}$ is a $\sigma-\bar{\psi}$-bisimulation.\\
Let us consider some pair of series in ${\cal S}$. If the given pair is $(\emptyset,\emptyset)$, points (1)(2) of definition \ref{def_sigma_bisimulation} are clearly fulfilled.\\
Otherwise, it has the form 
$(\theta(v) \odot u, \theta (v') \odot u')$, where $(u,u') \in \bar{\psi}$ and
$(v \odot_\Gamma u, v' \odot_\Gamma u') \in {\cal R}$ .\\
1.1-Let $x \in X$.\\
{\bf case 1.1.1}: $\theta (v) \odot ux \neq \emptyset$. \\
$$\Language(\theta (v) \odot ux) \neq \emptyset$$ (because the grammar $G$ is reduced)
, hence , using lemma \ref{language_equivalence}, 
$$\Language(v \odot_\Gamma ux)= \Language(v) \bullet ux
= \Language(\theta (v))\bullet ux \neq \emptyset.$$
It follows that
$$v \odot_\Gamma ux \neq \bot.
$$
As ${\cal R}$ is a $\bar{\psi}$-simulation, there must exists some 
$x' \in \bar{\psi}(x)$ such that
$$ (v \odot_\Gamma ux, v' \odot_\Gamma u'x') \in {\cal R}.$$
Hence 
$$ (\theta(v )\odot ux, \theta(v') \odot u'x') \in {\cal S}.$$ 
{\bf case 1.1.2}: $\theta (v) \odot ux = \emptyset$.\\
In this case, by lemma \ref{language_equivalence} and the fact that $\Gamma$ is bi-rooted,
$v \odot_\Gamma ux$ must be equal to $\bot$. As $\Gamma$  is $\bar{\psi}$-saturated, it follows that 
$$\forall x' \in \bar{\psi}(x), v \odot_\Gamma ux'= \bot$$
As ${\cal R}^{-1}$ is a $\bar{\psi}^{-1}$-simulation, it must also be true that
$$\forall x' \in \bar{\psi}(x), v' \odot_\Gamma u'x'= \bot$$
choosing some particular $x' \in \bar{\psi}(x)$, and using again lemma \ref{language_equivalence} we obtain:
$$\theta(v') \odot u'x'= \emptyset$$
In both cases, as $v,v'$ are playing symmetric roles,  property (1) of definition \ref{def_sigma_bisimulation} has been verified.
If the starting pair in ${\cal S}$ is $(\emptyset,\emptyset)$, property (1)
is again verified.\\
1.2- Let us suppose that $\theta(v) \odot u=\epsilon$.\\
This means that 
$$\Language(\theta(v)) \bullet u= \epsilon,$$ hence, using lemma \ref{language_equivalence} that 
$$\Language(v \odot_\Gamma u) = \epsilon,$$
hence 
$$v \odot_\Gamma u= \bar{q}.$$
As $\Gamma$ is bi-rooted, $\bar{q}$ is the only vertex having no outgoing edge
(see \S \ref{graphs}).
As ${\cal R}$ is a $\bar{\psi}$-bisimulation, $v' \odot_\Gamma u'$ we must
also have no outgoing edge,hence
$$v' \odot_\Gamma u'= \bar{q},$$
and by the same arguments, used backwards now,
$$\Language(\theta(v')) \bullet u'= \epsilon,$$
which, as the grammar $G$ is proper and reduced, implies
$$\theta(v') \odot u'= \epsilon.$$
As $(v,v')$ are playing symmetric roles, property (2) of 
definition \ref{def_sigma_bisimulation} has been verified.\\
2-Let us suppose that $(\theta(v),\theta(v')) \in {\cal S}$, where ${\cal S}$
is some $\sigma-\bar{\psi}$-bisimulation.\\
Let ${\cal R}=\{(v \odot_\Gamma u, v' \odot_\Gamma u'\mid
(u,u') \in \bar{\psi},(\theta(v) \odot u, \theta(v') \odot u') \in {\cal S}-\{(\emptyset,\emptyset)\}\} \cup \{ (c,c) \mid c \in V_\Gamma\}$. We show that
${\cal R}$ is a $\bar{\psi}$-bisimulation over $\Gamma$.\\
2.1-Using lemma \ref{language_equivalence}, we obtain:
$$
\theta(v) \odot u \neq \emptyset \Rightarrow 
v \odot_\Gamma u \neq \bot.
$$
Hence 
$$\dom({\cal R})\subseteq V_{\Gamma}.$$ 
Conversely, due to the term $\{ (c,c) \mid c \in V_\Gamma\}$,
$$\dom({\cal R})\supseteq V_{\Gamma}.$$
At end, point (1) of definition \ref{def_gamma_bisimulation} is fulfilled.\\
2.2-Due to the
term $\{ (c,c) \mid c \in V_\Gamma\}$, point (2) of definition \ref{def_gamma_bisimulation} is fulfilled.\\
2.3-Let us consider some pair of configurations in ${\cal R}$. It must have the form 
$(v \odot_\Gamma u, v' \odot_\Gamma u')$, where $(u,u') \in \bar{\psi}$ and
$(\theta(v) \odot u, \theta(v') \odot u') \in {\cal S}-
\{(\emptyset,\emptyset)\}$.\\
By the same arguments as in case 1.1.1 above, one can show that,
for every $x\in X$, such that 
$$v \odot_\Gamma ux \neq \bot,$$ there exists
some $x' \in \bar{\psi}(x)$ such that
$$v' \odot_\Gamma u'x' \neq \bot.$$
Hence ${\cal R}$ fulfills the three points of definition \ref{def_gamma_bisimulation}. By same means, ${\cal R}^{-1}$ fulfills them too, so that ${\cal R}$
is a $\bar{\psi}$-bisimulation over the graph $\Gamma$.
\end{prooof}
\subsubsection{Extension to matrices}$\;\;$\\
Let $\delta,\lambda \in \bbbn-\{0\}$.
We extend the binary relation $\sim$ from vectors in $\bbbD \bbbR \bbbB_{1,\lambda}\pa{V}$ to
matrices in $\bbbD \bbbR \bbbB_{\delta,\lambda}\pa{V}$ as follows:
for every $T,T' \in \bbbD \bbbR \bbbB_{\delta,\lambda}\pa{V}$,
\begin{equation}
T \sim T' \Leftrightarrow \forall i \in [1,\delta], T_{i,\ast} \sim T'_{i,\ast}.
\label{sim_for_matrices}
\end{equation}
We call {\em w-$\eta$-bisimulation} of order $n \in \bbbn \cup \{\infty\}$ with respect to 
$(T,T')$ every 
$${\cal R} = ({\cal R}_i)_{i \in [1,\delta]} \mbox{ such that }
\forall i \in [1,\delta], {\cal R}_i \in {\cal B}_n(T_{i,\ast}, T'_{i,\ast}).
$$
We denote by ${\cal B}_n(T, T')$ the set of w-$\eta$-bisimulations of order $n$ 
w.r.t. $(T,T')$ .\\
Some algebraic properties of this extended relation $\sim$ will be established in corollary
\ref{algebraic_version}.
\subsubsection{Operations on w-bisimulations}$\;\;$\\
\label{operations_on_wbisimulations}
The following operations on word-$\bar{\psi}$-bisimulations turn out to be useful.\\
{\bf right-product}:\\
Let $\delta, \lambda \in \bbbn-\{0\},  S, S' \in \bbbD \bbbR \bbbB_{1,\delta}\pa{V},
T \in \bbbD \bbbR \bbbB_{\delta,\lambda}\pa{V}$. For every $n \in \bbbn \cup\{\infty\}$ and
${\cal R} \in {\cal B}_n(S,S')$ we define:
\begin{eqnarray}
<S |{\cal R}> &=& [\{ (u,u') \in {\cal R} \mid \forall v \preceq u, \forall i \in [1,\delta],
 S \odot v \neq \epsilon^{\delta}_i\}\\
 & \cup &
\{ (u \cdot w, u' \cdot w) \mid (u,u') \in {\cal R}, w\in X^*, \exists i \in [1,\delta],
S \odot u = \epsilon^{\delta}_i\}] \cap X^{\leq n} \times X^{\leq n}.\nonumber\\
&\;\;&
\label{right_quotient_R}
\end{eqnarray} 
One can check that $<S |{\cal R}> \in {\cal B}_n(S \cdot T, S'\cdot T)$.\\
{\bf left-product}:\\
Let $\delta, \lambda \in \bbbn-\{0\},  S \in \bbbD \bbbR \bbbB_{1,\delta}\pa{V},
T, T' \in \bbbD \bbbR \bbbB_{\delta,\lambda}\pa{V}$. For every $n \in \bbbn \cup \{\infty\}$ and
${\cal R} \in {\cal B}_n(T,T')$ we define:
\begin{eqnarray}
<S,{\cal R}> &=& [\{ (u,u) \mid u \in X^*, \forall v \preceq u, \forall i \in [1, \delta],
 S \odot v \neq \epsilon^{\delta}_i\}\\ 
& \cup &
\{ (u \cdot w, u \cdot w') \mid u \in X^*, \exists i \in [1,\delta],
S \odot u = \epsilon ^{\delta}_i, (w,w') \in {\cal R}_{i}\}]
\cap X^{\leq n} \times X^{\leq n} \nonumber\\
&\;\;&
\label{left_product_R}
\end{eqnarray}
One can check that $<S ,{\cal R}> \in {\cal B}_n(S \cdot T, S \cdot T')$.\\
{\bf star}:\\
Let $\lambda \in \bbbn-\{0\}, S_1 \in \bbbD \bbbR \bbbB_{1,1}\pa{V}, S_1 \neq \epsilon, (S_1,S) \in \bbbD \bbbR \bbbB_{1,\lambda+1}\pa{V}, T \in \bbbD \bbbR \bbbB_{1,\lambda}\pa{V}$. For every $n \in 
\bbbn \cup \{\infty\}$ and ${\cal R} \in {\cal B}_n (S_1 \cdot T+ S, T)$ we define:
\begin{eqnarray}
{\cal R}_{0} &=&{\cal R}\\
{\cal S}_{0} &=& \left( \begin{array}{c}
{\cal R}_{0}\\ \vdots \\ {\cal R}_{0} 
\end{array} \right)
\end{eqnarray}
\begin{eqnarray}
\forall k \geq 0, {\cal R}_{k+1} &=& <(S_1,S),{\cal S}_{k}> \circ {\cal R}_{0}\\
{\cal S}_{k} &=& \left( \begin{array}{c}
{\cal R}_{k}\\ \vdots \\ {\cal R}_{k} 
\end{array} \right)
\end{eqnarray}
and finally 
\begin{equation}
{\cal R}^{<S_1,*>} = \bigcup_{k \geq 0} {\cal R}_{k} \cap X^{\leq k} \times X^{\leq k}. 
\label{star_R}
\end{equation}
One can check that, for every $k \geq 0$:
\begin{eqnarray}
{\cal R}_k & \in & {\cal B}_n ( S_1^{k+1} + \sum^k_{i=0}S_1^i \cdot S,T)\\
{\cal S}_k & \in &         {\cal B}_n ( 
\left( \begin{array}{c}
S_1^{k+1} + \sum^k_{i=0}S_1^i \cdot S \\ {\rm I}_\lambda  
\end{array} \right),
\left( \begin{array}{c}
T\\ {\rm I}_\lambda  
\end{array} \right)
),
\end{eqnarray}
and finally ${\cal R}^{<S_1,*>} \in {\cal B}_n (S^{*}_1 \cdot S, T)$.
\begin{rem} In fact operations could be more adequately defined on ``pointed'' w-bisimulations,
i.e. on binary relations with sets of ``terminal pairs of words'' of type $i \in [1,\delta]$ 
corresponding to the pairs $(u,u')$ such that $S \odot u = \epsilon^\delta_i, S' \odot u' = \epsilon^\delta_i$.  The two different external operations $<S,{\cal R}>, <S|{\cal R}>$ could then be 
replaced by only one binary operation $<{\cal R}_1,{\cal R}_2>$ over ``pointed'' w-bisimulations.  
\end{rem}
\subsection{Deterministic spaces}
\label{spaces}
We adapt here the key-idea of \cite{Mei89,Mei92} to bisimulation of vectors.

\subsubsection{Definitions}
Let $(W,\sim)$ be some structured alphabet. 
A vector $U = \sum^{n}_{i = 1} \gamma_i \cdot U_i$  where $\vec{\gamma} \in 
\bbbD \bbbR \bbbB_{1,n}\pa{W}$, $ U_i \in \bbbD \bbbR \bbbB_{1,\lambda}\pa{W}$ 
is called a {\em linear combination} of the $U_i$'s. We call {\em deterministic space} 
of rational vectors ( d-space for short) any subset $\bbbV$ of 
$\bbbD \bbbR \bbbB_{1,\lambda}\pa{W}$ which is closed under finite linear combinations. 
Given any set  ${\cal G}=\{U_i | i \in I\} \subseteq \bbbD \bbbR \bbbB_{1,\lambda}\pa{W}$, one can check that the set $\bbbV$ of
 all (finite) linear combinations of elements of $\cal G$ is a d-space 
(by lemma \ref{product_is_deterministic}) and that it is the smallest
d-space containing $\cal G$. Therefore we call $\bbbV$ the d-space {\em generated} by $\cal G$ and
we call $\cal G$ a {\em generating set} of $\bbbV$ ( we note $\bbbV = \bbbV (\{U_i | i \in I\})$).
( Similar definitions can be given for {\em families} of vectors).\\

\subsubsection{Linear independence}
We let now $W = V$. Following an analogy with classical linear algebra, we develop now a notion corresponding to a 
kind of {\em linear independence} of the classes $\pmod{\sim}$ of the given vectors.
Let us extend the equivalence relation $\sim$ to d-spaces by: 
if $\bbbV_1,\bbbV_2$ are d-spaces , 
$$ \bbbV_1 \sim \bbbV_2 \Leftrightarrow \forall i,j \in \{1,2\},
\forall S \in \bbbV_i, \exists S' \in \bbbV_j, S \sim S'.$$

\begin{lem}
Let $S_1, \ldots, S_j, \ldots, S_m \in \bbbD \bbbR \bbbB_{1,\lambda}\pa{V}$.
The following are equivalent
\begin{enumerate}
\item $ \exists \vec{\alpha}, \vec{\beta} \in \bbbD \bbbR\bbbB_{1,m}\pa{V}, \vec{\alpha} \not\sim 
\vec{\beta}$,
such that
$$ \sum^m_{j=1} \alpha_j \cdot S_j \sim
   \sum^m_{j=1} \beta_j \cdot S_j$$
\item $\exists j_0 \in [1,m], \exists \vec{\gamma} \in \bbbD \bbbR \bbbB_{1,m}\pa{V}, 
\vec{\gamma} \not\sim \epsilon^m_{j_0}$, such that
$$ S_{j_0} \sim
   \sum^m_{j=1} \gamma_j \cdot S_j$$
\item $\exists j_0 \in [1,m], \exists \vec{\gamma}' \in \bbbD\bbbR \bbbB_{1,m}\pa{V}, \gamma'_{j_0} \sim
\emptyset$, such that 
$$S_{j_0} \sim
   \sum^m_{j=1} \gamma'_j \cdot S_j$$ 

\item $\exists j_0 \in [1,m]$, such that
$$\bbbV ((S_j)_{1 \leq j \leq m}) \sim \bbbV ((S_j)_{1 \leq j \leq m, j \neq j_0}).$$
\end{enumerate}
\label{fundamental_lemma2}.
\end{lem}
The equivalence between (1),(2) and (3) was first proved in \cite{Mei89,Mei92}, in the case where
the $S_j$'s are configurations $q_j \omega$, with the same $\omega$ and
$\bar{\psi} = {\rm Id}_{X^*}$ hence $\sim$ is just the language equivalence realtion $\equiv$. 
This is the key-idea around which we 
have developed the notion of d-spaces.\\
\begin{prooof}
\noindent{\bf $(1) \Rightarrow (2)$:}\\
Let us consider
${\cal R} \in {\cal B}_\infty(\vec{\alpha} \cdot S,\vec{\beta} \cdot S )$,
$\nu = \Div(\vec{\alpha},\vec{\beta})$ and
\begin{equation} 
(u,v) = \mi \{(w,w') \in {\cal R} \cap X^{\leq \nu} \times X^{\leq \nu} \mid
\exists j \in [1, m], (\vec{\alpha} \odot w = \epsilon^m_j)
\Leftrightarrow (\vec{\beta} \odot w' \neq \epsilon^m_j) \}.
\label{choice_vv'}
\end{equation}
Let us suppose , for example, that $\vec{\alpha} \odot u = \epsilon^m_{j_0}$ while
$\vec{\beta} \odot v \neq \epsilon^m_{j_0}$ and let 
$\vec{\gamma} = \vec{\beta} \odot u$.
As $ (u,v) \in {\cal R} \in {\cal B}_\infty(\vec{\alpha} \cdot S,\vec{\beta} \cdot S )$
\begin{equation}
(\vec{\alpha} \cdot S) \odot u \sim (\vec{\beta} \cdot S) \odot v.
\label{equiv1}
\end{equation}
Using lemma \ref{RA_odot} we obtain:
\begin{equation}
(\vec{\alpha} \cdot S) \odot u = S_{j_0}.
\label{value_lhs}
\end{equation}
Let us examine now the righthand-side of equality (\ref{equiv1}).
Let $(u',v')\prec (u,v)$ with $|u'| = |v'|$. By condition (4) in 
definition \ref{def_word_bisimulation} $(u',v') \in {\cal R}$
\footnote{here is the main place where this condition (4) is used}
and by minimality of $v$, $\vec{\beta} \odot v'$ is 
a unit iff 
$\vec{\alpha} \odot u'$ is a unit. But if $\vec{\alpha} \odot u'$ is a unit, then 
$\vec{\alpha} \odot u = \emptyset$, which is false. Hence $\vec{\beta} \odot v'$ is not a unit.
Hence, 
$\forall v' \prec v$, $\vec{\beta} \odot v'$ is not a unit . By lemma \ref{RA_odot}
\begin{equation}
(\vec{\beta} \cdot S) \odot v = (\vec{\beta} \odot v) \cdot S.
\label{value_rhs}
\end{equation}
Let us plug equalities (\ref{value_lhs}) and (\ref{value_rhs}) in equivalence (\ref{equiv1}) and
let us define $\vec{\gamma} = \vec{\beta} \odot v$. We obtain:
$$ S_{j_0} \sim \vec{\gamma} \cdot S, \vec{\gamma} \neq \epsilon^m_{j_0}.$$

\noindent{\bf $(2) \Rightarrow (3)$:}\\
$$S_{j_0} \sim \gamma_{j_0} \cdot S_{j_0} + (\sum_{j \neq j_0} \gamma_j \cdot S_j),\;\; 
\gamma_{j_0} \neq \epsilon.$$
By corollary \ref{algebraic_version}, point C1, we can deduce that
$$S_{j_0} \sim \sum_{j \neq j_0} \gamma^*_{j_0} \gamma_j \cdot S_j=
\nabla^*_{j_0}(\gamma) \cdot S.$$
Taking $\gamma' = \nabla^*_{j_0}(\gamma)$ we obtain
$$ S_{j_0} \sim \gamma' \cdot S \mbox{ where } \gamma'_{j_0} = \emptyset.$$
\noindent{\bf $(3) \Rightarrow (4)$:}\\
Let us denote by $\hat{S}$ the vector 
$(S_1, \ldots, S_{j_0-1},\emptyset,S_{j_0 + 1} \ldots, S_m) \in \bbbD \bbbB_{m,1}\pa{V}$.
If $T= \vec{\alpha} \cdot S$ then
$T \sim  (\vec{\alpha} \nabla_{j_0} \vec{\gamma'}) \cdot \hat{S}.$\\
\noindent{\bf $(4) \Rightarrow (1)$:}\\
Let us suppose (4) is true for some integer $j_0$. The element $S_{j_0}$ is clearly
equivalent (mod $\sim$) to two linear combinations of the $S_j$'s with non-equivalent vectors 
of coefficients (mod $\sim$). Hence (1) is true.\\
\end{prooof} 
\subsection{Derivations}
\label{stacking_derivations}
For  every $u \in X^*$ we define the binary relation $\uparrow(u)$ over 
$\bbbD \bbbB_{1,\lambda}\pa{V}$ by:
for every $S,S' \in \bbbD \bbbB_{1,\lambda}\pa{V}, S \uparrow(u) S' \Leftrightarrow
\exists q  \in \bbbn,\exists E_1, \ldots,E_k, \ldots, E_q \in V, \Phi \in \bbbD \bbbB_{q,\lambda}\pa{V}$ such that
$$ S = \sum_{k=1}^q E_k \cdot \Phi_k, S' = \sum_{k=1}^q (E_k \odot u ) \cdot \Phi_k,$$
and $\forall k \in [1,q], E_1 \smile E_k, E_k \odot u \notin \{ \emptyset, \epsilon\}$.\\
It is clear that if $S \uparrow(u) S'$ then $S \odot u = S'$ and that the converse is  not
true in general.
A sequence of deterministic row-vectors $S_0, S_1, \ldots, S_n$ is a {\em derivation}
iff there exist $x_1, \ldots, x_n \in X$ such that $S_0\odot x_1= S_1, \ldots, 
S_{n-1} \odot x_n = S_n$. The {\em length} of this derivation is $n$. If 
$u= x_1\cdot x_2 \cdot  \ldots \cdot x_n$ we call $S_0, S_1, \ldots, S_n$ the derivation 
{\em associated} with $(S,u)$. We denote this derivation by $S_0 \reduito{u}{} S_n$.\\
A derivation $S_0, S_1, \ldots, S_n$ is said to be {\em stacking} iff it is the derivation 
associated to a pair $(S,u)$  such that $S = S_0$ and $S_0 \uparrow(u) S_n$.
A derivation $S_0, S_1, \ldots, S_n$ is said to be a {\em sub-derivation} of a derivation 
$S'_0, S'_1, \ldots, S'_m$ iff there exists some $i \in [0,m]$ such that,
$ \forall j \in [1,n], S_j = S'_{i+j}$. 
\begin{defi}
A vector $S \in \bbbD \bbbR \bbbB_{1,\lambda}\pa{V}$ is said {\em loop-free} if and only if
for every $v \in V^+$, $S \bullet v \neq S$.
\end{defi}
Let us notice that every polynomial is loop-free. 
The two following lemmas give other examples of loop-free vectors.
\begin{lem}
Let $\alpha \in \bbbD \bbbB_{1,n}\langle V \rangle, \Phi \in \bbbB_{n,\lambda}\pa{V}$, such that
$\infty > \|\alpha \cdot \Phi \| > \|\Phi\|$. Then $\alpha \cdot \Phi $ is loop-free.
\label{poly_and_large}
\end{lem}
\begin{prooof}
Let $\alpha,\Phi$ fulfill the hypothesis of the lemma and suppose,
for sake of contradiction, that there exists some $v \in V^+$ such that:
$$
(\alpha \cdot \Phi) \bullet v = \alpha \cdot \Phi
\label{sake_of_contradict}
$$
By induction, for every $n \geq 0$:
\begin{equation}
(\alpha \cdot \Phi)\bullet v^n= \alpha \cdot \Phi
\label{inductively_true}
\end{equation}
As $\alpha$ is a polynomial, there exists some $n_0 \geq 0$ such that
$|v^{n_0}|$ is greater than the greatest length of a monomial of $\alpha$.
Using lemma \ref{residual_linear_combination}, equality (\ref{inductively_true}) for such an integer 
$n_0$ means that there exists some $k \in [1,n], v''$ suffix of $v^{n_0}$
such that:
\begin{equation}
\Phi_k \bullet v''= \alpha \cdot \Phi
\label{terribly_true}
\end{equation}
Using the hypothesis of the lemma we conclude that:
$$
\| \Phi \| \geq \| \Phi_k \bullet v''\|= \|\alpha \cdot \Phi\| > \| \Phi \|
$$
which is contradictory.
\end{prooof}

\begin{lem}
Let $S \in \bbbD \bbbR \bbbB_{1,\lambda}\pa{V}, u \in X^*$, such that
$\|S \odot u\| > \|S\|$. Then $S \odot u$ is loop-free.
\label{increasing_action}
\end{lem}
\begin{prooof}
Let us consider $S,u$ fulfilling the hypothesis of the lemma and let us consider the 3 possible
forms of $S \odot u$ proposed by lemma \ref{bullet_and_odot}. The forms (1) or (2) are incompatible
with the inequality $\|S \odot u\| > \|S\|$. Hence $S \odot u$ has the 
form (3):
$$
u = u_1 \cdot u_2, S \odot u_1 = S \bullet v_1 = \sum_{k=1}^q E_k \cdot \Phi_k,\;\;
S \odot u = \sum_{k=1}^q (E_k \odot u_2) \cdot \Phi_k,\mbox{ and }
$$
$$\forall k \in [1,q], E_k \smile E_1, E_k \odot u_2 \notin \{ \epsilon, \emptyset\}.$$
Hence $S \odot u =\alpha \cdot \Phi$ for some polynomial $\alpha \in \bbbD \bbbR \bbbB_{1,q}\langle V \rangle$.
As for every $k$, $\Phi_k= S \bullet (v_1E_k)$, we obtain that $\|S\| \geq \|\Phi\|$. Finally
$$\infty > \| S \odot u\| = \|\alpha \cdot \Phi \| > \|S\| \geq \|\Phi\|,$$ 
and by lemma \ref{poly_and_large}, $S \odot u$ is loop-free.
\end{prooof}
\begin{lem}
Let $S \in \bbbD \bbbR \bbbB_{1,\lambda}\pa{V}, w \in X^*$, such that\\
1- $S$ is loop-free\\
2- $\forall u \preceq w, \|S \odot u\| \geq \|S\|$.
Then the derivation $S \deriv{w} S \odot w $ is stacking.
\label{flat_is_stacking}
\end{lem}
\begin{prooof}
$S$ is left-deterministic. If it has type $\emptyset$ or $(\epsilon,j)$,
the lemma is trivially true. Otherwise
$$
S = \sum_{k=1}^q E_k \cdot \Phi_k,
$$
for some class of letter $[E_1]_\smile = \{ E_1,\ldots,E_q\}$ and some
matrix $\Phi \in \bbbD \bbbR \bbbB_{q,\lambda}\pa{V}$.
Suppose that for some prefix $u \preceq w$ and   $k \in [1,q]$,
\begin{equation}
E_k \odot u = \epsilon.
\label{epsilon_absurdum}
\end{equation}

Then, $S \odot u= \Phi_k$ so that $\|S \odot u\| \leq \| \Phi\| \leq \|S\|$
which shows that $ S = S \odot u$ while $u \neq \epsilon$.
This would contradict the hypothesis that $S$ is loop-free, hence
(\ref{epsilon_absurdum}) is impossible.\\
Let us apply now lemma \ref{RA_odot} to the expression 
$(E \cdot \Phi) \odot w$:
case (2) is impossible , hence 
$$ (E \cdot \Phi) \odot w=(E \odot w) \cdot \Phi,$$
which is equivalent to
$$S \uparrow(w) S \odot w.$$
\end{prooof}

\begin{lem}
Let $S\in \bbbD \bbbR \bbbB_{1,\lambda}\pa{V}, w \in X^*, k \in \bbbn$, such that\\ 
$\|S \odot w\| \geq \|S\| +  k \cdot K_0 + 1$.\\
Then the derivation $S \deriv{w} S \odot w $ 
contains some stacking sub-derivation of length $k$.
\label{increasing_contains_stacking}
\end{lem}
\begin{sketch}
Let $S=S_0, \ldots, S_i,\ldots, S_n$ be the derivation associated to $(S,w)$.
Let $i_0=\ma\{ i \in [0,n] \mid \|S_i\|=\mi\{ \|S_j\| \mid 0 \leq j \leq n \} \}$
and 
$i_1=\ma\{ i \in [i_0+1,n] \mid \|S_i\|=\mi\{ \|S_j\| \mid i_0+1 \leq j \leq n \} \}$.
 Let $w = w_0 w_1 w'$ where $|w_0|= i_0, |w_0 w_1|= i_1$.\\ 
As $\|S\odot w_0 w_1\| > \|S\odot w_0\|$, by lemma \ref{increasing_action}
$S\odot w_0 w_1 = S_{i_1}$ is loop-free. Using lemma \ref{norm_of_action}:
$$\|S_n\|-\|S_{i_1}\| \geq \|S_n\|-\|S_{i_0}\| - (\|S_{i_1}\|-\|S_{0}\|)
\geq (k-1) \cdot K_0 +1. $$ 
Using lemma \ref{norm_of_action} we must have $|w'|\geq k$. 
Let $w' = w_2 w_3$ with $|w_2|=k$.
By definition of $i_1$, $\forall i \in [i_1+1,i_1 + k], \|S_i\| \geq \|S_{i_1}\| +1$.\\
By lemma \ref{flat_is_stacking}, the sub-derivation $S_{i_1}, \ldots, S_{i_1 + k}$
(associated to  $(S_{i_1}, w_2)$) is stacking.
\end{sketch}
\begin{lem}
Let $S,S' \in \bbbD \bbbR \bbbB_{1,\lambda}\pa{V}, w \in X^*, k,d,d' \in \bbbn$, such that $S$ is
$d$-marked and:
\begin{description}
\item{(1)} the derivation 
$S \deriv{w} S' $ contains no  stacking  sub-derivation of length $k$.
\item{(2)} $|w| \geq d \cdot k$.
\end{description}
Then $S' $ is unmarked.
\label{no_stacking_removes_the_marks}
\end{lem}
\begin{prooof}
By hypothesis
$$S = \sum_{k=1}^q \alpha_k \cdot \Phi_k$$
for some $\alpha \in \bbbD \bbbR \bbbB_{1,q}\langle~V~\rangle, 
\Phi \in \bbbD \bbbR \bbbB_{q,\lambda}\pa{V}$, $\| \alpha\| \leq d, \Phi$  unmarked.\\
Let $S \deriv{w} S' = (S_0, \ldots, S_n)$.
By induction on $\ell$, using hypothesis (1) and lemma
\ref{flat_is_stacking} ( on polynomials , which are particular cases of loop-free series)
one can show that: 
for every $\ell \in[0,d]$, there exists some prefix $w_\ell$ of $w$ , with length
$|w_\ell| \leq k \cdot \ell$ such that
either
\begin{equation}
S \odot w_\ell = \sum_{k=1}^q (\alpha_k\odot w_\ell) \cdot \Phi_k,
\mbox {  with  }
\| \alpha_\odot w_\ell \| < \|\alpha\| - \ell 
\label{small_alpha}
\end{equation}
or there exists an integer $k \in [1,q]$ such that
\begin{equation}
S \odot w_\ell = \Phi_k.
\label{no_alpha}
\end{equation}
Let us apply this property to $\ell = d$:
inequality (\ref{small_alpha}) is not possible for this value of $\ell$ because, by hypothesis (2)
of the lemma  $\|\alpha\| - \ell \leq 0$.
Hence (\ref{no_alpha}) is true and, as $\Phi$ is unmarked, $\Phi_k$ is unmarked , so that
$S \odot w$ is unmarked.
\end{prooof}

\section{Deduction systems}
\label{glogics4}
\subsection{General formal systems}
We  follow here the general philosophy of \cite{Har79,Cou83}.
Let us call {\em formal system} any triple ${\cal D} = <{\cal A}, H, \deducedir>$
where $\cal A$ is a denumerable set called the {\em set of assertions},
 $H$, the {\em cost function} is a mapping ${\cal A} \rightarrow \bbbn \cup \{\infty\}$ and 
$\deducedir$, the {\em deduction relation} is a subset 
of ${\cal P}_f({\cal A}) \times {\cal A}$ ; ${\cal A}$ is given with a fixed bijection with 
$\bbbn$ (an
``encoding'' or ``G\"odel numbering'') so that the notions of recursive subset,
recursively enumerable subset, recursive function, ... over ${\cal A},  {\cal P}_f ({\cal A})$,
... are defined, up to this fixed bijection ; we assume that $\cal D$ satisfies the
following axiom:\\
(A 1) $\forall (P, A) \in  \deducedir$, (min $\{H(p), p \in P\} < H(A))$ or $(H(A) = \infty)$.\\
(We let $\mi(\emptyset) = \infty$).
We call ${\cal D}$ a {\em deduction system} iff ${\cal D}$ is a formal system satisfying the 
additional axiom:\\
(A 2) $\deducedir$ is recursively enumerable.\\
In the sequel we use the notation $P \deducedir A$ for $(P, A) \in \deducedir$. We call
{\em proof} in the system $\cal D$, {\em relative to the set of hypotheses} ${\cal H}
\subseteq \cal A$, any subset $P \subseteq \cal A$ fulfilling : 
$$\forall p \in P, (\exists Q \subseteq P, Q \deducedir p) \mbox{ or }  (p \in \cal H).$$ 
We call $P$ a {\em proof} iff
$$\forall p \in P, (\exists Q \subseteq P, Q \deducedir p)$$ 
(i.e. iff $P$ is a proof relative to $\emptyset$).\\
 Let us define the total map $\chi : {\cal A} \rightarrow \{0,1\}$
and the partial map $\overline{\chi} : {\cal A} \rightarrow \{0, 1\}$ by :\\
$$\chi(A) = 1 \mbox{ if } H(A) = \infty , \chi(A) = 0 \mbox{ if } H(A) < \infty ,$$
$$ \overline{\chi}(A) = 1 \mbox{ if } H(A) = \infty, \overline{\chi} \mbox{ is undefined if }
H(A) < \infty.$$
($\chi$ is the ``truth-value function'', $\overline{\chi}$ is the ``1-value
function'').
\begin{lem}
Let $P$ be a proof relative to ${\cal H} \subseteq H^{-1}(\infty)$ and $A \in P$. Then
$\chi(A) = 1$.
\label{D_is_sound}
\end{lem}
In other words : if an  assertion is provable from true hypotheses, then it is true.
\begin{prooof} Let $P$ be a proof. We prove by induction on $n$ that, 
$${\cal P}(n) : \forall p \in P, H(p) \geq n.$$ It is clear that, $\forall p \in P,
H(p) \geq 0$. Suppose that ${\cal P}(n)$ is true. Let $p \in P - {\cal H} :
\exists Q \subseteq P, Q \deducedir p$. By induction hypothesis,  
$\forall q \in Q, H(q) \geq n$ and by (A1),$ H(p) \geq n+1$. It follows that : $\forall p \in 
P -{\cal H}, H(p) = \infty$. But by hypothesis, $\forall p \in {\cal H}, H(p) = \infty$.
\end{prooof}

A formal system $\cal D$  will be said {\em complete} iff, conversely,
$\forall A \in {\cal A}, \chi(A) = 1 \Longrightarrow$ there exists some {\em
finite} proof $P$ such that $A \in P$. (In other words, $\cal D$ is complete iff every true 
assertion is ``finitely'' provable).

\begin{lem} : If $\cal D$ is a complete deduction system, $\overline{\chi}$ is a recursive partial
map.
\label{complete_implies_pre}
\end{lem}

\begin{prooof} 
Let $i \mapsto P_i$ be some recursive function whose domain is $\bbbn$ and whose image is 
${\cal P}_f({\cal A})$.
Let $h : ({\cal P}_f({\cal A}) \times {\cal A} \times \bbbn) \rightarrow \{0,
1\}$ be a total recursive function such that :
$$P \deducedir A \mbox{ iff }\exists n \in \bbbn, h(P, A, n) = 1
$$
(such an $h$ exists, because the r.e. sets are the projections of the
recursive sets, see \cite{Rog67}).\\

The following (informal) semi-algorithm computes $\overline{\chi}$ on the
assertion A :
{\tt
\begin{enumerate}
\item $i$ := $0$ ; $n$ := $0$ ; $s$ := $i+n$;
\item $P$ := $P_i$;
\item $b$ := $\mi_{p \in P} \{\ma_{Q \subseteq P}\{ h(Q,p,n)\}\}$;
\item $c$:= $(A \in P);$
\item if $(b \wedge c)$ then $(\overline{\chi}(A) = 1$ ; stop);
\item if $i = 0$ then ($i:= s+1$ ; $n := 0$; $s := i+n$)\\
                 else ($i := i-1$ ; $n := n+1)$ ;
\item goto 2 ;
\end{enumerate}
}
\end{prooof}

In order to define deduction relations from more elementary ones, we set the following
definitions.

Let $\deducedir \subseteq {\cal P}_f({\cal A}) \times {\cal A}$. For every $P, Q \in {\cal P}_f({\cal
A})$ we set : 

\begin{itemize}
\item $P {\stackrel{[0]}{\deducedir}} Q$ iff $P \supseteq Q$ \nonumber\\
\item $P {\stackrel{[1]}{\deducedir}}Q $ iff $\forall q \in Q, \exists R \subseteq P, R
\deducedir q$\\
\item $P {\stackrel{<0>}{\deducedir}} Q$ iff $P {\stackrel{[0]}{\deducedir}} Q$\\
\item $P {\stackrel{<1>}{\deducedir}} Q$ iff $\forall q \in Q,(\exists R \subseteq P, R
\deducedir q)$ or $(q \in P)$\\
\item $P {\stackrel{<n+1>}{\deducedir}} Q$ iff $\exists R \in {\cal P}_f(A),P
{\stackrel{<1>}{\deducedir}} R \mbox{ and } R {\stackrel{<n>}{\deducedir}} Q$ 
(for every $n \geq 0$).\\
\item ${\stackrel{<*>}{\deducedir}} = \bigcup_{n\geq 0}{\stackrel{<n>}{\deducedir}}$.
\end{itemize}
Given ${\deducedir}_1, {\deducedir}_2 \subseteq {\cal P}_f({\cal A}) \times {\cal
P}_f({\cal A})$, for every $P, Q \in {\cal P}_f({\cal A})$ we set :\\
$$P ({\deducedir}_1 \circ {\deducedir}_2) Q\mbox{ iff } 
\exists R \subseteq {\cal A}, (P {\deducedir}_1 R) \wedge (R {\deducedir}_2 Q).$$

\subsection{System ${\cal B}_0$}

Let us define here a particular formal system ${\cal B}_0$ ``Taylored for the $\sigma$-
$\bar{\psi}$-bisimulation problem for deterministic series''.\\
Let us fix two finite alphabets $X,Y$, a surjection $\psi: X \rightarrow Y$ (which induces a surjection $X^* \rightarrow Y^*$ denoted by the same symbol $\psi$) and its kernel
$\bar{\psi}= {\rm Ker}\psi \subseteq X^* \times X^*$ ( see section \ref{sigma_bisimulation}).
We also fix a  dpda ${\cal M}$ over the terminal alphabet $X$ and consider the
variable alphabet $V$ associated to ${\cal M}$ (see section \ref{series}) and the sets ${\sf
D R B}_{\delta,\lambda} \pa{V}$ (the sets of Deterministic Rational Boolean matrices over $V^*$, 
with $\delta$ rows and $\lambda$ columns ). The set
of assertions is defined by :
$${\cal A} = \bigcup_{\lambda \geq 1}\bbbn \times {\sf D R B}_ {1,\lambda}\pa{V} \times {\sf D R B}_{1,\lambda}\pa{V}$$
i.e. an assertion is here a { \em weighted equation} over ${\sf D R B}_{1,\lambda} \pa{V}$ for some
integer $\lambda$.\\
For every $n \geq 0$ we define
\begin{equation}
\bar{{\cal B}}_n = \{ {\cal R} \subseteq \bar{\psi} \mid {\cal R} \mbox{ fulfills conditions } 
 (1'),(2') \mbox{ and }(4) \mbox{ of definition } \ref{def_word_bisimulation} \}.
\end{equation}
We call the elements of $\bar{{\cal B}}_n$ the {\em admissible} relations of order $n$ over
$X^* \times X^*$.
For every pair $(S,S') \in {\sf D R B}_{1,\lambda} \pa{V} \times {\sf D R B}_{1,\lambda} \pa{V}$,
and $n \in \bbbn \cup \{\infty\}$ we define:
\begin{equation}
{\cal B}_n(S,S') = \{ {\cal R} \subseteq \bar{\psi} \mid {\cal R} \mbox{ is a } 
w-\bar{\psi}-\mbox{bisimulation of order n w.r.t. } (S,S') \}.
\end{equation}
The ``cost-function'' $H : {\cal A} \rightarrow \bbbn \cup\{\infty\}$ is defined by :
$$H(n, S, S') = n+2\cdot \Div (S, S'),$$
where $\Div(S, S')$ is  the {\em divergence} between $S$ and $S'$ (definition \ref{divergence}). 
We recall it  is defined by :
$$\Div(S, S') = \borninf \{ n \in \bbbn \mid {\cal B}_n(S,S') = \emptyset\}.$$
(We recall  $\borninf( \emptyset) = \infty$).\\

Let us notice that, by lemma \ref{sigma_reducesto_w} :
$$\chi(n, S, S') = 1 \Longleftrightarrow S \sim S'.$$

We define a binary relation ${\ededucedir} \subseteq {\cal P}_f({\cal A}) \times
{\cal A}$, the {\em elementary deduction relation},  as the set of all the 
pairs having one of the following forms:

\begin{description}

\item[(R0)]$$\; \; \{(p, S, T)\} {\ededucedir} (p +1, S, T)$$ 
for $p \in \bbbn, \lambda  \in \bbbn-\{0\},  S, T \in \bbbD \bbbR \bbbB_{1,\lambda}\pa{V}$,
\item[(R1)]$$ \; \; \{(p, S, T)\}  {\ededucedir} (p, T, S)$$
for $p \in \bbbn, \lambda  \in \bbbn-\{0\}, S, T \in \bbbD \bbbR \bbbB_{1,\lambda}\pa{V}$,
\item[(R2)]$$ \; \; \{(p, S, S'), (p, S', S'')\}  {\ededucedir}  (p, S, S'')$$
for $p \in \bbbn, \lambda  \in \bbbn-\{0\},  S, T \in \bbbD \bbbR \bbbB_{1,\lambda}\pa{V}$,
\item[(R3)]$$ \; \; \emptyset   {\ededucedir} (0, S, S)$$
for $ S \in \bbbD \bbbR \bbbB_{1,\lambda}\pa{V}$,
\item[(R'3)]$$ \;\; \emptyset  {\ededucedir} (0, S, \rho_e(S))$$
for $S \in \bbbD \bbbR \bbbB_{1,1}\pa{V},$
\item[(R4)]$$ \; \; \{(p+1, S \odot x, T \odot x') \mid (x,x') \in {\cal R}_1 \}  {\ededucedir} (p, S, T)$$
for $p \in \bbbn, \lambda  \in \bbbn-\{0\},  S, T \in \bbbD \bbbR \bbbB_{1,\lambda}\pa{V},
(S  \neq \epsilon \wedge T \neq \epsilon)$ and ${\cal R}_1 \in \bar{{\cal B}}_1$,
\item[(R5)]$$  \; \; \{(p, S, S')\} {\ededucedir} (p+ 2, S \odot x, S' \odot x')$$
for $p \in \bbbn, \lambda  \in \bbbn-\{0\},  S, T \in \bbbD \bbbR \bbbB_{1,\lambda}\pa{V},
(x,x') \in \bar{\psi}, S \sim S' \wedge S \odot x \sim S' \odot x'$,
\item[(R6)]$$ \; \{(p, S_1 \cdot T+ S, T)\} {\ededucedir}  (p, S_1^* \cdot S , T)$$
for $p \in \bbbn, \lambda \in \bbbn-\{0\}, S_1 \in \bbbD \bbbR \bbbB_{1,1}\pa{V}, S_1 \neq \epsilon, (S_1,S) \in \bbbD \bbbR \bbbB_{1,\lambda+1}\pa{V}, T \in \bbbD \bbbR \bbbB_{1,\lambda}\pa{V}$,
\item[(R7)]$$ \; \; \{(p, S, S')\} {\ededucedir} (p, S \cdot T, S' \cdot T)$$
for $p \in \bbbn, \delta, \lambda \in \bbbn-\{0\},  S, S' \in \bbbD \bbbR \bbbB_{1,\delta}\pa{V},
T \in \bbbD \bbbR \bbbB_{\delta,\lambda}\pa{V}$,
\item[(R8)]$$ \; \; \{(p, T_{i,*}, T'_{i,*}) \mid 1 \leq i \leq \delta\} {\ededucedir} 
(p, S \cdot T, S \cdot T')$$
for $p \in \bbbn, \delta, \lambda \in \bbbn-\{0\},  S \in \bbbD \bbbR \bbbB_{1,\delta}\pa{V},
T, T' \in \bbbD \bbbR \bbbB_{\delta,\lambda}\pa{V}$.
\end{description}
\begin{rem}
We do not claim that this formal sytem is {\em recursively enumerable}: due to rule (R5), 
establishing 
this property is as difficult as to solve the general bisimulation problem for equational 
graphs of finite out-degree. This difficulty will be overcome in section \ref{elimination} by an 
{\em elimination} lemma .\\
\end{rem}
\begin{lem} : Let $P \in {\cal P}_f({\cal A}), A \in {\cal A}$ such that $P {\ededucedir}
A$. Then $min\{H(p) \mid p \in P\} \leq H(A)$.
\label{H_increasing}
\end{lem}
Let us introduce a notation: for every $n \in \bbbn \cup \{ \infty \}, \lambda \in \bbbn-\{0\}, 
S,S' \in \bbbD \bbbR \bbbB_{1,\lambda}\pa{V}$,
$$ S \sim_n S' \Leftrightarrow {\cal B}_n(S,S') \neq \emptyset. $$
\begin{prooof} Let us check this property for every type of rule.\\
{\bf R0}: $p + 2\cdot \Div(S,T) \leq p+1+2 \cdot \Div (S,T).$\\
{\bf R1}: $p+2 \cdot \Div(S,T) = p+2 \cdot \Div(T,S).$\\
{\bf R2}: as the weight $p$ is the same in all the considered equations, we are reduced to
prove that :\\
$\forall n \in {\bbbn}, S \sim_n S' \wedge S' \sim_n S" \Longrightarrow S \sim_n
S"$. This is true because, if ${\cal R} \in {\cal B}_n(S,S')$ and ${\cal R}' \in 
{\cal B}_n(S',S")$, then ${\cal R} \circ {\cal R}' \in {\cal B}_n(S,S").$ \\
{\bf R3}: Let us notice that ${\rm Id}_{X^*} \subseteq \bar{\psi}$. It follows that
 $\infty =\Div(S,S).$\\
{\bf R'3}: The definition of $G$ from $G_0$ is such that, $S \equiv \rho_e(S)$, hence
$S \sim \rho_e(S)$ and $\infty =\Div(S,\rho_e(S)).$\\
{\bf R4}: Let $n \in \bbbn$ such that:
$$\forall (x,x') \in {\cal R}_1, n \leq \Div(S \odot x, S' \odot x') .$$
Let us choose, for every $(x,x') \in {\cal R}_1$, some ${\cal R}_{x,x'} \in 
{\cal B}_n(S \odot x,S' \odot x')$. Let us define then
$$ {\cal R} = \bigcup_{(x,x') \in {\cal R}_1} (x,x')\cdot {\cal R}_{x,x'}.$$
${\cal R}$ belongs to ${\cal B}_{n+1}(S,S')$. It follows that
$$ \min \{\Div ( S \odot x, S' \odot x') \mid (x,x') \in {\cal R}_1\}+1 \leq \Div ( S , S')$$  
hence that 
$$ \min \{H (p+1, S \odot x, S' \odot x') \mid (x,x') \in {\cal R}_1\} \leq H(p,  S , S') - 1.$$
{\bf R5}: By hypothesis, $H(p+ 2, S \odot x, S' \odot x') = \infty.$\\
{\bf R6}: Let $n \in \bbbn$ such that:
$$n \leq \Div(S_1 \cdot T+ S, T) .$$
Let ${\cal R} \in {\cal B}_n(S_1 \cdot T+ S, T)$.
Let ${\cal R}' = {\cal R}^{<S_1,*>}$ (see definition (\ref{star_R}) in \S 
\ref{operations_on_wbisimulations}).
As we have
$$ {\cal R}' \in {\cal B}_n (S_1^* \cdot S , T),$$
we get the inequality : $ \Div(S_1 \cdot T+ S, T) \leq \Div (S_1^* \cdot S, T)$.\\
{\bf R7}: 
Let $ n \leq \Div((S , S')$ and ${\cal R} \in {\cal B}_n(S,S')$. Let us consider:
${\cal R}'= <S |{\cal R}>$ (see definition (\ref{right_quotient_R}) in \S 
\ref{operations_on_wbisimulations}).
As we have 
${\cal R}' \in {\cal B}_n (S \cdot T,S' \cdot T)$, the required inequality is proved.\\
{\bf R8}: 
Let $ n \leq \min\{\Div( T_{i,*}, T'_{i,*}) \mid 1 \leq i \leq \delta\}$ and, for every 
$i \in [1,\delta]$, let ${\cal R}_i \in {\cal B}_n(T_{i,*}, T'_{i,*})$. Let us consider 
${\cal R}' = <S,{\cal R}>$ (see definition (\ref{left_product_R}) in 
\S \ref{operations_on_wbisimulations}). As we know that
$${\cal R}' \in {\cal B}_n (S \cdot T,S \cdot T'),$$ the required inequality is proved.
\end{prooof}
Let us define ${\deducedir}$ by : for every $ P \in {\cal P}_f({\cal A}), A \in {\cal A}$,\\
$$P {\deducedir} A  \Longleftrightarrow P \ededuce{*} \circ
{\stackrel{[1]}{\ededucedir}}_{0,3,4} \circ \ededuce{*} \{A\}.$$\\
where $\ededucedir_{0,3,4}$ is the relation defined by
$R_0,R_{3},R'_{3},R_4$ only. We let 
$${\cal B}_0 = <{\cal A}, H, \deducedir>.$$
\begin{lem} : ${\cal B}_0$ is a formal system.
\end{lem}
\begin{prooof} 
Using lemma \ref{H_increasing}, one can show by induction on $n$ that :
$$P \ededuce{n} Q \Longrightarrow \forall q \in Q, \mi\{H(A)\mid A \in P\} \leq
H(q).$$
The proof of lemma \ref{H_increasing} also reveals that :
$$
P \ededucedir_{\{0,3,4\}} q \Longrightarrow (\mi \{H(p) \mid p \in P\} < H(q)) \mbox{ or }
H(q) = \infty.
$$
It follows that, for every $m, n \geq 0$ :
$$
P \ededuce{n} Q \stackrel{[1]}{\ededucedir}_{0,3,4} R \ededuce {m} {q}
\Longrightarrow (\mi\{H(p) \mid p \in P\} < H(q)) \mbox{ or } H(q) = \infty.
$$
Hence  axiom $(A1)$ is fulfilled.
\end{prooof}
Let us remark the following algebraic corollaries of lemma \ref{H_increasing}.
\begin{cor}$\;\;$\\
\begin{description}
\item[(C1)]$ \; \forall \lambda \in \bbbn-\{0\}, S_1 \in \bbbD \bbbR \bbbB_{1,1}\pa{V}, S_1 \not\equiv \epsilon, (S_1,S) \in \bbbD \bbbR \bbbB_{1,\lambda+1}\pa{V}, 
T \in \bbbD \bbbR \bbbB_{1,\lambda}\pa{V}$,\\
$$S_1 \cdot T+ S \sim T  {\Longrightarrow}  S_1^* \cdot S \sim T$$
\item[(C2)]$\forall \delta, \lambda \in \bbbn-\{0\},  S, S' \in \bbbD \bbbR \bbbB_{1,\delta}\pa{V},
T \in \bbbD \bbbR \bbbB_{\delta,\lambda}\pa{V}$,\\
$$ \; \; S \sim S' {\Longrightarrow} S \cdot T \sim S' \cdot T $$
\item[(C3)]$\forall \lambda \in \bbbn-\{0\},  S, S' \in \bbbD \bbbR \bbbB_{1,1}\pa{V},
T \in \bbbD \bbbR \bbbB_{1,\lambda}\pa{V}$,\\
$$ \; \; [S \cdot T \sim S' \cdot T \mbox{ and } T \neq \emptyset^{\lambda}]\;{\Longrightarrow} S \sim S'$$ 
\item[(C4)]$\forall \delta, \lambda \in \bbbn-\{0\},  S \in \bbbD \bbbR \bbbB_{1,\delta}\pa{V},
T, T' \in \bbbD \bbbR \bbbB_{\delta,\lambda}\pa{V}$,\\
$$ \; \; T \sim T' {\Longrightarrow} S \cdot T \sim S \cdot T'.$$
\end{description}
\label{algebraic_version}
\end{cor}
\begin{prooof}
Statement (C$i$) (for $1 \leq i \leq 4$) is a direct corollary of the fact that the value of $H$ 
at the left-hand side of some rule (R$j$) is smaller or equal to the value of $H$ at the right-hand side of rule (R$j$): (C1) is justified by  (R6), (C2) by (R7),(C4) by (R8).\\  
Let us prove (C3):
suppose that $\lambda \in \bbbn-\{0\},  S, S' \in \bbbD \bbbR \bbbB_{1,1}\pa{V},
T \in \bbbD \bbbR \bbbB_{1,\lambda}\pa{V}$ and 
\begin{equation}
S \cdot T \sim S' \cdot T \mbox{ and } S \not\sim S'.
\label{bisimilarity_hypo}
\end{equation}
Let ${\cal R} \in {\cal B}_\infty(S \cdot T, S' \cdot T)$ and let $$(u,u') =
\mi\{ (v,v') \in {\cal R} \mid (\rho_{\epsilon}(S \odot v) = \epsilon)
\Leftrightarrow (\rho_{\epsilon}(S'\odot v') \neq \epsilon) \} .$$
From the hypothesis that ${\cal R} \in {\cal B}_\infty(S \cdot T, S' \cdot T)$, we get that
$$\forall (v,v') \in {\cal R},\;\; (S \cdot T)\odot v  \sim (S' \cdot T)\odot v',$$
and by the choice of $(u,u')$ we obtain that:
$$ T \sim (S' \odot u') \cdot T \mbox{ or } (S \odot u) \cdot  T \sim T,$$
 which, by $C1$, implies:
$$ T \sim (S' \odot u')^* \cdot \emptyset^\lambda \mbox{ or } 
(S \odot u)^* \cdot \emptyset^\lambda \sim T,$$
i.e. $T \sim \emptyset^\lambda$, which implies (because $G$ is a reduced grammar) that
\begin{equation}
T = \emptyset^\lambda.
\label{emptyness_conclusion}
\end{equation}
We have proved that (\ref{bisimilarity_hypo}) implies (\ref{emptyness_conclusion}), hence (C3). 
\end{prooof}
\subsection{Congruence closure}
\label{congruence_closure}
Let us consider the subset ${\cal C}$ of the rules of ${\cal B}_0$, consisting of all the 
instances of the metarules  R0,R1,R2,R3,R'3,R6,R7,R8. We also denote by 
$\ededucedir_{\cal C} \subseteq {\cal P}_f({\cal A}) \times {\cal A}$ the set of all instances of 
these meta-rules. We are 
interested here ( and later in section \ref{B1_is_complete}) in special subsets of ${\cal A}$ 
which express an ordinary weighted equation 
$(p,S,S')$ together with an admissible binary relation ${\cal R}$ of finite order (which is a
{\em candidate} to be a w-$\bar{\psi}$-bisimulation w.r.t. $(S,S')$).\\
For every $p,n  \in \bbbn, \lambda  \in \bbbn-\{0\},  S, S' \in \bbbD \bbbR \bbbB_{1,\lambda}\pa{V},
{\cal R} \in \bar{\cal B}_n$, we use the notation:
\begin{equation}
[p,S,S',{\cal R}] = \{ (p+ |u|, S \odot u, S' \odot u') \mid (u,u') \in {\cal R} \}.
\label{quadruples_def}
\end{equation}
One can check the following properties.\\
{\bf composition}:\\
for every $p, n \in \bbbn, \lambda  \in \bbbn-\{0\},  S, T \in \bbbD \bbbR \bbbB_{1,\lambda}\pa{V},
{\cal R}_1,{\cal R}_2 \in \bar{\cal B}_n$,
$$[p,S,S',{\cal R}_1] \cup [p,S',S'',{\cal R}_2] \ededuce{*}_{\cal C} 
[p,S,S'',{\cal R}_1 \circ {\cal R}_2]$$
{\bf star}:\\
for every $p,n \in \bbbn, \lambda \in \bbbn-\{0\}, S_1 \in \bbbD \bbbR \bbbB_{1,1}\pa{V}, S_1 \not\equiv \epsilon, (S_1,S) \in \bbbD \bbbR \bbbB_{1,\lambda+1}\pa{V}, T \in \bbbD \bbbR \bbbB_{1,\lambda}\pa{V},{\cal R} \in \bar{\cal B}_n$,
$$[p,S_1 \cdot T+ S ,T ,{\cal R}] \ededuce{*}_{\cal C}[p, S_1^* \cdot S, T, {\cal R}^{<S_1,*>}]$$
{\bf right-product}:\\
for every $p,n \in \bbbn, \delta, \lambda \in \bbbn-\{0\},  S, S' \in 
\bbbD \bbbR \bbbB_{1,\delta}\pa{V}, T \in \bbbD \bbbR \bbbB_{\delta,\lambda}\pa{V}, {\cal R}\in \bar{\cal B}_n$,
$$[p,S,S',{\cal R}] \ededuce{*}_{\cal C} [p,S \cdot T, S' \cdot T,<S |{\cal R}>]$$
 {\bf left-product}:\\
for every $p,n \in \bbbn, \delta, \lambda \in \bbbn-\{0\},  S \in \bbbD \bbbR \bbbB_{1,\delta}\pa{V},
T, T' \in \bbbD \bbbR \bbbB_{\delta,\lambda}\pa{V}, {\cal R}_1, \ldots, {\cal R}_\delta \in \bar{\cal B}_n$,
$$\bigcup_{1 \leq i \leq \delta} [p,T_{i,*}, T'_{i,*},{\cal R}_i] \ededuce{*}_{\cal C}
[p,S \cdot T, S \cdot T',<S ,{\cal R}>].$$
Given a subset $P \in {\cal P}_f({\cal A})$, we call {\em congruence closure} of $P$, denoted by
$\CO(P)$, the set: 
\begin{equation}
 \CO(P) = \{ A \in {\cal A} \mid P \ededuce{*}_{\cal C} \{A\} \}
\label{congruence_closure_def}
\end{equation}
As well, for every integer $q \geq 0$ we define:
\begin{equation}
 \CO_q(P) = \{ A \in {\cal A} \mid P \ededuce{q}_{\cal C} \{A\} \}
\label{congruence_closureq_def}
\end{equation}
\subsection{Strategies}
One  key-step of this work is the statement that ${\cal B}_0$ is complete (theorem 
\ref{B1_complete}). We prove this completeness result by exhibiting a ``strategy'' $\Str$ which, 
for every true
assertion $(p, S, S')$, constructs a finite ${\cal B}_0$-proof of this assertion.
Let ${\cal D} = <{\cal A}, H, \deducedir >$ be a formal system. We call  a {\em
strategy} for ${\cal D}$ any map $\Str : {\cal A}^+ \rightarrow {\cal P}({\cal A}^*)$ such that:\\
(S1) if $B_1 \cdots B_m \in \Str(A_1A_2 \cdots A_n)$ then $\exists Q \subseteq
\{A_i \mid 1 \leq i \leq n-1\}$ such that\\
$$\{B_j \mid 1 \leq j \leq m \} \cup Q \deducedir A_n,$$
(S2) if $B_1 \cdots B_m \in \Str(A_1A_2 \cdots A_n)$ then 
$$\mi \{H(A_i) \mid 1
\leq i \leq n \} = \infty \Longrightarrow \mi \{H(B_j) \mid 1 \leq j \leq m \} =
\infty.$$
\begin{rem}
It may happen that $\epsilon \in \Str(A_1A_2 \cdots A_n)$ ( and correspondingly, that $m=0$ in 
the above conditions): it just means that $\{A_1,\ldots,A_{n-1}\} \deducedir A_n$. 
It may also happen that $\Str(A_1A_2 \cdots A_n) = \emptyset$: it means , intuitively, that $\Str$ 
``does not know'' how to extend a proof ( with hypothesis), with the only information that the 
given proof  contains the assertions $A_1,A_2 ,\cdots A_n$.
\end{rem}
\begin{rem}
Axiom (A1) on systems is similar to the ``monotonicity'' condition of \cite{Har79} or axiom 
(2.4.2') of \cite{Cou83}.
Axiom (S2) on strategies is similar to the ``validity'' condition of \cite{Har79} or property 
(2.4.1') of \cite{Cou83}. 
\end{rem}
Given a strategy $\Str$, we define ${\cal T}(\Str, A)$, the set of proof-trees
associated to the strategy $\Str$ and the assertion $A$ as the set of all the trees $t$ 
fulfilling the following properties:\\
\begin{equation}
\varepsilon \in dom(t),\;\;\; t(\varepsilon) = A,
\label{root_condition}
\end{equation}
and, for every path $x_0x_1, \cdots x_{n-1}$ in $t$, with labels $t(x_i) = A_{i+1}$
(for $0 \leq i \leq n-1$) if $x_{n-1}$ has $m$ sons $x_{n-1}\cdot 1, \cdots. x_{n-1}\cdot m
\in \dom(t)$ with labels $t(x_{n-1}\cdot j) = B_j$ (for $1 \leq j \leq m)$ then\\
\begin{equation}
(B_1 \cdots B_m) \in \Str(A_1 \cdots A_n) \mbox{ or }
m=0.
\label{node_condition}
\end{equation} 

The proof-tree $t$ is said {\em closed} iff it fulfills the additional condition: for every path 
$x_0x_1, \cdots x_{n-1}$ in $t$, with labels $t(x_i) = A_{i+1}$
(for $0 \leq i \leq n-1$) if $x_{n-1}$ has $m$ sons $x_{n-1}\cdot 1, \cdots, x_{n-1}\cdot m
\in \dom(t)$ with labels $t(x_{n-1}\cdot j) = B_j$ (for $1 \leq j \leq m)$ then\\
\begin{equation}
m=0 \Rightarrow ((\exists i \in [1,n-1], A_i = A_n) \mbox{ or }
(\varepsilon \in \Str(A_1 \cdots A_n)) )
\label{leaf_condition}
\end{equation}
A node $x \in \dom(t)$ is said {\em closed} iff it is an internal node or it is a leaf fulfilling
property (\ref{leaf_condition}) above.\\
The proof-tree $t$ is said {\em repetition-free} iff,
for every $x,x' \in \dom(t)$,
$$
[ x \preceq x' \mbox{and } t(x) = t(x')]\;\; \Rightarrow \;\;x = x' \mbox{ or } x' \mbox{ is a leaf }.
$$
For every tree $t$ let us define:
$$\leaves(t)= \{ t(x) | \forall y \in \dom(t), x \preceq y \Rightarrow x = y \},\;\;
  \internal(t)= \{ t(x) | \exists y \in \dom(t), x \prec y \}.$$
(Here $\leaves$ stands for ``leaves'' and $\internal$ stands for ``internal nodes'').

\begin{lem} If $\Str$ is a strategy for the deduction-system $\cal D$ then,
for every true assertion $A$ and every $t \in {\cal T}(\Str, A)$\\
(1) the set of labels of $t$ is a $\cal D$-proof, relative to
the set $\leaves(t) - \internal(t)$.\\
(2) every label of a leaf is true.
\label{trees_are_proofs}
\end{lem}
\begin{prooof} Let us suppose that $H(A) = \infty$. Let $t \in {\cal T}(\Str,
A), P = \im(t)$ (the set of labels of $t$), ${\cal H} = \leaves(t) - \internal(t)$.\\
Using (S2), one can prove by induction on the depth of $x \in \dom(t)$ that, $H(t(x)) =
\infty$. Point (2) is then proved. Let $x$ be an internal node of $t$, with sons
$x\cdot 1, x\cdot 2, \cdots, x \cdot m$ ($m \geq 1$), and with ancestors 
$y_1,y_2, \cdots, y_{n-1}, y_n = x$ ($n \geq 1$), such
that 
$$t(y_1) \cdots t(y_n) = A_1 \cdots A_n,\;\;\; t(x_1) \cdots t(x_m) = B_1
\cdots B_m .$$ 
By definition of ${\cal T}(\Str,A)$, 
$$B_1 \cdots B_m \in \Str(A_1 \cdots A_n)$$ and by condition (S1):
$$\exists Q \subseteq \{A_i \mid i \leq n-1\},  \mbox{ such that }
\{B_j \mid 1 \leq j \leq m \} \bigcup Q {\deducedir} A_n.$$
It follows that for every $p \notin {\cal H}, \exists R \subseteq P, R {\deducedir} p$, hence
$$\forall p \in P, (\exists R \subseteq P, R {\deducedir} p) \mbox{ or } p  \in {\cal
H}.$$ 
Point (1) is proved.
\end{prooof}
For every ${\cal D}$-strategy $\Str$, we use the notation:
$${\cal T}(\Str) = \bigcup_{A \in H^{-1}(\infty)} {\cal T}(\Str,A).$$ 
We call  a {\em global strategy} w.r.t. ${\cal S}$ any total map 
$\hat{\Str} : {\cal T}(\Str) \rightarrow {\cal T}(\Str)$ such that:\\
\begin{equation}
\forall t \in {\cal T}(\Str) , t \preceq \hat{\Str}(t).
\label{global_strategy}
\end{equation}
$\hat{\Str}$ is a {\em terminating} global strategy iff:
\begin{equation}
\forall A_0 \in H^{-1}(\infty), \exists n_0 \in \bbbn,
\hat{\Str}^{n_0}(A_0)=\hat{\Str}^{n_0 +1}(A_0),
\label{terminating_global_strategy}
\end{equation}
$\hat{\Str}$ is a {\em closed} global strategy iff:
\begin{equation}
\forall A_0 \in H^{-1}(\infty), \forall n \in \bbbn,
\hat{\Str}^n(A_0) \mbox{ is closed} \Longleftrightarrow \hat{\Str}^n(A_0) = \hat{\Str}^{n+1}(A_0),
\label{closed_global_strategy}
\end{equation}
(where the assertion $A_0$ is identified with the tree reduced to one node whose label is $A_0$).
\begin{lem}: Let ${\cal D}$ be a formal system, $\Str$ a strategy for ${\cal D}$ and $\hat{\Str}$
a global strategy w.r.t. $\Str$. If $\hat{\Str}$ is terminating and $\hat{\Str}$ is closed, then 
${\cal D}$ is complete.
\label{term_closed_implies_complete}
\end{lem}
\begin{prooof}
Let $A_0 \in {\cal A}$. Under the hypothesis of the lemma, $\exists n_0 \in \bbbn$ such that
(\ref{terminating_global_strategy}) and (\ref{closed_global_strategy}) are both true.
Hence $t_\infty=\hat{\Str}^{n_0}(A_0)$ is a closed proof-tree for ${\cal S}$. By lemma 
\ref{trees_are_proofs} $\im(t_\infty)$ is a ${\cal D}$-proof relative to the set $\leaves(t_\infty) - \internal(t_\infty)$. Let $x$ be a leaf such that $t_\infty(x) \in \leaves(t_\infty) - \internal(t_\infty)$.
Let $A_0,A_1,\ldots,A_n=t_\infty(x)$ be the word labelling the path from the root to $x$.
As $x$ is closed and $t_\infty(x) \in \leaves(t_\infty) - \internal(t_\infty)$ by (\ref{leaf_condition}), 
$\varepsilon \in \Str(A_1 \cdots A_n)$ hence $\{A_1,\ldots,A_{n-1}\} \deducedir t_\infty(x)$.
It follows that $\im(t_\infty)$ is a ${\cal D}$-proof .
\end{prooof}

\section{Triangulations}
\label{triangulations}
Let $S_1, S_2, \cdots, S_d$ be a family of deterministic row-vectors  over the structured
alphabet $V$ ( i.e. $S_i \in \bbbD \bbbR \bbbB_{1,\lambda}\pa{V}$ where $\lambda \in \bbbn-\{0\}$). 
We recall $V$ is the alphabet associated with some dpda ${\cal M}$ as defined in 
section \ref{grammars}.\\
Let us consider a sequence $\Sys$ of $n$ ``weighted'' linear equations :
\begin{equation}
({\cal E}_i) : p_i, \sum^{d}_{j=1} \alpha_{i,j} S_j \;,\;
\sum^{d}_{j=1}\beta_{i,j} S_j
\label{THE_system}
\end{equation}
where $p_i \in \bbbn-\{0\}$, and $A = (\alpha_{i,j}), B = (\beta_{i,j})$ are deterministic rational
matrices of dimension $(n,d)$, with indices $m \leq i \leq m+n-1, 1 \leq j \leq d$.\\

For any weighted equation, ${\cal E} = (p, S, S')$, we recall the ``cost'' of
this equation is : $H({\cal E}) = p + 2 \cdot \Div(S, S')$.\\

Let us define an {\em oracle } on deterministic vectors as a mapping
$\OR: \bigcup_{\lambda \geq 1}\bbbD \bbbR \bbbB_{1,\lambda}\pa{V} \times \bbbD \bbbR \bbbB_{1,\lambda}\pa{V} \rightarrow {\cal P}(X^* \times X^*)$ such that:
$$ \forall (S,S') \in \bbbD \bbbR \bbbB_{1,\lambda}\pa{V} \times \bbbD \bbbR \bbbB_{1,\lambda}\pa{V},
S \sim S' \Rightarrow \OR(S,S') \in {\cal B}_\infty(S,S').$$
In other words, an oracle is a {\em choice} of $w$-$\bar{\psi}$-bisimulation for every pair
of equivalent vectors ( modulo $\sim$). Let us denote by $\Omega$ the set of all oracles.
Let us fix an oracle $\OR$ throughout this section.

We associate to every system (\ref{THE_system}) another equation, $\INV(\Sys)$,
which ``translates the equations of $\Sys$ into equations
over the coefficients $(\alpha_{i,j}, \beta_{i,j})$ only''\footnote{The function INV defined
in \cite{Sen97} was an ``elaborated version'' of the {\em inverse} systems defined in 
\cite{Mei89,Mei92} in the case of a single equation. We consider here a {\em relativization} of 
this notion to some oracle \OR.}.
The general idea of the construction of $\INV$ consists in iterating the transformation
used in the proof of $(1) \Rightarrow (2) \Rightarrow (3)$ in lemma \ref{fundamental_lemma2}, i.e.
the classical idea of {\em triangulating} a system of linear equations. Of course we must deal with
the weights and relate the construction with the deduction system ${\cal B}_0$.
 
We assume here that 
\begin{equation}
\forall j \in [1,d], S_j \neq \emptyset^{\lambda}.
\label{non_null_series}
\end{equation}

Let us define $\INV(\Sys)$, $\W(\Sys) \in \bbbn \cup \{\bot\}, \D(\Sys) \in \bbbn $
by induction on $n$. $\W(\Sys)$ is the {\em weight} of $\Sys$. $\D(\Sys)$
is the {\em weak codimension} of $\Sys$.\\
{\bf Case 1} : $\alpha_{m,*} \sim \beta_{m,*}$.\\
$$
\INV(\Sys) = (\W(\Sys), \alpha_{m,\ast}, \beta_{m,\ast}) ,
\W(\Sys) = p_m-1,\; \D(\Sys) = 0.
$$
{\bf Case 2} : $\alpha_{m,*} \not\sim \beta_{m,*}, n \geq 2,
p_{m+1} - p_m \geq 2 \cdot \Div(\alpha_{m,*}, \beta_{m,*}) +1.$\\
Let us consider
${\cal R} = \OR (\sum^{d}_{j=1} \alpha_{m,j} S_j,\sum^{d}_{j=1}\beta_{m,j} S_j)$,
$\nu = \Div(\alpha_{m,*},\beta_{m,*})$ and
\begin{equation} 
(u,u') = \mi \{(v,v') \in {\cal R} \cap X^{\leq \nu} \times X^{\leq \nu} \mid
\exists j \in [1, d], (\alpha_{m,*}\odot v = \epsilon^\lambda_j)
\Leftrightarrow (\beta_{m,*} \odot v' \neq \epsilon^\lambda_j) \}.
\label{choose_uu'}
\end{equation} 
Let us consider the integer $j_0 \in [1,d]$ such that 
$(\alpha_{m,*}\odot u = \epsilon^\lambda_{j_0})
\Leftrightarrow (\beta_{m,*} \odot u' \neq \epsilon^\lambda_{j_0})$.\\
{\bf Subcase 1} : $\alpha_{m,j_0} \odot u = \varepsilon, \beta_{m, j_0} \odot u' \neq \varepsilon$.\\
Let us consider the equation
$$({\cal E}'_m) :p_m + 2 \cdot |u|, S_{j_0} \;,\; \sum^d_{\stackrel{j=1}{j \neq j_0}} (\beta_{m,j_0} \odot u')^* (\beta_{m,j} \odot u')
S_j$$
and define a new system of weighted equations $\Sys' = ({\cal E}'_i)_{m+1
\leq i \leq m+n-1}$ by :
$$({\cal E}'_i) : p_i, \sum^{}_{j \neq j_0} [(\alpha_{i,j} + \alpha_{i,j_0}
(\beta_{m,j_0} \odot u')^* (\beta_{m,j} \odot u')] S_j \;,\; \sum^{}_{j \neq j_0}
[(\beta_{i,j} + \beta_{i,j_0} (\beta_{m,j_0} \odot u')^*  (\beta_{m,j} \odot u')]
S_j$$\\
where the above equation is seen as as an equation between two linear combinations of the $S_i$'s,
$1 \leq i \leq d$,
where the $j_0$-th coefficient is $\emptyset$ on both sides. We then define :
$$\INV(\Sys) = \INV(\Sys'), \W(\Sys) = \W(\Sys'), \D(\Sys) =\D(\Sys') + 1.
$$
{\bf Subcase 2} : $\alpha_{m,j_0} \odot u \neq \varepsilon, \beta_{m, j_0} \odot u' =
\varepsilon$.\\
(analogous to subcase 1).\\
{\bf Case 3} : $\alpha_{m,*} \not\sim \beta_{m,*}, n = 1$.\\
We then define:
$$\INV(\Sys) = \bot, \W(\Sys) = \bot, \D(\Sys) = 0,$$
where $\bot$ is a special symbol which can be understood as meaning ``undefined''.\\
{\bf Case 4} : $\alpha_{m,*} \not\sim \beta_{m,*}, n \geq 2,
p_{m+1} - p_m \leq 2 \cdot \Div(\alpha_{m,*}, \beta_{m,*}).$\\
We then define:
$$\INV(\Sys) = \bot, \W(\Sys) = \bot, \D(\Sys) = 0.$$ 

\begin{lem} : Let $\Sys$ be a system of weighted linear equations with deterministic rational coefficients.
 If $\INV(\Sys) \neq \bot$ 
then, $\INV(\Sys)$ is a weighted linear equation with deterministic rational coefficients.
\label{T0}
\end{lem}
\begin{prooof} Follows from lemmas \ref{Boxj},\ref{Boxj*} and the formula defining $\Sys'$
from $\Sys$.
\end{prooof}
From now on, and up to the end of this section, we simply write ``linear equation'' to mean ``
weighted linear equations with deterministic rational coefficients''.
\begin{lem} : Let $\Sys$ be a system of weighted linear equations with deterministic rational coefficients.
 If $\INV(\Sys) \neq \bot$ 
then:
\begin{enumerate}
\item $ \{\INV(\Sys)\} \cup 
\{{\cal E}_i \mid m \leq i \leq m+\D(\Sys)-1\} \deducedir {\cal E}_{m+\D(\Sys)}$\\
\item $\mi \{H({\cal E}_i ) \mid m \leq i \leq m+ \D(\Sys)\} = \infty \Longrightarrow
H(\INV(\Sys)) = \infty$.
\end{enumerate}
\label{T1}
\end{lem}
\begin{prooof} 
See on figure \ref{sche_T1} the ``graph of the deductions'' we use for
proving point (1). 
\begin{figure}[h]
\input{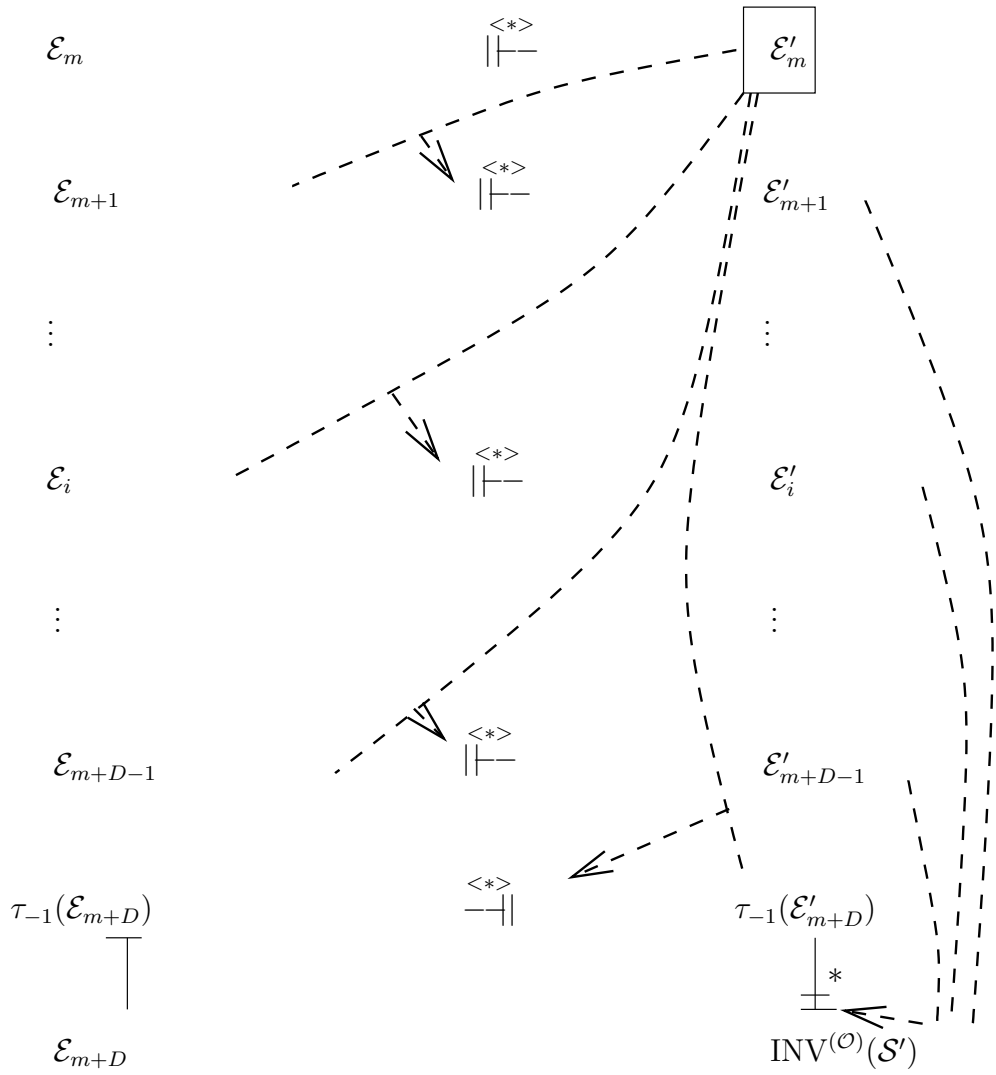}
\caption{Proof of lemma 5.2}
\label{sche_T1}
\end{figure} 
Let us prove by induction on $\D(\Sys)$ the following strengthened version of point (1): 
\begin{equation}
\{\INV(\Sys)\} \cup 
\{{\cal E}_i \mid m \leq i \leq m+\D(\Sys)-1\} \ededuce{*} \tau_{-1}({\cal E}_{m+\D(\Sys)})
\label{one_strengthened}
\end{equation}
where, for every integer $k \in \bbbz$, $\tau_{k}: \{ (p,S,S') \in {\cal A}\mid p \geq -k\} 
\rightarrow {\cal A}$ is the {\em translation}
map on the weights: $\tau_{k}(p,S,S') = (p+k,S,S')$.\\
{\bf if $\D(\Sys) = 0$} : as $\INV(\Sys) \neq \bot, \Sys$
must fulfill the hypothesis of case 1.
$${\cal E}_m = (p_m, \sum^{d}_{j=1} \alpha_{m,j} S_j, \sum^d_{j=1}
\beta_{m,j} S_j) = {\cal E}_{m+\D(\Sys)}$$
$$\INV(\Sys) = (p_m-1, \alpha_{m,*}, \beta_{m,*}).$$
Using rules (R7) we obtain :\\
$$\INV(\Sys) \ededuce{*} (p_{m}-1, \sum^d_{j=1} \alpha_{m,j}
S_j,\sum^d_{j=1}, \beta_{m,j} S_j)
= \tau_{-1}({\cal E}_m).$$
{\bf if $\D(\Sys) = n+1, n \geq 0$} : $\Sys$ must fulfill case 2.\\
$\bullet$ Suppose {\bf case 2, subcase 1} occurs.\\
As the relation ${\cal R}$ used in the construction of ${\cal E'}_m$ from ${\cal E}_m$ is 
a $w$-$\bar{\psi}$-bisimulation w.r.t. the pair of sides of equation ${\cal E}_m$, using (R5) 
and then (R6) ,(this is possible because $ \beta_{m,j_0} \odot u' \neq 
\epsilon$),
we obtain a deduction :
\begin{equation}
{\cal E}_m \ededuce{2 \cdot |u|+1} {\cal E'}_m.
\label{EmE'm}
\end{equation}
Using (R2,R8) we get that, for every $i \in [m+1,m+\D(\Sys)]$
$$
\{{\cal E}_i,  {\cal E'}_m\} \ededuce{*}(\ma
\{p_i, p_m + 2 \mid u \mid\}, \sum_{j \neq j_0}(\alpha_{i,j} +
\alpha_{i,j_0}(\beta_{m,j} \odot u')) S_j,  \sum_{j \neq j_0}\beta_{i,j} +
\beta_{i,j_0}(\beta_{m,j} \odot u')) S_j)
$$
but the hypothesis of case 2 implies that $\ma \{p_{m+1},  p_m + 2 \mid u
\mid\} = p_{m+1}$ and the fact that $\INV(\Sys')$ is defined implies
that $\forall i \in [m+1, m+\D(\Sys)], p_i \geq p_{m+1}$, hence,
 $\ma\{p_i, p_m + 2 \mid u \mid\} = p_i$ and the right-hand side of the
above deduction is exactly ${\cal E'}_i$. Hence,
\begin{equation}
\forall i \in [m+1,m+\D(\Sys)], \;\;\{{\cal E}_i,  {\cal E'}_m\} \ededuce{*} {\cal E'}_i.
\label{EiE'i}
\end{equation}

Using deductions (\ref{EmE'm}) and (\ref{EiE'i}),  we obtain that:
\begin{equation}
\{{\cal E}_i \mid m \leq i \leq m+\D(\Sys) - 1\} \ededuce{*} 
\{{\cal E'}_i \mid m \leq i \leq m+\D(\Sys) - 1\}.
\label{fromE_to_E'}
\end{equation}
By induction hypothesis :
$$
\INV(\Sys') \cup \{{\cal E}'_i \mid m+1 \leq i \leq m+1+\D(\Sys') -
1 \} \ededuce{*} \tau_{-1}({\cal E'}_{m+1+\D(\Sys')})
$$
which is equivalent to
\begin{equation}
\INV(\Sys) \cup \{{\cal E}'_i \mid m+1 \leq i \leq m+\D(\Sys)-1\}
\ededuce{*} \tau_{-1}({\cal E'}_{m+\D(\Sys)}).
\label{byinduction_hypo}
\end{equation}
As $ p_m + 2 \cdot |u| \leq p_{m+1} -1 \leq p_{m+\D(\Sys)}-1$, we have also the following inverse
deduction ( which is similar to deduction (\ref{EiE'i})):
\begin{equation}
\{{\cal E'}_m, \tau_{-1}({\cal E'}_{m+\D(\Sys)})\} \ededuce{*} \tau_{-1}({\cal E}_{m+\D(\Sys)}).
\label{fromE'_to_E}
\end{equation}
Combining together deductions (\ref{fromE_to_E'}) (\ref{byinduction_hypo}) and (\ref{fromE'_to_E}), we have proved 
(\ref{one_strengthened}). Using rule (R0), this last deduction leads to point (1) of the lemma.\\
$\bullet$ Suppose now that {\bf case 2, subcase 2} occurs.\\
This case can bet treated in the same way as subcase 1, just by exchanging the roles of 
$\alpha,\beta$.\\

\noindent Let us prove statement (2) of the lemma.\\
We prove by induction on $\D(\Sys)$ the statement: 
\begin{equation}
\mi \{H({\cal E}_i ) \mid m \leq i \leq m+ \D(\Sys)\} = \infty \Longrightarrow
H(\INV(\Sys)) = \infty.
\label{H_equals_infty}
\end{equation}
{\bf if $\D(\Sys) = 0$} : as $\INV(\Sys) \neq \perp$, case 1 must occur.
$\alpha_{m,*} \sim \beta_{m,*}$ implies that  $H(\INV(\Sys)) = \infty$,
 hence the statement is true.\\
%
{\bf if $\D(\Sys) = p+1, p \geq 0$} : as $\D(\Sys) \geq 1$ and $\INV(\Sys) \neq \perp$, case 2 
must occur.\\ 
Using deductions (\ref{EmE'm}) and (\ref{EiE'i}) established above we obtain that :\\
$$
\{{\cal E}_i \mid m \leq i \leq m+\D(\Sys)\} \ededuce{*}
\{{\cal E}'_i \mid m+1 \leq i \leq m+1+\D(\Sys')\},
$$
which proves that
\begin{equation}
\mi \{H({\cal E}_i ) \mid m \leq i \leq m+ \D(\Sys)\} \leq
\mi \{H({\cal E}'_i ) \mid m+1 \leq i \leq m+1+ \D(\Sys')\}.
\label{eqT1}
\end{equation}
As $\D(\Sys') = \D(\Sys)-1$, we can use the induction hypothesis:
\begin{equation}
\mi \{H({\cal E}'_i ) \mid m+1 \leq i \leq m+1+D(\Sys')\} = \infty \Longrightarrow
H(\INV(\Sys'))= \infty.
\label{eqT2}
\end{equation}
As $\INV(\Sys) = \INV(\Sys')$, (\ref{eqT1},\ref{eqT2}) imply statement (\ref{H_equals_infty}).  
\end{prooof}

\begin{lem} : Let $\Sys$ be a system of linear equations satisfying the
hypothesis of case 2. Then, $\forall i \in [m+1, m+n-1]$,\\ 
$\parallel \alpha'_{i,*} \parallel \leq \parallel
\alpha_{i,*} \parallel + \parallel \beta_{m,*} \parallel + K_0 \mid u \mid,
\parallel \beta'_{i,*} \parallel \leq \parallel  \beta_{i,*} \parallel +
\parallel \beta_{m,*} \parallel + K_0 \mid u \mid.$
\label{T2}
\end{lem}
\begin{prooof}The formula defining ${\cal S}'$ from ${\cal S}$ show that:
$$
\alpha'_{i,\ast} = \alpha_{i,\ast}\Box_{j_0}(\Box^*_{j_0} (\beta_{m,\ast} \odot u'));\;\;
\beta'_{i,\ast} = \beta_{i,\ast}\Box_{j_0}(\Box^*_{j_0} (\beta_{m,\ast}\odot u') ).
$$
From these equalities and lemmas \ref{Boxj},\ref{Boxj*}, \ref{norm_of_action} the 
inequalities on the norm follow.
\end{prooof}

Let us consider the function $F$ defined by :\\
$$F(d,n) = \ma\{\Div(A,B) \mid A, B \in {\sf
DRB}_{1,d}\pa{V}, \parallel A \parallel \leq n, \parallel B \parallel \leq n, 
A \not\sim B\}.$$

For every integer parameters $K_0,K_1,K_2,K_3,K_4 \in \bbbn -\{0\}$, we
define integer sequences $(\delta_i, \ell_i, L_i, s_i, S_i,
\Sigma_i)_{m \leq i \leq m+n-1}$ by :\\
\begin{equation}
\delta_m = 0, \ell_m = 0, L_m = K_2,
s_m = K_3 \cdot K_2+ K_4, S_m = 0, \Sigma_m = 0,
\label{6_initialisation}
\end{equation}
\begin{equation}
\left\{ \begin{array}{ll}
\delta_{i+1} & = 2 \cdot F(d,s_i + \Sigma_i) + 1 \\
\ell_{i+1} & = 2 \cdot \delta_{i+1}+ 3 \\ 
L_{i+1} & = K_1 \cdot(L_i + \ell_{i+1}) + K_2 \\
s_{i+1} & = K_3 \cdot L_{i+1} + K_4 \\
S_{i+1} & = s_i + \Sigma_i + K_0 F (d,s_i + \Sigma_i) \\
\Sigma_{i+1} & = \Sigma_i + S_{i+1}
\end{array}
\right.
\label{6_recurrences} 
\end{equation}
for $ m \leq i \leq m+n-2$.

These sequences are intended to have the following meanings when $K_0,K_1,K_2,K_3,K_4$ are chosen 
to be the constants defined in section \ref{constants} and the equations $({\cal E}_i)$ are 
labelling nodes of a B-stacking sequence (see section \ref{NSS}):
\begin{description}
\item $\delta_{i+1} \leq$ increase of weight between ${\cal E}_i,
{\cal E}_{i+1}$
\item $\ell_{i+1} \geq$ increase of depth between ${\cal E}_i,
{\cal E}_{i+1}$
\item $\L_{i+1} \geq$ increase of depth between ${\cal E}_m,
{\cal E}_{i+1}$
\item $s_{i+1} \geq$ size of the coefficients of ${\cal E}_{i+1}$
\item $S_{i+1} \geq$ size of the coefficients of ${\cal E}^{(i+1-m)}_{i+1}$ 
( these systems are introduced below in the proof of lemma \ref{T3})
\item $\Sigma_{i+1} \geq$ increase of the coefficients between ${\cal E}^{(i-m)}_k, 
{\cal E}^{(i+1-m)}_k (\mbox{ for } k \geq i+1)$.
\end{description}

For every linear equation ${\cal E} = (p,
\sum^{d}_{j=1} \alpha_j S_j \;,\; \sum^{d}_{j=1} \beta_j S_j)$, we
define 
$$\mid\parallel {\cal E} \mid \parallel = max\{\parallel (\alpha_1,\ldots,\alpha_d)
\parallel, \parallel (\beta_1,\ldots, \beta_d) \parallel\}.$$
\begin{lem}
Let $\Sys = ({\cal E}_i)_{m \leq i \leq m+d-1}$ be a system of $d$
linear equations such that $H({\cal E}_i)= \infty$ ( for every $i$) and :
\begin{description}
\item (1) $\forall i \in [m, m+d-1], \mid\parallel {\cal E}_i
\mid \parallel \leq s_i$
\item (2) $\forall i \in [m, m+d-2], {\rm W}({\cal E}_{i+1}) - {\rm W}({\cal E}_i)
\geq \delta_{i+1}.$
\end{description}
Then 
\begin{description}
\item (3) $\INV(\Sys) \neq \perp$,
\item (4) $\D(\Sys) \leq d-1, $
\item (5) $\mid\parallel \INV(\Sys) \mid\parallel \leq \Sigma_{m+\D(\Sys)} + s_{m+\D(\Sys)}$.
\end{description}
\label{T3}
\end{lem}
\begin{prooof}
(Figure \ref{sche_T2} might help the reader to follow the definitions below). 
\begin{figure}
\input{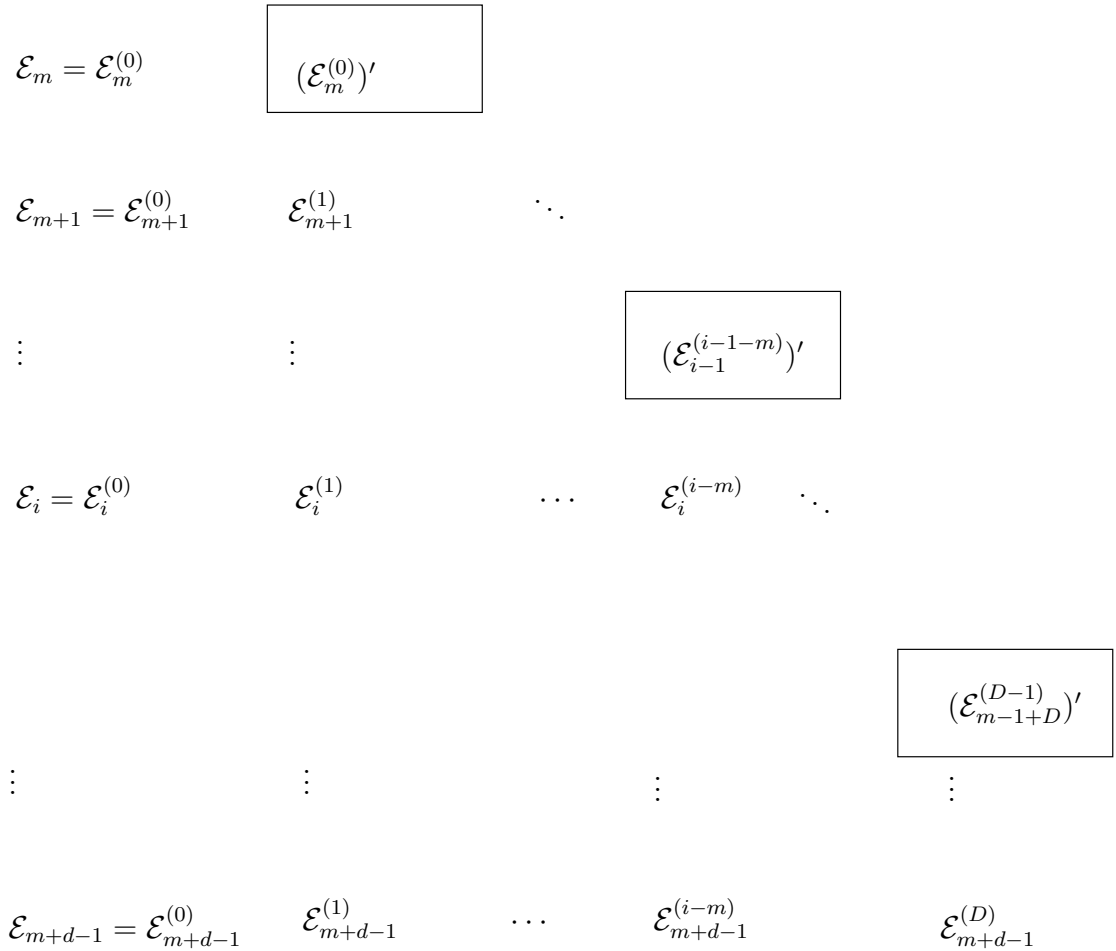}
\caption{Proof of lemma 5.4}
\label{sche_T2}
\end{figure}
Let us define a sequence of systems $\Sys^{(i-m)}= 
({\cal E}^{(i-m)}_k)_{m \leq i \leq k \leq m + d-1 }$, where $i \in [m, m + \D(\Sys)]$, 
by induction :
\begin{itemize}
\item ${\cal E}^{(0)}_k = {\cal E}_k$ for $m \leq k \leq m + d-1$
\item if case 1 or case 3 or case 4 is realized, $\D(\Sys) = 0$, hence $\Sys^{(i-m)}$
is well-defined for $m \leq i \leq m+\D({\Sys})$
\item if case 2 is realized then we set :
$\forall i \geq m+1, {\cal E}^{(i-m)}_k = ({\cal E'}_k)^{(i-m-1)}$, for
$m+1 \leq k \leq m+ d-1.$
\end{itemize}
Let us prove by induction on $i \in [m, m + \D(\Sys)]$ that, $\forall k \in [i, m+ d-1]$ :
\begin{equation}
\mid\parallel {\cal E}^{(i-m)}_k \mid \parallel \leq s_k + \Sigma_i.
\label{inequationI}
\end{equation}
{\bf $i= m$ } : in this case
$$\mid\parallel{\cal E}^{(i-m)}_k\mid\parallel = \mid\parallel {\cal E}_k
\mid\parallel \leq s_k = s_k + \Sigma_m.$$
{\bf $i+1 \leq m+\D(\Sys)$}: in this case, by lemma \ref{T2}, 
$$\mid\parallel {\cal E}^{(i+1-m)}_k \mid\parallel \leq
\mid\parallel {\cal E}^{(i-m)}_k\mid \parallel + \mid \parallel  
{\cal E}^{(i-m)}_i  \mid\parallel + K_0 \mid u \mid$$

\noindent where 
${\cal R} = \OR (\sum^{d}_{j=1} \alpha^{(i-m)}_{i,j} S_j,\sum^{d}_{j=1}\beta^{(i-m)}_{i,j} S_j)$,
$\nu = \Div(\alpha^{(i-m)}_{i,*},\beta^{(i-m)}_{i,*}),$  and
$(u,u') = \mi \{(v,v') \in {\cal R} \cap X^{\leq \nu} \times X^{\leq \nu} \mid
\exists j \in [1, d], (\alpha^{(i-m)}_{i,*}\odot v = \epsilon^\lambda_j)
\Leftrightarrow (\beta^{(i-m)}_{i,*} \odot v' \neq \epsilon^\lambda_j) \}.$
\\
By definition of $F$ and the induction hypothesis :
$$
\mid u \mid \leq F(d,\mid\parallel {\cal E}^{(i-m)}_i \mid \parallel)
\leq F(d,s_i + \Sigma_i).
$$
Hence
\begin{eqnarray*}
\mid\parallel {\cal E}^{(i+1-m)}_k \mid\parallel \leq 
(s_k + \Sigma_i) + (s_i + \Sigma_i)+
K_0 F(d,s_i + \Sigma_i) &=& 
(s_k + \Sigma_i) + S_{i+1} \\
&=& s_k + \Sigma_{i+1}.
\end{eqnarray*}
Let us notice that $\D(\Sys)$ is always an integer and that this proof is valid for 
$m \leq i \leq m+\D(\Sys), i \leq k \leq m + d-1$.\\
Let us prove now that $\INV(\Sys) \neq \perp$. Let us
consider the system $({\cal E}^{(\D(\Sys))}_k)_{m+\D(\Sys) \leq k \leq
m+d-1}$.\\
{\bf If $\D(\Sys) = d-1$}, $({\cal E}^{(\D(\Sys))})$ fulfills either case 1 or case 3 of 
the definition of $\INV$ (just because this system consists of a single equation).\\
Using the successive  deductions (\ref{EmE'm})(\ref{EiE'i}) established in the proof of 
lemma \ref{T1} we get that:
$$
\{{\cal E}_i \mid m \leq i \leq m+d - 1\} \ededuce{*} 
\{{\cal E}^{(d-1)}_{m+d-1}\}.
$$
Using now the hypothesis that $H({\cal E}_i) = \infty$ ( for $m \leq i \leq m+d - 1$), we obtain:
\begin{equation}
H({\cal E}^{(d-1)}_{m+d-1}) = \infty.
\label{Eq_infty}
\end{equation}
For any system of equations $\Sys$ , let us define the {\em support} of the system as
$$\support(\Sys) = \{j \in [1,d] \mid \sum^{m+n-1}_{i=m}
\alpha_{i,j} + \beta_{i,j} \neq \emptyset\}.
$$
Let us consider $ \delta = \card(\support(\Sys^{(d-1)})$. One can prove by induction on $i$ 
that:
$$
\card(\support(\Sys^{(i-m)}) \leq d - i + m,
$$
hence
$$
\delta = \card(\support(\Sys^{(d-1)}) \leq d - (d-1) = 1.
$$
\begin{itemize}
\item If $\delta = 1, \support(\Sys^{(d-1)}) = \{j_0\}$, for some $j_0 \in [1,d]$.\\
By corollary \ref{algebraic_version} point C3 and hypothesis (\ref{non_null_series}), the 
implication 
$$[(\alpha^{(d-1)}_{m+d-1, j_0} S_{j_0} \sim \beta^{(d-1)}_{m+d-1,j_0} S_{j_0}) 
\Longrightarrow 
\alpha^{(d-1)}_{m+d-1,j_0} \sim \beta^{(d-1)}_{m+d-1,j_0}]$$ holds. Hence , by (\ref{Eq_infty}),
$\alpha^{(d-1)}_{m+d-1,j_0} \sim \beta^{(d-1)}_{m+d-1,j_0}$, i.e. $\Sys^{(d-1))}$ fulfills case 1, so that
$$ \INV( \Sys) = \INV(\Sys^{(d-1)}) \neq \bot.$$
\item If $\delta = 0,\support(\Sys) = \emptyset$.\\
Then $\alpha^{(d-1)}_{m+d-1,\ast}=\beta^{(d-1)}_{m+d-1,\ast} = \emptyset^d$. 
Here also $\Sys^{(d-1)}$ fulfills case 1.
\end{itemize}
{\bf If $\D(\Sys) < d-1$}, by hypothesis :
$${\rm W}({\cal E}_{m+ \D(\Sys)+1}) - {\rm W}({\cal E}_{m + \D(\Sys)}) \geq
\delta_{m + \D(\Sys)+1} = 2 F(d,s_{m+\D(\Sys)} +
\Sigma_{m+\D(\Sys)})+1.$$
If $\;$ $\alpha^{\D(\Sys)}_{m + \D(\Sys), \ast} \sim \beta^{\D(\Sys)}_{m + \D(\Sys), \ast}$,
then ${\cal E}^{(\D(\Sys))}_{m + \D(\Sys)}$ fulfills case 1 of the definition of $\;\INV$, hence
$\INV(\Sys) \neq \perp$.\\ Otherwise, let us consider:
$${\cal R} = \OR (\sum^{d}_{j=1} \alpha^{(\D(\Sys))}_{m + \D(\Sys),j} S_j,\sum^{d}_{j=1}\beta^{(\D(\Sys))}_{m+ \D(\Sys),j} S_j),$$
$$\nu = \Div(\alpha^{(\D(\Sys))}_{m+ \D(\Sys),*},\beta^{(\D(\Sys))}_{m+ \D(\Sys),*}),
\mbox{ and }$$ 
\begin{eqnarray*} 
(u,u') &=& \mi \{(v,v') \in {\cal R} \cap X^{\leq \nu} \times X^{\leq \nu} \mid
\exists j \in [1, d], (\alpha^{(\D(\Sys))}_{m+ \D(\Sys),*}\odot v = \epsilon^\lambda_j)
\Leftrightarrow\\
& & (\beta^{(\D(\Sys))}_{m+ \D(\Sys),*} \odot v' \neq \epsilon^\lambda_j) \}.
\end{eqnarray*}
By definition of $F$  and inequality (\ref{inequationI}),
$$\mid u \mid \leq F(d,\mid\parallel {\cal E}^{(\D(\Sys))}_{m+\D(\Sys)}
\mid\parallel) \leq F(d,s_{m+\D(\Sys)} + \Sigma_{m+\D(\Sys)}).$$
Hence $p_{m+\D(\Sys)+1} - p_{m+\D(\Sys)} \geq 2 \mid u \mid + 1$ i.e.
the hypothesis of case 2 is realized. This proves that $ \D(\Sys^{(\D(\Sys))}) \geq 1$
 while in fact, $ \D(\Sys^{(\D(\Sys))}) = 0$. This contradiction shows that this last case
($ \D(\Sys) < d-1$ and ${\cal E}^{(\D(\Sys))}_{m+\D(\Sys)}$ not fulfilling case 1 of definition
of $\INV$) is impossible. We have proved point (3) of the lemma.
\end{prooof}

\section{Constants}
\label{constants}
Let us {\em fix} a birooted dpda $\cal M$, a s.r.  morphism $\bar{\psi}$  and an initial equation 
$A_0=(\Pi_0,S^-_0,S^+_0) \in \bbbn \times {\bbbD \bbbR \bbbB}_{1,\lambda_0}\pa{V} \times 
{\bbbD \bbbR \bbbB}_{1,\lambda_0}\pa{V}$ in the corresponding set of assertions.
This short section is devoted to the definition of some integer {\em constants}: these integers
are constant in the sense that they are depending only on this triple : $({\cal M},\bar{\psi},A_0)$.
The
{\em motivation} of each of these definitions will appear later on, in different places for the 
different constants. The equations below provide merely an overview of the dependencies betweens 
these constants and allow to check that the definitions are sound ( i.e. there is no hidden loop 
in the dependencies).
\begin{equation}
k_0= \ma\{ \nu(v) \mid v \in V\},\;\;
k_1= \ma\{ 2 k_0 + 1 , 3 \},
\label{k01}
\end{equation}
\begin{eqnarray}
K_0 &=& \ma\{ \| (E_1,E_2, \ldots,E_n) \odot x\|\mid (E_i)_{1 \leq i \leq n} 
\mbox{ is a bijective numbering } \nonumber\\
&& \mbox{ of some class in } V/\smile, x \in X\}.
\label{K0recall}
\end{eqnarray} 
$K_0$ serves as an upper-bound on the possible increase of norm under the right-action of a single letter
$x \in X$ , see lemma \ref{norm_of_action}.
\begin{equation}
D_1= k_0 \cdot K_0 + |Q| + 2,\;\;
k_2= D_1 \cdot k_1 \cdot K_0 + 2 \cdot k_1 \cdot K_0 + K_0.
\label{k2D1}
\end{equation}
$k_1$ is used in the definition of strategy $T_B$ (section \ref{strategies_B0}), $D_1$ appears 
as an upper-bound on the marked part of series and $k_2$ is used
in lemma \ref{prolog}.
\begin{equation}
k_3= k_2 + k_1\cdot K_0,\;\; 
k_4= (k_3 +1)\cdot K_0 + k_1.
\label{k3k4}
\end{equation}
$k_3$ appears in 
in lemma \ref{argTB_swim}, $k_4$ is used in the definition (\ref{V0}) of the d-space $V_0$.
\begin{equation}
K_1 = k_1 \cdot K_0 + 1 ,\;\; K_2 = k_1^2\cdot D_1 \cdot K_0 + k_1^2 \cdot K_0 + 
2\cdot k_1 \cdot K_0 + D_1 \cdot k_1 + 2 \cdot k_1 +4.
\label{K12}
\end{equation}
These constants $K_1, K_2$ appear in lemma \ref{K1K2}.
\begin{equation}
K_3 = k_0 |Q|, \;\; K_4 = D_1.
\label{K34}
\end{equation}
These constants $K_3, K_4$ appear in lemma \ref{K3K4}.

\begin{equation}
d_0= \card(X^{\leq k_4}).
\label{d0}
\end{equation}
$d_0$ appears as an upper-bound on the dimension of the d-space $V_0$ defined by equation
(\ref{V0}) and used in lemma \ref{K1K2}.
We consider now the integer sequences $(\delta_i, \ell_i, L_i, s_i, S_i,
\Sigma_i)_{m \leq i \leq m+n-1}$ defined by the relations (\ref{6_recurrences}) of section
\ref{triangulations} where the parameters $K_0,K_1,\ldots,K_4$ are chosen to be the above constants
and $m=1,n=d=d_0$. Equivalently, they are defined by:
\begin{equation}
\delta_1 = 0, \ell_1 = 0, L_1 = K_2,
s_1 = K_3 \cdot K_2+ K_4, S_1 = 0, \Sigma_1 = 0,
\label{6_initialisation_constants}
\end{equation}
\begin{equation}
\left\{ \begin{array}{ll}
\delta_{i+1} & = 2 \cdot F(d_0,s_i + \Sigma_i) + 1 \\
\ell_{i+1} & = 2 \cdot \delta_{i+1}+ 3 \\ 
L_{i+1} & = K_1 \cdot(L_i + \ell_{i+1}) + K_2 \\
s_{i+1} & = K_3 \cdot L_{i+1} + K_4 \\
S_{i+1} & = s_i + \Sigma_i + K_0 \cdot F (d_0,s_i + \Sigma_i) \\
\Sigma_{i+1} & = \Sigma_i + S_{i+1}
\end{array}
\right.
\label{6_recurrences_constants} 
\end{equation}
for $ 1 \leq i \leq d_0-1$. The function $F$ is defined in section \ref{triangulations} and depends on 
the pair $({\cal M}, \bar{\psi})$ only.
\begin{equation}
D_2 = \ma\{\Sigma_{d_0} + s_{d_0}, \|S^{-}_0\|,\|S^{+}_0\|\},
\label{D2}
\end{equation}
$\Sigma_{d_0} + s_{d_0}$ appears in the conclusion of lemma \ref{T3} when we take $d=d_0$ 
in the hypothesis and suppose that $\D(\Sys)$ has its maximal possible value i.e.  
$\D(\Sys)= d_0-1$.
It is used as an upper-bound on the norm of vectors at the root of the trees $\tau$ analysed 
in part \ref{trees} (inequation (\ref{unmarked})).\\
\begin{equation}
\lambda_2 = \ma\{\lambda_0,d_0\},
\label{lambda2}
\end{equation}
The integer $\lambda_2$ is used as an upper-bound on the length of vectors at the root of the trees $\tau$ analysed in part \ref{trees} (inequation (\ref{lambda_small})).\\
\begin{equation}
N_0 = 1 + k_3 + D_2.
\label{N0}
\end{equation}
$N_0$ appears as a lower bound for the norm in the definition of a B-stacking sequence (section
\ref{NSS}, condition (\ref{B_stacking})).\\

\section{Strategies for ${\cal B}_0$}
\label{strategies_B0}
Let us define strategies for the particular system ${\cal B}_0$.\\
\subsection{Strategies}
\label{local_strategies}
We shall define first auxiliary strategies $T_{cut},T_{\emptyset}, T_{\varepsilon}$, and then
for every oracle $\OR \in \Omega$ auxiliary strategies
$T^{(\OR)}_A, T^{(\OR)}_B, T^{(\OR)}_C$, we define the strategies $T_A, T_B, T_C$ and finally 
the  ``compound'' strategies $\Str^{(\OR)}_{AB}, \Str^{(\OR)}_{ABC},\Str_{AB}, \Str_{ABC}$. 
Let us fix here some total ordering on $X : x_1 < x_2 < \cdots <x_{\alpha}$.
\begin{description}
\item{$T_{cut}$}:\\
$B_1 \cdots B_m \in T_{cut} (A_1 \cdots A_n)$ iff $\exists i
\in [1, n-1], \exists S, T,$ 
$$A_i = (p_i, S, T), A_n = (p_n, S, T), p_i < p_n \mbox{ and } m=0
\footnote{ i.e. $B_1 \cdots B_m = \epsilon$} .$$
\item{$T_{\emptyset}$:}\\
$B_1 \cdots B_m \in T_{\emptyset}(A_1 A_2 \cdots A_n)$ iff $\exists S, T,$
$$A_n = (p,S,T), p \geq 0, S=T=\emptyset^{\lambda} \mbox{ and } m = 0.$$
\item{$T_{\varepsilon}$:}\\
 $B_1 \cdots B_m \in T_{\varepsilon} (A_1 \cdots A_n)$ iff
$$A_n = (p, S, T), p \geq 0, S=T=\varepsilon^{\lambda}_i 
(\mbox{ for some } i \in [1,\lambda]) \mbox{ and }m = 0.$$
\end{description}
Let us consider an oracle $\OR \in \Omega$.
\begin{description}
\item{$T^{(\OR)}_A$}:\\
$B_1 \cdots B_m \in T^{(\OR)}_A(A_1 \cdots A_n)$ iff
$$A_n = (p, S, T), \mid X \mid \leq m \leq  \mid X \mid^2, B_1 = (p+1, S \odot x_1, T \odot x'_1),
\cdots, B_m = (p+1, S \odot x_m, T \odot x'_m),$$ 
where $S \not\equiv \varepsilon, T \not\equiv \varepsilon, \OR(S,T)\cap X \times X = 
\{(x_1,x'_1),\ldots, (x_i,x'_i), \ldots,(x_m,x'_m)\}. $
\item{$T^{(\OR),+}_B$}:\\
$B_1 \cdots B_m \in T^+_B(A_1 \cdots A_n)$ iff
$n \geq k_1+1, A_{n-k_1} = (\pi, \overline{U}, U')$, (where $\overline
U$ is unmarked)
$$U' = \sum_{k=1}^{q}E_k \cdot \cdot \Phi_k\;\;\;\mbox{ for some } q \in \bbbn, E_k \in V,$$
$(E_k)_{1 \leq k \leq}$ bijective numbering of a class in $V/\smile$, $\Phi_k \in {\sf DRB}_{1,\lambda}\pa{V}
$\\
$A_i = (\pi + k_1 + i - n, U_i, U'_i)$ for $n-k_1 \leq i \leq n$,
$(U_i)_{n-k_1\leq i\leq n}$ is a derivation,\\
$(U'_i)_{n-k_1\leq i\leq n}$ is a ``stacking derivation'' (see definitions in \S 
\ref{stacking_derivations}),
$$U'_n =\sum_{k=1}^{q} (E_k \odot u) \cdot \Phi_k,\;\;
\mbox{ for some } u \in X^*,$$
$m = 1, B_1 = (\pi + k_1 - 1, V, V'), V = U_n$,
$$V' = \sum_{k=1}^{q} \bar{\rho_e}(E_k \odot u)\cdot (\overline{U} \odot u_k)$$
where
$\forall k \in [1,q], u'_k = \mi(\varphi (E_k))$, and 
 if ${\cal R}=\OR(S,T),  \forall k \in [1,q], u_k = \mi\{{\cal R}^{-1} (u'_k)\}.$
\item{$T^{(\OR),-}_B$}:\\
$T^{(\OR),-}_B$ is defined in the same way as $T^{(\OR),+}_B$ by exchanging the left series
$(S^-)$ and right series $(S^+)$ in every assertion $(p, S^-, S^+)$.\\
\item{$T_C^{(\OR)}$}:\\
 $B_1 \cdots B_m \in T^{(\OR)}_C (A_1 \cdots A_n)$ iff there exists
$d \in [1,d_0], D \in [0,d-1], \lambda \in \bbbn -\{0\}, S_1, S_2, \cdots, S_d \in 
{\sf DRB}_{1,\lambda}\pa{V}-\{ \emptyset^{\lambda}\}, 
1 \leq \kappa_1 < \kappa_2 < \cdots < \kappa_{D+1} =n,$ such that, 
\begin{description}
\item{(C1)} every equation ${\cal E}_i = A_{\kappa_i}=(p_{\kappa_i}
S^-_{p_{\kappa_i}}, S^+_{p_{\kappa_i}})$ is a weighted equation over $S_1, S_2, \cdots,
S_d$, with $p_{\kappa_i} \geq 1$,
\item{(C2)}$\D(\Sys)=D$ (where $\Sys = ({\cal E}_i)_{1 \leq i \leq D+1}$),
\item{(C3)}$ \INV(\Sys) \neq \bot,  \NN{\INV(\Sys)} \leq \Sigma_{d_0} + s_{d_0}$  ,
\item{(C4)} $m=1$ and $B_1= \rho_e(\INV(\Sys))$ ( where $\rho_e$ is the obvious extension of 
$\rho_e$ to weighted pairs of deterministic row-vectors; in other words the 
result of $T^{(\OR)}_C$ is $\INV(\Sys)$ where the marks have been removed).\\
\end{description}
\end{description}
We then set, for every $W \in {\cal A}^+$:
$$T_A(W) = \bigcup_{\OR \in \Omega} T^{(\OR)}_A(W),$$
$$T^{+}_B(W) = \bigcup_{\OR \in \Omega}T^{(\OR),+}_B(W),\;\;
T^{-}_B(W) = \bigcup_{\OR \in \Omega}T^{(\OR),-}_B(W),$$  
$$T_C(W) = \bigcup_{\OR \in \Omega} T_C^{(\OR)}(W).$$
\begin{lem} : $T_{cut},T_{\emptyset}, T_{\varepsilon},T_A$ are ${\cal B}_0$-strategies.
\label{B0_strategies}
\end{lem}
\begin{prooof}\\
$T_{cut}$ : (S1) is true by rule R0. (S2) is trivially true.\\
$T_{\emptyset}$: (S1) is true by rule R'3. (S2) is trivially true.\\
$T_{\varepsilon}$: (S1) is true by rule R'3. (S2) is trivially true.\\ 
$T_A$ : by rule (R4), $\{B_j \mid 1 \leq j \leq m\} {\ededucedir}_4\; A_n$, which
proves (S1).
Suppose $H(A_n) = \infty$ i.e. $S \sim T$. Then, $\forall j \in [1,m], S \odot x_j
\sim T \odot x'_j$, so that $\mi\{H(B_j) \mid 1 \leq j \leq m \} = \infty$. (S2)
is proved.\\
\end{prooof}

\begin{lem} : $T^+_B, T^-_B$ are ${\cal B}_0$-strategies.
\label{TB_are_strategies}
\end{lem}
\begin{prooof} Let us show that $T^+_B$ is a ${\cal B}_0$-strategy.\\
Let us use the notation of the definition of $T^{(\OR),+}_B$.
Let ${\cal H} = \{(\pi, \overline{U}, U'), (\pi + k_1-1, V, V')\}$.
Let us show that
\begin{equation} 
{\cal H} \ededuce{*}_{{\cal B}_0} (\pi + k_1-1, U_n, U'_n).
\label{num0}
\end{equation}
Using rule (R5) we obtain: $\forall k \in [1,q]$,
\begin{eqnarray}
\{(\pi, \overline{U}, U')\} = 
\{(\pi, \overline{U}, \sum_{j=1}^q E_j \cdot \Phi_j)\} &\ededuce{*}_{R5}& 
(\pi + 2\cdot \mid u_k \mid, \overline{U} \odot u_k, U' \odot u'_k) \nonumber \\
&\ededuce{*}_{R0}  &
(\pi + 2\cdot k_0, \overline{U} \odot u_k, U' \odot u'_k )\nonumber\\
&=&
(\pi + 2\cdot k_0, \overline{U} \odot u_k, \Phi_k).
\label{num1}
\end{eqnarray}

Using rule (R'3), 
\begin{equation}
\emptyset \ededucedir_{R'3} (0, (\rho_e(E_1 \odot u),\ldots,\rho_e(E_q \odot u)),
(E_1,\ldots,E_q)).
\label{num4}
\end{equation}
Using (\ref{num4}),(\ref{num1}) and rules (R3),(R7),(R8), we obtain :
\begin{eqnarray}
\{(\pi, \overline{U}, U')\} &\ededuce{*}_{{\cal B}_0}& (\pi + 2 k_0, 
\sum_{k=1}^q (E_k \odot u) \cdot \Phi_k, 
\sum_{k=1}^q \rho_e(E_k \odot u) \cdot (\bar{U} \odot u_k)) \nonumber\\
&=& \{(\pi, \overline{U}, U')\} \ededuce{*} (\pi + 2 k_0, U'_n, V').
\label{num7}
\end{eqnarray}
Let us recall that $U_n = V$. Hence, by ($R0,R1,R2$)\\
\begin{equation}
\{(\pi + k_1-1, V, V'), (\pi + 2 k_0, U'_n, V')\}\ededuce{*}_{\cal C} 
(\pi + k_1-1, U_n, U'_n).
\label{num8}
\end{equation}

By (\ref{num7},\ref{num8}),(\ref{num0}) is proved. Using now (\ref{num0}) and rule ($R0$), 
we obtain:
\begin{equation}
{\cal H} \ededuce{*}_{{\cal B}_0} 
(\pi + k_1-1, U_n, U'_n)\deducedir_{R0} (\pi+k_1, U_n, U'_n).
\label{num9}
\end{equation}
i.e. $T^+_B$ fulfills (S1).\\
Let us suppose now that $\forall i \in [n-k_1,n], U_i \sim U'_i$.
Then, by (\ref{num7}), $U'_n \sim V'$ and by hypothesis $V = U_n \sim U'_n$.
Hence $V \sim V'$. This shows that $T^+_B$ fulfills (S2).\\
An analogous proof can obviously be written for $T^-_B$.
\end{prooof}

\begin{lem}
Let $(p,S,S')$ be a weighted equation , i.e. $p \in \bbbn, \lambda \in \bbbn - \{0\},
S,S' \in \bbbD \bbbR \bbbB_{1,\lambda}\pa{V}$.
Then $ \{(p,S,S')\} \ededuce{*}_{\cal C} \{(p,\rho_e(S),\rho_e(S'))\}$ and 
$\{(p,\rho_e(S),\rho_e(S'))\} \ededuce{*}_{\cal C}\{(p,S,S')\}$.
\label{rhoe_ornot_rhoe}
\end{lem}
\begin{prooof}
Follows easily from (R1),(R2),(R'3).
\end{prooof}
\begin{lem}For every $\OR \in \Omega$, $T^{\OR}_C$ is a ${\cal B}_0$-strategy.
\label{TC}
\end{lem}
\begin{prooof}
By lemma \ref{T1}, point (1), combined with lemma \ref{rhoe_ornot_rhoe}, (S1) is proved.
By lemma \ref{T1}, point (2), combined with lemma \ref{rhoe_ornot_rhoe}, (S2) is proved.
\end{prooof}

Let us define the strategy $\Str_{ABC}$ by :
 for every $W = A_1 A_2 \cdots A_n$,
\begin{description}
\item{(0)} if $T_{cut}(W) \neq \emptyset$,  then $\Str_{ABC}(W) = T_{cut}(W)$
\item{(1)} elsif $T_{\emptyset}(W) \neq \emptyset$, 
then $\Str_{ABC}(W) =T_{\emptyset}(W)$
\item{(2)} elsif $T_{\varepsilon}(W) \neq \emptyset$, 
then $\Str_{ABC}(W) = T_{\varepsilon}(W)$
\item{(3)} elsif  $T_B^+(W) \cup T_B^-(W) \neq \emptyset$, 
then $\Str_{ABC}(W) = T_B^+(W) \cup T_B^-(W) \cup T_C(W)$
\item{(4)} else  $\Str_{ABC}(W) =T_A(W) \cup T_C(W)$
\end{description}
The strategy $\Str_{AB}$ is obtained from $\Str_{ABC}$ by removing the occurence of $T_C$ in cases (3)(4).
\subsection{Global strategy}
Let us define a global strategy $\hat{\Str}_{ABC}$ w.r.t. the strategy ${\Str}_{ABC}$. 
Let us fix (until the end of this article) a total well-ordering $\sqsubseteq$ over the set 
of oracles $\Omega$. We need now three technical definitions.
\begin{defi}
Let $P \in {\cal P}_f({\cal A}), \OR \in \Omega$ and $\bar{\pi} \in \bbbn \cup \{\infty\}$.
$\OR$ is said {\em $\bar{\pi}$-consistent with $P$} iff, for every $(\pi,S,S') \in \CO(P)$,
and every $n \in \bbbn$, if
$$\pi + n - 1 < \bar{\pi},$$ 
then, the binary relation ${\cal R}_n = \OR(S,S')\cap X^{\leq n} \times X^{\leq n}$ fulfills
$$[\pi,S,S',{\cal R}_n] \subseteq \CO(P).$$
\label{consistency_for_oracles}
\end{defi}
We use the notation:
$$\Omega(\bar{\pi},P) = \{ \OR \in \Omega \mid  \OR \mbox{ is }\bar{\pi}-\mbox{consistent with }
P \}.$$
\begin{defi}
Let $P$ be a finite subset of ${\cal A}$,  and let $\bar{\pi} \in \bbbn \cup \{ \infty\}$.
$P$ is said {\em $\bar{\pi}$-consistent} iff, there exists some oracle $\OR \in 
\Omega$, which is $\bar{\pi}$-consistent with $P$.
\label{consistency_for_subsets}
\end{defi} 
For every proof tree $t \in {\cal T}(\Str_{ABC})$, we denote by $\bar{\Pi}(t)$ the integer:
\begin{equation}
\bar{\Pi}(t) =
\mi\{ \pi \in \bbbn \mid \exists x \in \dom(t), x \mbox{ is not closed for }\Str_{ABC}, 
\exists S,S',
t(x) = (\pi,S,S') \}.
\label{tree_weight}
\end{equation}
( we admit here that $\mi(\emptyset) = \infty$.)    
\begin{defi}
Let $t$ be a finite proof-tree for the strategy $\Str_{ABC}$, $t \in {\cal T}(\Str_{ABC})$.
$t$ is said {\em consistent} iff, $\im(t)$ is  $\bar{\Pi}(t)$-consistent.
\label{consistency_for_trees}
\end{defi}
Let us consider some tree $t \in {\cal T}(\Str_{ABC})$ which is consistent and not closed.
Let $\bar{\pi}= \bar{\Pi}(t)$, let $x$ be the smallest 
unclosed node of weight $\bar{\pi}$. 
Let 
\begin{equation}
W = A_1 \cdots A_n
\label{THE_word_W}
\end{equation}
 be the word labelling the path  from the root to $x$ in t.
(One can notice that, as $x$ is not closed, $T_{cut}(W) \cup T_{\emptyset}(W) \cup T_{\varepsilon}(W)= \emptyset).$
We define a tree of height one, $\hat{\Delta}(t)$ as follows:
\begin{description}
\item{(0)} if $\exists \OR \in \Omega(\bar{\pi}, \im(t)), T^{(\OR)}_C(W) \neq \emptyset$ then
$$\OR_0 = \mi \{\OR \in \Omega(\bar{\pi}, \im(t)), T^{(\OR)}_C(W) \neq \emptyset\},\;\;
\hat{\Delta}(t)= A_n(T^{(\OR_0)}_C(W)),$$
\item{(1)} elsif  $T_B^+(W)\neq \emptyset$ then
$$\OR_0 = \mi (\Omega(\bar{\pi}, \im(t))),\;\;
\hat{\Delta}(t)= A_n(T^{(\OR_0),+}_B(W)),$$
\item{(2)} elsif  $T_B^-(W)\neq \emptyset$ then
$$\OR_0 = \mi (\Omega(\bar{\pi}, \im(t))),\;\;
\hat{\Delta}(t)= A_n(T^{(\OR_0),-}_B(W)),$$ 
\item{(3)} else 
$$\OR_0 = \mi (\Omega(\bar{\pi}, \im(t))),\;\;
\hat{\Delta}(t)= A_n(T^{(\OR_0)}_A(W)).$$
\end{description}
( In the above definition by $A(W')$, where $A \in {\cal A}, W' \in {\cal A}^+$ we mean the tree
of height one with root labelled by $A$ and whose  sequence of leaves is the word $W'$).

\begin{equation}
\hat{\Str}_{ABC}(t) = t[\hat{\Delta}(t)/x],
\label{StrABC_def1}
\end{equation}
i.e. $\hat{\Str}_{ABC}(t)$ is obtained from $t$ by substituting $\hat{\Delta}(t)$ at the leaf $x$.
\begin{lem} For every $t \in {\cal T}(\Str_{ABC})$, if $t$ is consistent, 
then $\hat{\Delta}(t)$ is defined. 
\label{SABC_closed}
\end{lem}
\begin{prooof} By the definition of consistency the oracle $\OR_0$ is always defined ( i.e.
$\Omega(\bar{\pi}, \im(t))\neq \emptyset$), and for the word $W$ defined above
$T_{\varepsilon}(W)= \emptyset \Rightarrow \forall \OR \in \Omega, T^{(\OR)}_A(W) \neq \emptyset$,
 hence one of cases (0-3) must occur.
\end{prooof}
If $t$ is not consistent or is closed then we define:
\begin{equation}
 \hat{\Str}_{ABC}(t) = t.
\label{StrABC_def2}
\end{equation}
\begin{lem}
$\hat{\Str}_{ABC}$ is a global strategy for ${\Str}_{ABC}$.
\label{hat_is_strategic}
\end{lem}
\begin{sketch}
By lemma \ref{SABC_closed} $\hat{\Str}_{ABC}$ is defined on every $t \in {\cal T}(\Str_{ABC})$.
It suffices to check that, in every case, the word constitued by the leaves of $\hat{\Delta}(t)$
belongs to ${\Str}_{ABC}(W)$ (where $W$ is the word considered in (\ref{THE_word_W})).  
\end{sketch}

\section{Tree analysis}
\label{trees}
This section is devoted to the analysis of the proof-trees $\tau$ produced by the strategy 
${\cal S}_{AB}$ defined in section \ref{strategies_B0}. The main results are lemma 
\ref{TC_is_defined} and \ref{K5K6} whose combination asserts that if some branch 
of $\tau$ is infinite, then there exists some finite prefix on 
which $T_C$ has a non-empty value. This key technical result 
will ensure termination of the global strategy $\hat{{\cal S}}_{ABC}$ ( see section 
\ref{B0_is_terminating}).

We fix throughout this section a tree 
$\tau \in {\cal T}({\cal S}_{AB}, (\pi_0,U^{-}_0,U^{+}_0))$ (i.e. $\tau$ is a proof tree 
associated to
the assertion $(\pi_0,U^{-}_0,U^{+}_0))$ by the strategy ${\cal S}_{AB}$). We suppose that
\begin{equation}
\|U^{-}_0\| \leq D_2,\;\;\; \|U^{+}_0\| \leq D_2, \;\;\; U^{-}_0,U^{+}_0
\mbox{ are both unmarked },
\label{unmarked}
\end{equation}
\begin{equation}
U^{-}_0,  U^{+}_0 \in {\bbbD \bbbR \bbbB}_{1,\lambda}\pa{V} \mbox{  with  } \lambda \leq \lambda_2.
\label{lambda_small}
\end{equation}
\begin{equation}
U^{-}_0 \equiv U^{+}_0 
\label{root_is_true}
\end{equation}

We recall that, formally,  $\tau$ is a map $\dom(\tau) \rightarrow 
\bbbn \times \bbbD \bbbR \bbbB_{1,\lambda}\pa{V} \times \bbbD \bbbR \bbbB_{1,\lambda}\pa{V}$ 
such that 
$\dom(\tau) \subseteq \{ 1,\ldots,|X|^2\}^*$ is closed under prefix and under ``left-brother'' 
( i.e. $w \cdot (i+1) \in \dom(\tau) \Rightarrow w \cdot i \in \dom(\tau)$).
We denote by $pr_{2,3}: \bbbn \times \bbbD \bbbR \bbbB_{1,\lambda}\pa{V} \times \bbbD \bbbR \bbbB_{1,\lambda}\pa{V} \rightarrow
\bbbD \bbbR \bbbB_{1,\lambda}\pa{V} \times \bbbD \bbbR \bbbB_{1,\lambda}\pa{V}$ the projection
$(\pi, U,U') \mapsto (U,U')$. By $\tau_s$ we denote the tree obtained from $\tau$ by forgetting 
the weights: $\tau_s = \tau \circ pr_{2,3}$.

\subsection{Depth and weight}
In this paragraph we check that the {\em weight} and the {\em depth} of a given node are closely
related. Let us say that the strategy $T$ ``occurs at''  node $x$ iff, 
$$\tau(x) \in T(\tau(x[0]) \cdot \tau(x[1]) \cdots \tau(x[|x|-1])),$$
i.e. the label of $x$ belongs to the image of the path from $\epsilon$ (included) to $x$ 
(excluded) by the strategy $T$. 

\begin{lem}
Let $\alpha \in \{-,+\}, A_1,\ldots,A_n \in {\cal A}$ such that 
$T^{\alpha}_B(A_1 \cdots A_n) \neq \emptyset$.
Then, $\forall i \in [n-k_1+1,n], A_i \notin T_B(A_1 \cdots A_{i-1})$.
\label{TB+and-_impossible}
\end{lem}
In other words: if $T_B$ occurs at  node $x$ of $\tau$, it cannot occur at any of its $k_1$
above immediate ancestors.\\
\begin{prooof}

Suppose that $\exists i \in [n-k_1+1,n], A_i \in T_B(A_1 \cdots A_{i-1})$.
Hence $\pi_i = \pi_{i-1} -1 < \pi_{n-k_1}+i$, contradicting one of the hypothesis
under which $T_B(A_1 \cdots A_{n})$ is not empty.
\end{prooof}
Lemma \ref{TB+and-_impossible} ensures that, in every branch $(x_i)_{i \in I}$ and for every 
interval $[n+1,n+4] \subseteq I$, at most one integer $j$ is such that $T_B$ occurs at $j$.

\begin{lem} : Let ${\tau}$ be a proof-tree associated to the strategy $\Str_{AB}$. 
Let $x, x' \in dom({\tau}), x \preceq x'$. Then $\mid W(x') - W(x) \mid \leq |x'| - |x| 
\leq 2 \cdot(W(x') - W(x)) + 3$.
\label{depth_eq_weight}
\end{lem}
(We recall the {\em depth} of a node $x$ is just its length $|x|$). We denote by $W(x)$ the 
weight of $x$ which we define as the first component of ${\tau}(x)$ i.e.
the weight of the equation labelling $x$). \\
\begin{prooof}
Let $x,x'$ be such that $ |x'| = |x|+1$. Then $ W(x') - W(x) \in \{-1, +1\}$, hence the
inequality $\mid W(x') - W(x) \mid \leq |x'| - |x|$ is fulfilled by such
nodes. The general case follows by induction on $(|x'| - |x|)$.\\
Let us prove now the other inequality. We distinguish two cases.\\
{\bf Case 1} : $|x'| - |x| \leq 3$.\\
Then $|x'| - |x| \leq 2 \cdot (W(x') - W(x)) + 3$
(because there is at most one $T_B$ step in a sequence of length
$\leq 3$).\\
{\bf Case 2}:  $|x'| - |x| \geq 4$.\\
Let $x = x_0, x_1, \cdots, x_q, x'$ be the sequence of nodes such that $|x'| -
|x| = 4 \cdot q + r, 0 \leq r < 4$ and 
$\forall i \in [0, q-1], |x_{i+1}| - |x_i| = 4$.\\

By lemma \ref{TB+and-_impossible}, in every set $\{y \in \dom({\tau}) \mid x_i \prec y \preceq x_{i+1}\}$ at most one node
$z$ is such that $T_B$ occurs at $z$. Hence
$W(x_{i+1}) - W(x_i) \geq 2$.\\
It follows that :
\begin{eqnarray*}
|x'| - |x| &= &\sum^{q-1}_{i=0} [|x_{i+1}| - |x_i|] + |x'| - |x_q|\\
& \leq & \sum^{q-1}_{i=0} 2(W(x_{i+1} - W(x_i)) + |x'| - |x_q|\\
& \leq & 2(W(x_q) - W(x)) + 2(W(x') - W(x_q))+3\;\; (\mbox{ by the first case })\\
& \leq & 2(W(x') - W(x)) + 3.
\end{eqnarray*}
\end{prooof}

Let us recall the values of some constants (defined in section \ref{constants}):
$$
\begin{array}{lll}
k_0= \ma\{ \nu(v) \mid v \in V\},&
k_1= \ma\{ 2 k_0 + 1 , 3 \},\;\;&
D_1= k_0 \cdot K_0 + |Q| + 2,\\
k_2= D_1 \cdot k_1 \cdot K_0 + 2 \cdot k_1 \cdot K_0 + K_0,&\;\;
k_3= k_2 + k_1\cdot K_0,&\;\; 
k_4= (k_3 +1)\cdot K_0 + k_1,\\
d_0= \card(X^{\leq k_4}),&\;\;
N_0 = 1 + k_3 + D_2.&
\end{array}
$$

\subsection{$B$-stacking sequences}
\label{NSS}

We establish here that every infinite branch must contain an infinite suffix 
( a ``B-stacking sequence'') where at least $d_0$ labels $(U,U')$ are belonging
to the same d-space $V_0$ of dimension $\leq d_0$ with coordinates not greater than 
$s_{d_0}$ ( over some fixed generating family of cardinality $\leq d_0$).

Let $\sigma = (x_i)_{i \in I}$ be a path in $\tau$,
where $I = [i_0, \infty[$ and let $(x_i)_{i \geq 0}$ the unique branch of $\tau$ containing 
$\sigma$.
Let us note $\tau(x_i) = ( \pi_i,U^-_i, U^+_i)$.

We call $\sigma$ a
{\em B-stacking sequence} iff: there exists some $\alpha_0 \in \{-,+\}$ such that
\begin{equation}
T_B^{\alpha_0} \mbox{  occurs at } x_{i_0+k_1+1}
\label{beginning_of_stacking}
\end{equation}

and, for every $i \in I, \alpha \in \{-,+\}$, if $T_B^\alpha$ occurs at $x_{i+k_1+1}$ then

\begin{equation} 
 \|U_{i}^{-\alpha}\| \geq  \|U_{i_0}^{-\alpha_0}\|\geq N_0.
\label{B_stacking}
\end{equation}  
From now on and until lemma \ref{K5K6} , we fix a B-stacking sequence $\sigma =  (x_i)_{i \in I}$
and we denote by $S_0$ the series $U_{i_0}^{-\alpha_0}$.

\begin{lem}
There exists some word $u_0 \in X^*$ and some sign $\alpha'_0 \in \{-,+\}$ such that
$S_0 = U_{0}^{\alpha'_0} \odot u_0$.
\label{S0U0}
\end{lem}
\begin{prooof}
One can prove by induction on $i \in \bbbn$ that, for every $\alpha \in \{-,+\}$, 
$U_i ^{\alpha}$ has one of the two following forms:\\
1- $U_i ^{\alpha}= U_{0}^{\alpha'} \odot u$ for some $\alpha' \in \{-,+\}, |u| \leq i$,\\
2- $U_i ^{\alpha}= \sum_{k=1}^q \beta_k \cdot (U_{0}^{\alpha'} \odot uu_k)$,\\
for some deterministic rational vector $\beta$, $\alpha' \in \{-,+\}, |u\cdot u_k| \leq i,
|u_k| \leq k_0.$
\end{prooof}

\begin{lem}
Suppose that $i_0 \leq j < i$, no $T_B$ occurs in $[j+1,i]$,  
$U_j^{-\alpha}$ is $D_1$-marked and  $U_j^{\alpha}$ is unmarked.
Then, for every $j' \in [j,i]$, $\|U_{j'}^{\alpha}\| \geq \|U_i^{\alpha}\| - k_2$.
\label{prolog}
\end{lem}

\begin{prooof}
Let $i,j$ fulfill the hypothesis of the lemma.\\
1-Let us treat first the case where $j' = j$.\\
If $(i-j) \leq (D_1+1)k_1$ then, by lemma \ref{norm_of_action}
$$\|U_i^{\alpha}\| \leq  \|U_j^{\alpha}\|+ (D_1+1)\cdot k_1 \cdot K_0 \leq k_2$$
hence the lemma is true.\\
Let suppose now that $(i-j) \geq (D_1+1)k_1 +1$. We can then define the integers
$j < i_1 < i_2 < i$ by:
$$ i_1 = j+ D_1 \cdot k_1, i_2 = i-k_1-1. $$
By lemma \ref{norm_of_action} we know that:
\begin{equation}
\|U_{i_1}^{\alpha}\| \leq \|U_{j}^{\alpha}\| + D_1 \cdot k_1 \cdot K_0 \mbox{ and }
\|U_{i}^{\alpha}\| \leq \|U_{i_2}^{\alpha}\| + (k_1 +1) \cdot K_0.
\label{extremities}
\end{equation}
If there was some stacking subderivation of length $k_1$ in 
$U_j^{-\alpha} \rightarrow U_{i_1}^{-\alpha}$, as all the $U_k^{\alpha}$ (for $k \in [j,i]$) 
are unmarked, $T_B$ would occur at some integer in $[j+k_1 +1, i_1+1]$, which is untrue.
Hence there is no such stacking subderivation, and by lemma \ref{no_stacking_removes_the_marks}
$U_{i_1}^{-\alpha}$ is unmarked.\\
If there was some stacking subderivation of length $k_1$ in 
$U_{i_1}^{\alpha} \rightarrow U_{i_2}^{\alpha}$, as all the $U_k^{-\alpha}$ (for $k \in [i_1,i]$) 
are unmarked, $T_B$ would occur at some integer in $[i_1+k_1 +1, i]$, which is untrue.
Hence there is no such stacking subderivation, and by lemma \ref{increasing_contains_stacking}
\begin{equation}
\|U_{i_2}^{\alpha}\| \leq \|U_{i_1}^{\alpha}\| + k_1 \cdot K_0.
\label{middle}
\end{equation}
Adding inequalities (\ref{extremities},\ref{middle}) we obtain:
$$
\|U_{i}^{\alpha}\| \leq \|U_{j}^{\alpha}\| + (D_1\cdot k_1  + 2 \cdot k_1 +1) \cdot K_0
= \|U_{j}^{\alpha}\| + k_2,$$
which was to be proved. \\
2-Let us suppose now that $j \leq j'\leq i$.\\
If $(i-j) \leq (D_1+1)k_1$, the same inequality is true for $i-j'$ and the conclusion 
is true for $j'$.\\
Otherwise, if $j' \leq i_1$, (\ref{extremities}, \ref{middle}) are still true for $j'$ instead of $j$, hence the conclusion too.\\
Otherwise, by the arguments of part 1, $U_{j'}^{-\alpha},U_{j'}^{\alpha} $ are both unmarked.
Hence the hypothesis of part 1 are met by $(j',i)$ instead of $(j,i)$, hence the conclusion
is met too.
(We illustrate our argument on figure \ref{too_small}).
\begin{figure}[h]
\input{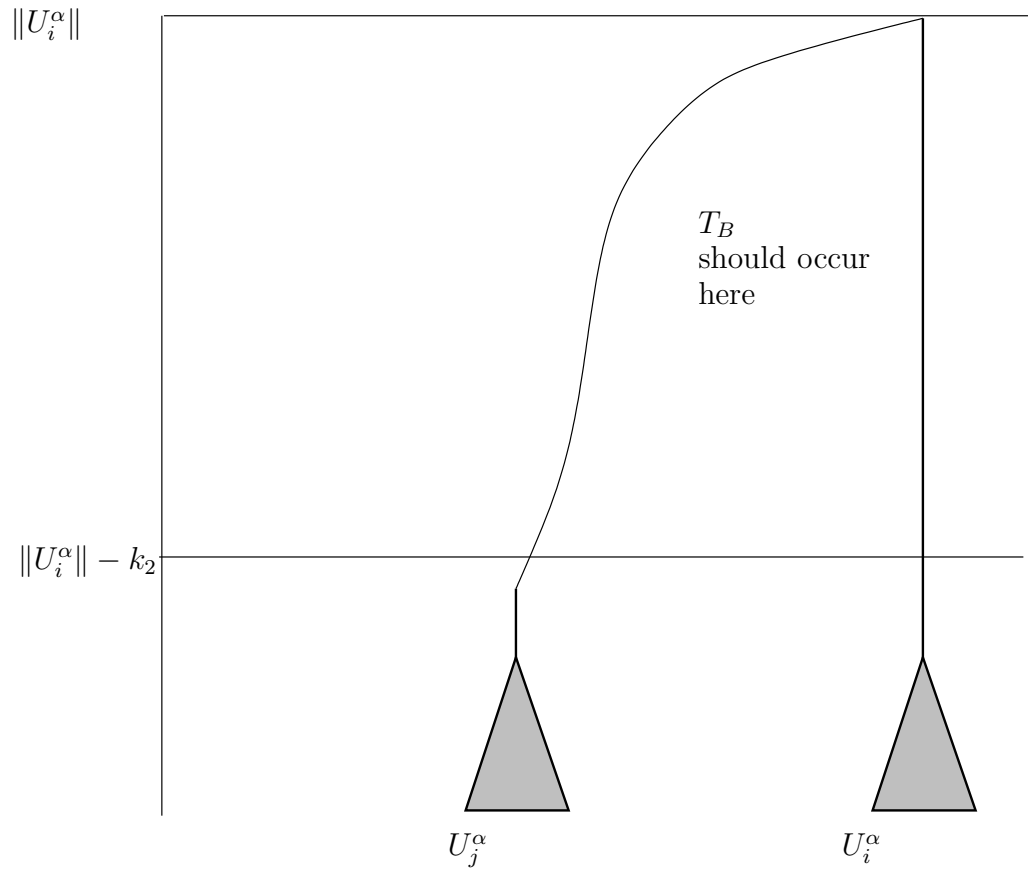}
\caption{$\|U^{\alpha}_{j}\|$ too small is impossible.}
\label{too_small}
\end{figure}
\end{prooof}
\begin{lem}
Let $i \in I,  \alpha \in \{ -,+\}$ such that $T_B^{\alpha}$ occurs at $i+k_1+1$.
Then, there exists $u \in X^*, |u| \leq (i-i_0), U^{-\alpha}_i = S_0 \odot u$
and, for every prefix $w \preceq u$, 
$$\|S_0 \odot w \| \geq \|S_0\| - k_3.$$
\label{argTB_swim}
\end{lem}

\begin{prooof}
We prove the lemma by induction on $i\in [i_0,\infty[$.\\
{\bf Basis}: $i=i_0$.\\
Choosing $u=\epsilon$, the lemma is true.\\
{\bf Induction step}: $i_0 \leq i' < i$, $T_B^{\alpha'}$ occurs at $i'+k_1+1$,
$T_B^{\alpha}$ occurs at $i+k_1+1$ and $T_B$ does not occur in $[i'+k_1+2,i+k_1]$.\\
By induction hypothesis, there exists some $u' \in X^*, |u'| \leq (i'-i_0)$ 
fulfilling 
\begin{equation}
U^{-\alpha'}_{i'} = S_0 \odot u'
\label{conclusion1_i'}
\end{equation}
\begin{equation}
\forall w' \preceq u', 
\|S_0 \odot w' \| \geq \|S_0\| - k_3.
\label{conclusion2_i'}
\end{equation}
Let us define  $j = i'+k_1 +1$.\\
Let $\bar{u} \in X^*$ be the word such that
\begin{equation} 
U_j^{-\alpha} \deriv{\bar{u}} U_i^{-\alpha}
\label{from_j_to_i}
\end{equation}
is the derivation described by the $-\alpha$ component of the path from $x_j$ to $x_i$.\\
{\bf Case 1}: $\alpha' = \alpha$.\\
$$U_j^{-\alpha}= U_{i'}^{-\alpha'} \odot u_1$$
for some $u_1 \in X^*, |u_1| = k_1$ and $U_j^{\alpha}$ is $D_1$-marked.
Let us choose $u = u' \cdot u_1 \cdot \bar{u}$.
Hence
\begin{equation}
U^{-\alpha}_i = S_0 \odot u.
\label{conclusion1_i}
\end{equation}
Let us consider some prefix $w$ of $u$.\\
{\bf subcase 1}: $w \preceq u'$.\\
By (\ref{conclusion2_i'}) we know that $\|S_0 \odot w \| \geq \|S_0\| - k_3.$\\
{\bf subcase 2}: $w = u' \cdot u_1 \cdot u''$, for some $u'' \preceq \bar{u}$.\\
By lemma \ref{prolog} we know that 
$\|S_0 \cdot w\| \geq \|U_i^{\alpha}\| - k_2 $, and by definition of a B-stacking sequence we also
know that $\|U_i^{\alpha}\| \geq \|S_0\|$. Hence 
$$\|S_0 \odot w\| \geq \|S_0\| - k_2 .$$
{\bf subcase 3}: $w = u' \cdot u'_1$, where $u'_1$ is a prefix of $u_1$.\\
 Then, by lemma \ref{norm_of_action}
and the above inequality we get:
$$\|S_0 \odot w\| \geq \|S_0 \odot u'u_1\| - k_1\cdot K_0 \geq \|S_0\| - k_3 .$$
{\bf Case 2}: $-\alpha' = \alpha$.\\
$$U_j^{-\alpha}= \sum_{k=1}^q \beta_k \cdot (U_{i'}^{\alpha} \odot u_k)$$
where $\beta$ is a polynomial which is fully marked and every $|u_k| \leq k_0$.\\
By lemma \ref{RA_odot} either 
$U_i^{-\alpha}= \sum_{k=1}^q (\beta_k \odot \bar{u}) \cdot (U_{i'}^{\alpha} \odot u_k)$
or there exists a decomposition
\begin{equation}
\bar{u} = \bar{u}_1 \cdot \bar{u}_2
\label{baru_decomposed}
\end{equation}
and an integer $k \in [1,q]$ such that
\begin{equation}
U_i^{-\alpha}= U_{i'}^{\alpha} \odot u_k \bar{u}_2.
\label{Ui_from_Ui'uk}
\end{equation}
But, as  $U_i^{-\alpha}$ is unmarked (by definition of $T_B^{\alpha}$),
the first formula is impossible unless $\beta \odot {\bar{u}}$ is unitary or nul.
Hence (\ref{baru_decomposed},\ref{Ui_from_Ui'uk}) is the only possibility.\\
Let us choose $u = u' \cdot u_k \cdot \bar{u}_2$.
It is clear from (\ref{Ui_from_Ui'uk}) that $U^{-\alpha}_i = S_0 \odot u.$\\
Let us consider some prefix $w$ of $u$.\\
{\bf subcase 1}: $w \preceq u'$.\\
Same arguments as in case1 , subcase1.\\
{\bf subcase 2}: $w = u' \cdot u_k \cdot u''$, for some $u'' \preceq \bar{u}_2$.\\
By lemma \ref{prolog} applied on the interval $[j+|\bar{u}_1|+1,i]$, we can conclude that
$$\|S_0 \odot w \| \geq \|S_0\| - k_3.$$
{\bf subcase 3}: $w = u' \cdot u'_k$, where $u'_k$ is a prefix of $u_k$.\\
Same arguments as in case1 , subcase3.
\end{prooof}
Let us define now the following families of vectors and d-spaces of vectors 
\begin{equation}
{\cal G}_0 = \{S_0\odot u |\; u \in X^*,|u| \leq k_4\},
\label{G0}
\end{equation}
\begin{equation}
V_0 = \bbbV({\cal G}_0).
\label{V0}
\end{equation}
\begin{lem} Let $i \geq i_0$ such that $T_B$ occurs at $i$. Then, 
$U^{-}_i , U^{+}_i \in V_0$.
\label{TB_in_V0}
\end{lem}
\begin{prooof}
Let us suppose that $T^{\alpha}_B$ occurs at $i$.
By lemma \ref{argTB_swim},
$U^{-\alpha}_{i-k_1-1} = S_0 \odot u$ and, for every prefix $w \preceq u$, 
$$\|S_0 \odot w \| \geq \|S_0\| - k_3.$$
By lemma \ref{bullet_and_odot}, $\exists u_1, u_2 \in X^*, v_1 \in V^*, E_1, \ldots, E_k \in V,
E_1 \smile E_2 \ldots \smile E_k, \Phi \in \bbbD \bbbR \bbbB_{q,\lambda}\pa{V}$, such that
$u = u_1 \cdot u_2$,
\begin{equation}
S_0 \odot u_1 = S_0 \bullet v_1 = \sum_{k=1}^q E_k \cdot \Phi_k
\label{minimum_residual}
\end{equation}
\begin{equation}
S_0 \odot u = \sum_{k=1}^q (E_k \odot u_2) \cdot \Phi_k.
\label{result_wrt_minimum}
\end{equation}
Without loss of generality, we can suppose that $v_1$ is a minimal word realizing the equality
(\ref{minimum_residual}).
Let us notice that, as $G$ is a reduced grammar, for every $v \preceq v_1$, there exists
some $\bar{v} \in X^*$, such that $S_0 \bullet v = S_0 \odot \bar{v}$.
Hence, for every $v \preceq v_1$, 
$$S_0 \bullet v = U_0^{\alpha'_0} \odot u_0\cdot \bar{v} \mbox{ and }
\|U_0^{\alpha'_0} \odot u_0\cdot \bar{v}\| \geq \|S_0 \odot u_1\| > D_2 = \|U_0^{\alpha'_0}\|.$$
By lemma \ref{increasing_action}, all the vectors $S_0 \bullet v$ for $v \preceq v_1$
are loop-free. It follows that, for every $v \preceq v' \preceq v_1$
$$
v \prec v' \Rightarrow \|S_0 \bullet v \| > \|S_0 \bullet v' \|,
\label{norm_strict_incr}
$$
hence
$$ |v_1| \leq \|S_0\| -\|S_0 \bullet v_1\| \leq k_3.$$
The  formula (\ref{result_wrt_minimum}) can be rewritten
$$U^{-\alpha}_{i-k_1-1} = \sum_{k=1}^q (E_k \odot u_2) \cdot (S_0 \bullet v_1E_k)
=\sum_{k=1}^q (E_k \odot u_2) \cdot (S_0 \odot \bar{u}_k)$$
where $\bar{u}_k \in X^*, |\bar{u}_k| \leq (k_3 +1) \cdot K_0$.\\
Using lemmas \ref{RA_odot} and \ref{product_is_deterministic} we can deduce from the above form 
of $U^{-\alpha}_{i-k_1-1}$ that
$$ U^{\alpha}_{i} \in \bbbV(\{ S_0 \odot w \mid w \in X^*, 
|w| \leq (k_3 +1) \cdot K_0 + k_0\}),\;\;
U^{-\alpha}_{i} \in \bbbV(\{ S_0 \odot w \mid w \in X^*, 
|w| \leq (k_3 +1) \cdot K_0 + k_1\}),
$$
hence that both $U^{-\alpha}_i,U^{\alpha}_i$ belong to $V_0$.
\end{prooof}
We recall that:\\
$$ K_1 = k_1 \cdot K_0 + 1 ,\;\; 
K_2 = k_1^2\cdot D_1 \cdot K_0 + k_1^2 \cdot K_0 + 
2\cdot k_1 \cdot K_0 + D_1 \cdot k_1 + 2 \cdot k_1 +4.$$
\begin{lem} For every $L \geq 0$ there
 exists $i \in [i_0 + L, i_0 + K_1 \cdot L+ K_2]$ such that,
$U^{-}_i , U^{+}_i \in V_0$.
\label{K1K2}
\end{lem}
\begin{prooof}
Let us establish that
\begin{equation}
\exists i \in [i_0 + L, i_0 + K_1 \cdot L+ K_2-k_1-1], \exists \alpha \in \{-,+\},
T^{\alpha}_B \mbox{ occurs at } i+k_1+1.
\label{TB_in_I}
\end{equation}
Let $L \geq 0$ and let $i'\geq i_0$ be the greatest integer in $[i_0,i_0 +L]$ such that
$T_B$ occurs at $i' +k_1 +1$. Let $j = i' +k_1 +1$. We then have:
$$U_j^{\alpha'}= \sum_{k=1}^q \beta_k \cdot (U_{i'}^{-\alpha'} \odot u_k)$$
where $\| \beta\| \leq D_1$ and $U_j^{-\alpha'}$ is unmarked.\\
{\bf Case 1}: there exists $i \in [j, j+k_1\cdot D_1]$, such that $T_B$ occurs at $i+k_1 +1$.\\
In this case the small constants $K_1= 0, K_2= k_1 \cdot D_1+ k_1 +1$ would be sufficient to 
satisfy (\ref{TB_in_I}). A fortiori the given constants satisfy (\ref{TB_in_I}).\\
{\bf Case 2}: there exists no $i \in  [j, j+k_1\cdot D_1]$, such that $T_B$ occurs at $i+k_1 +1$.\\
Then, there is no stacking subderivation of length $k_1$ in 
$U_j^{\alpha'} \longrightarrow U_{j+k_1\cdot D_1}^{\alpha'}$. 
By lemma \ref{no_stacking_removes_the_marks} it follows that
both $U^{\alpha}_{j+D_1\cdot k_1}$ are unmarked.\\
1-Let $j_1 = j+D_1\cdot k_1$ and let us show that there exists some $i \ge j_1$ such that
$T_B$ occurs at $i+k_1 +1$.\\
If such an $i$ does not exist then, for every $\alpha \in \{-,+\}$, the infinite derivation 
$$U^{\alpha}_{j_1} \longrightarrow U^{\alpha}_{j_1+1} \longrightarrow \ldots $$
does not contain any stacking sequence of length $k_1$. 
By lemma \ref{increasing_contains_stacking} we would have:
$$\forall k \geq j_1, \|U^{\alpha}_k\| \leq \| U^{\alpha}_{j_1}\| +  k_1 \cdot K_0 .$$
As the set $\{\|U^{\alpha}_k\|, k \geq j_1, \alpha \in \{-,+\}\}$ is finite,
there would be a repetition 
$$(U^{-}_k,U^{+}_k) = (U^{-}_{k'},U^{+}_{k'}) \mbox{ with } j_1 \leq k < k'
\mbox{ and } \pi_k < \pi_{k'}$$,
so that $T_{cut}$ would have been defined on some finite prefix of the branch,
contradicting the hypothesis that the branch is infinite.\\
2-Let $i > i'$ be the smallest integer (in $[j_1, \infty[$) fulfilling point 1 above
and suppose that $T_B^{\alpha}$ occurs at $i+k_1 +1$.\\
By lemma \ref{prolog},
$$ \forall \ell \in [j_1,i],\| U_\ell ^{-\alpha} \| \geq N_0 -k_2 > D_2 .$$
Using lemma \ref{S0U0} , lemma \ref{increasing_action} and inequality (\ref{unmarked})
we conclude that 
$$ \forall \ell \in [j_1,i], U_\ell ^{-\alpha} \mbox{ is loop-free }.$$
By an argument analogous to that used in lemma \ref{S0U0} we see that
$U_{j_1}^{-\alpha}= S_0 \odot u$ for some $|u| \leq (j_1-i_0)$, and 
by lemma \ref{norm_of_action} we get
\begin{equation}
\|U_{j_1}^{-\alpha} \| \leq (j_1 - i_0) \cdot K_0 + \|S_0\|.
\label{Uj1_small}
\end{equation}
We also know that:
\begin{equation}
\|S_0\| \leq \|U_i^{-\alpha}\| \leq  \|U_{i-1}^{-\alpha}\| +K_0.
\label{Ui_big}
\end{equation}
As the derivation $U_{j_1}^{-\alpha} \longrightarrow U_{i-1}^{-\alpha}$ contains no
stacking sub-derivation of length $k_1$ and consists of loop-free series only, 
by lemma \ref{flat_is_stacking} we obtain:
\begin{equation}
\|U_{i-1}^{-\alpha}\| \leq \|U_{j_1}^{-\alpha}\|-(i-j_1-2)/k_1.
\label{Ui_minus_Uj}
\end{equation}
Combining the three inequalities (\ref{Uj1_small},\ref{Ui_big},\ref{Ui_minus_Uj})
we get successively:
$$\|S_0\|  \leq \|S_0\|+ (j_1 -i_0 +1) \cdot K_0 -(i-j_1 -2)/k_1,$$
$$(i-j_1 -2) \leq  (j_1 -i_0 +1)\cdot k_1 K_0.$$
\begin{eqnarray}
(i-i') &=&(i-j_1 -2) + (j_1 -i' +2) \leq  
(j_1-i_0 +1)\cdot k_1 \cdot K_0 + D_1 \cdot k_1 + k_1 +3 \nonumber\\
&=& (i' - i_0) \cdot k_1 \cdot K_0 + 
k_1^2 \cdot D_1 \cdot K_0 + k_1^2 \cdot K_0 +2 \cdot k_1 \cdot K_0 + D_1 \cdot k_1 +k_1 +3 \nonumber\\
&=& (K_1-1)(i'-i_0) + K_2-k_1-1.
\label{successive_TB}
\end{eqnarray}
3- By the choice of $i',i$, we know that $i' \leq i_0 + L \leq i$.
Using (\ref{successive_TB}) we obtain:
$$i \leq i' + (K_1-1)(i'-i_0) + K_2-k_1-1$$
$$i \leq i_0 +K_1 \cdot L + K_2-k_1-1.$$
Assertion (\ref{TB_in_I}) is now established for case 2 as well as for case 1.\\

From (\ref{TB_in_I}) and lemma \ref{TB_in_V0} the lemma follows.\\
(We illustrate our argument on figure \ref{fig_successive_TB}).
\begin{figure}[h]
\input{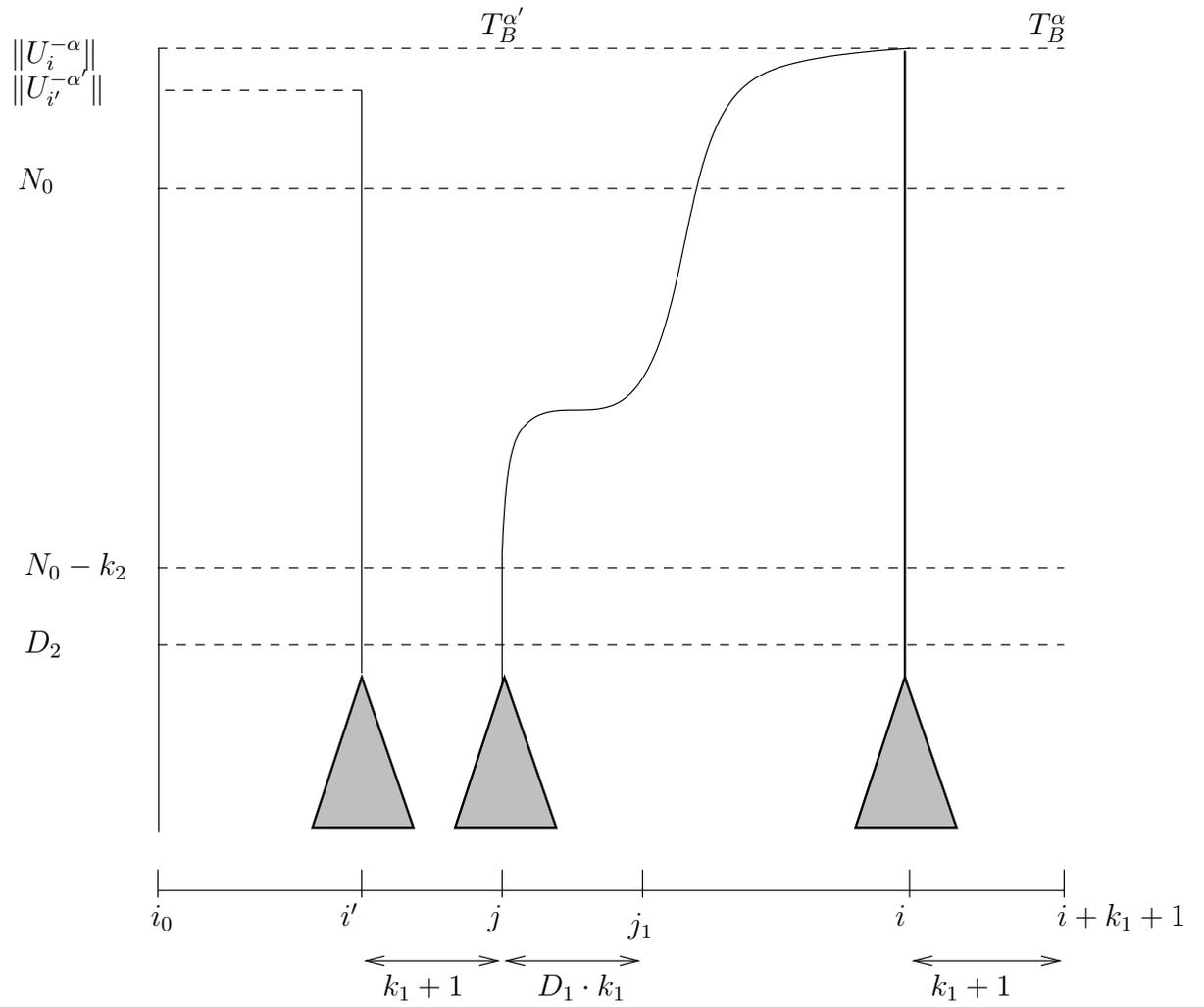}
\caption{Two successive $T_B$.}
\label{fig_successive_TB}
\end{figure}
\end{prooof}
Let us give now a stronger version of lemma \ref{K1K2} where we analyze the {\em size of the 
coefficients} of the linear combinations whose existence is proved in lemma \ref{K1K2}.
We recall that:
$$ K_3 = K_0 |Q|, \;\; K_4 = D_1.$$
Let us fix a total ordering on ${\cal G}_0$: 
$${\cal G}_0 = \{\theta_1,\theta_2,\ldots,\theta_d\}, \mbox{ where } d=\card({\cal G}_0).$$
Let us remark that $d \leq \card(X^{\leq k_4})=d_0.$
\begin{lem}Let $L \geq 0$. There exists 
$i \in [i_0 + L, i_0 + K_1 \cdot L+ K_2]$ and, for every 
$\alpha \in \{-,+\}$, there exists a deterministic rational family
$(\beta^{\alpha}_{i,j})_{1 \leq j \leq d}$ fulfilling
\begin{description}
\item{(1)} $U^{\alpha}_i = \sum^d_{j=1} \beta^\alpha_{i,j} \cdot \theta_j$
\item{(2)} $\| \beta^{\alpha}_{i,*}\| \leq K_3 \cdot (i - i_0) + K_4.$
\end{description}
\label{K3K4}
\end{lem}
\begin{prooof}
By lemma \ref{K1K2} there exists 
$i \in [i_0 + L, i_0 + K_1 \cdot L+ K_2]$ and $\alpha \in \{-,+\}$ such that 
$T^{\alpha}_B \mbox{ occurs at } i.$ Let us use the notation of the proof of 
lemma \ref{TB_in_V0} and compute upper-bounds on the coefficients of
$U_i^{-\alpha},U_i^{\alpha}$ expressed as linear combinations of the vectors of ${\cal G}_0$.\\
{\bf Coefficients of $U_i^{-\alpha}$}:\\
$U_i^{-\alpha}= U_{i-k_1-1}^{-\alpha} \odot u'$, for some $u' \in X^*,|u'| = k_1$. By lemma 
\ref{RA_odot}, $U_i^{-\alpha}$ can be expressed in one of the two following forms:
\begin{equation}
U_i^{-\alpha}= S_0 \odot (\bar{u}_ku'') \mbox{ where } u'' \mbox{ is a suffix of } u',
\label{first_form}
\end{equation}
\begin{equation}
U_i^{-\alpha}= \sum_{k=1}^q (E_k \odot u_2u') \cdot (S_0 \odot \bar{u}_k).
\label{second_form}
\end{equation}
In case (\ref{first_form}) we can choose as vector of coordinates :
$\beta^{-\alpha}_{i,\star}= \epsilon^d_{j_0}$. We then have
 $\| \beta_{i,\star} \| = 2 \leq K_4$.\\
In case (\ref{second_form}), we can choose: $\beta^{-\alpha}_{i,\star} = E \odot u_2u'$
(completed with $\emptyset$ in all the columns $j$ not corresponding to some vector 
$S_0 \odot \bar{u}_k$ of ${\cal G}_0$). We then have:
$$ \|\beta_{i,\star}\| =  \|E \odot u_2u'\| \leq K_0 \cdot (i-i_0) \leq K_3 \cdot (i-i_0).$$
{\bf Coefficients of $U_i^{\alpha}$}:\\
By definition of $T_B^{\alpha}$
\begin{equation}
U_i^{\alpha}= \sum_{\ell=1}^r \tau_\ell \cdot (U_{i-k_1-1}^{-\alpha}\odot  \bar{w}_\ell),
\label{formula_Ui}
\end{equation}
where $\| \tau\| \leq D_1, |\bar{w}_\ell| \leq k_0$.\\
Replacing $u'$ by $\bar{w}_\ell$ in the above analysis, we get:\\
\begin{equation}
\forall \ell \in [1,r], 
U_{i-k_1-1}^{-\alpha} \odot \bar{w}_\ell= 
\sum_{j=1}^d \gamma_{\ell,j} \cdot \theta_j,
\label{formula_action_wl}
\end{equation}
with $\|\gamma_{\ell,\star}\| \leq K_0 \cdot (i-i_0)$.\\
Equalities (\ref{formula_Ui},\ref{formula_action_wl}) show that:
$$U_i^{\alpha} = \tau \cdot \gamma \cdot \theta,$$
where $\tau , \gamma, \theta $ are deterministic rational matrices of dimensions respectively
$(1,r),(r,d),(d,1)$. Let us choose $\beta _{i,\star} = (\tau \cdot \gamma)$.\\
$$
\|\beta _{i,\star}\| \leq \|\tau\| + \| \gamma\| \leq D_1 + r \cdot K_0 \cdot (i-i_0) $$
$$\leq D_1 + |Q| \cdot K_0 \cdot (i-i_0) = K_3 \cdot (i-i_0) + K_4.$$
\end{prooof}
\begin{lem}
There exists $i_0 \leq \kappa_1 < \kappa_2 < \ldots < \kappa_d$ and deterministic rational
vectors $(\beta^{\alpha}_{i,j})_{1 \leq j \leq d}$ ( for every $i \in [1,d]$) such that
\begin{description}
\item{(0)} $W(\kappa_{1}) \geq 1$
\item{(1)} $\forall i, \forall \alpha, U^{\alpha}_{\kappa_i} = 
\sum^d_{j=1}\beta^{\alpha}_{i,j} \theta_j \in V_0$
\item{(2)} $\forall i, \forall \alpha, \|\beta^{\alpha}_{i,*}\| \leq s_i$
\item{(3)} $\forall i, W(\kappa_{i+1}) - W(\kappa_{i}) \geq \delta_{i+1}$
\end{description}
\label{TC_is_defined}
\end{lem}
where the sequences $(\delta_i, \ell_i, L_i, s_i, S_i,
\sigma_i)$ are those defined by relations (\ref{6_initialisation_constants}, \ref{6_recurrences_constants}) in section \ref{constants}.\\ 
\begin{prooof}
Let us consider the additional property\\
{\bf (4)} $\kappa_i - i_0 \leq L_i$.\\
We prove by induction on $i$ the conjunction $(1)  \wedge (2) \wedge (3) \wedge (4)$.\\
{\bf i=1}: \\
By lemma \ref{K3K4}, there exists $\kappa_1 \in [i_0,i_0 + K_2]$ such that $\forall \alpha
\in \{-,+\}, \exists$ a deterministic vector  $(\beta^{\alpha}_{1,j})_{1 \leq j \leq d}$,
such that 
$$U^{\alpha}_{\kappa_1} = 
\sum^d_{j=1}\beta^{\alpha}_{1,j} \theta_j$$ 
and in addition $\|\beta^{\alpha}_{1,*}\| \leq K_3 K_2 + K_4 = s_1$.\\
{\bf i $\rightarrow$ i+1}: \\
Suppose that $\kappa_1 < \kappa_2 < \ldots < \kappa_i$ are fulfilling $(1)  \wedge (2) \wedge (3) \wedge (4)$. By lemma \ref{K3K4}, there exists $\kappa_{i+1} \in 
[i_0 + L_i + \ell_{i+1}, i_0 + K_1 (L_i + \ell_{i+1}) + K_2]$ such that $\forall \alpha
\in \{-,+\}, \exists$ a deterministic vector  $(\beta^{\alpha}_{i+1,j})_{1 \leq j \leq d}$,
such that
\begin{equation} 
U^{\alpha}_{\kappa_{i+1}} = 
\sum^d_{j=1}\beta^{\alpha}_{i+1,j} \theta_j
\label{(0)}
\end{equation} 
and in addition 
\begin{eqnarray}
\|\beta^{\alpha}_{i+1,*}\| \leq K_3 ( K_1 (L_i + \ell_{i+1}) + K_2) + K_4 &=&
K_3 L_{i+1} + K_4 \nonumber\\
&=& s_{i+1}
\label{(1)}
\end{eqnarray}
By lemma \ref{depth_eq_weight}
$$2 (W(\kappa_{i+1}) - W(\kappa_{i})) + 3 \geq \kappa_{i+1}-\kappa_{i}
\geq \ell_{i+1} = 2 \delta_{i+1} + 3$$
hence
\begin{equation}
W(\kappa_{i+1}) - W(\kappa_{i}) \geq \delta_{i+1}.
\label{(2)}
\end{equation}
At last
\begin{equation}
\kappa_{i+1} - i_0 \leq K_1( L_i + l_{i+1}) + K_2 = L_{i+1}.
\label{(3)}
\end{equation}
The above properties (\ref{(0)}-\ref{(1)}-\ref{(2)}-\ref{(3)}) prove the required conjunction.\\
It remains to prove point (0): the integer $\kappa_1$ introduced by lemma \ref{K3K4} is such that
$T_B$ occurs at $\kappa_1$, hence
\begin{eqnarray*}
W(\kappa_1) &=& W(\kappa_1 -k_1-1) +k_1 -1\\
& \geq & W(\kappa_1 -k_1-1) + 2 \geq 1.
\end{eqnarray*}
\end{prooof}

\begin{lem}
Let $(x_i)_{i \in \bbbn}$ be an infinite branch of $\tau$. Then
there exists some $i_0 \in \bbbn$ such that 
$(x_i)_{i \geq i_0}$ is a B-stacking sequence.
\label{K5K6}
\end{lem}
\begin{prooof}
Let us distinguish, a priori, several cases , and see that only the case where $\tau$
admits a B-stacking sequence is possible.\\
{\bf Case 1}:$T_B$ occurs finitely often on $\tau$.\\
Let $j$ be the largest integer such that $T_B$ occurs at $j$.
By the arguments used in the proof of lemma \ref{TB_in_I}, Case 2, we know that
$U^-_{j+k_1 \cdot D_1},U^+_{j+k_1 \cdot D_1}$ are both unmarked, and that
$$ \forall k \geq j+k_1 \cdot D_1,\forall \alpha \in \{-,+\}, 
 \| U_k^\alpha \| \leq  \| U_{j+k_1 \cdot D_1}^\alpha \| + k_1 \cdot K_0.
$$
This would imply that the branch contains a finite prefix on which $T_{cut}$ is defined,
which is impossible on an infinite branch.\\
{\bf Case 2}:For some sign $\alpha$, there are infinitely many integers $i$ such that
[$T^\alpha_B$ occurs at $i+k_1 +1$ and $\|U_i^{-\alpha} \| < N_0$].\\
In this case there would exist an infinite sequence of 
integers  
$i_1 < i_2< \ldots < i_\ell < $ such that 
$$\forall \ell \geq 0, U_{i_1}^{-\alpha}= U_{i_\ell}^{-\alpha}.$$
For a given $U_i^{-\alpha}$, only a finite number 
of values are possible for the pair $(U_{i +k_1 +1}^{-},U_{i +k_1 +1}^{+})$.
Hence there exist integers $\ell < \ell'$ such that
$$ \ell < \ell', \pi_\ell < \pi_{\ell'} \mbox{ and }
(U_{\ell +k_1 +1}^{-},U_{\ell +k_1 +1}^{+})= (U_{\ell' +k_1 +1}^{-},U_{\ell' +k_1 +1}^{+}).
$$
Here again $T_{cut}$ would have a non-empty value on some prefix of $\tau$, which is impossible.\\
{\bf Case 3}:$T_B$ occurs infinitely often on $\tau$ and, for every sign $\alpha$,
there are only finitely many integers $i$ such that
[$T^\alpha_B$ occurs at $i+k_1 +1$ and $\|U_i^{-\alpha} \| < N_0$].\\
Let us consider the set $I_0$ of the integers  $i$ such that , there exists a sign $\alpha_i$ such that
$$[T^{\alpha_i}_B\mbox{  occurs at }i+k_1 +1 \mbox{ and }
\|U_{i}^{-\alpha_i} \| \geq N_0].$$
By the hypothesis of case 3, $I_0 \neq \emptyset$. 
Let $i_0$ such that 
$$\|U_{i_0}^{-\alpha_{i_0}} \| = \mi \{\|U_{i}^{-\alpha_i}\|\mid i \in I_0\}.$$
Then $(x_i)_{i \geq i_0}$ is a B-stacking sequence. 
\end{prooof}

\section{Termination}
\label{B0_is_terminating}
\begin{lem} :  $\hat{\Str}_{ABC}$ is terminating on every unmarked assertion $A_0$:
if $A_0 \in {\cal A}$ is unmarked, then, $\exists n_0 \in \bbbn, 
\hat{\Str}^{n_0+1}_{ABC}(A_0) = \hat{\Str}^{n_0}_{ABC}(A_0)$.
\label{SABC_terminating}
\end{lem}
\begin{prooof} Suppose $A_0 \in {\cal A}$, $A_0$ is true, $A_0$ is unmarked  and 
\begin{equation}
\forall n \in \bbbn,
\hat{\Str}^{n}_{ABC}(A_0) \prec  \hat{\Str}^{n+1}_{ABC}(A_0).
\label{hypo_to_kill}
\end{equation}
Let us consider all the constants associated to this precise $A_0$, the equivalence $\bar{\psi}$ and the dpda ${\cal M}$ in section \ref{constants}.
Let us note : $t_n = \hat{\Str}^{n}_{ABC}(A_0)$ ( for every $n \in \bbbn$) and let 
$$t_\infty = l.u.b. \{ t_n \mid n \in \bbbn \}.$$
Let us notice that, by definition (\ref{StrABC_def2}), the strict inequality (\ref{hypo_to_kill}) implies
that
\begin{equation}
\forall n \in \bbbn, t_n \mbox{ is consistent}.
\label{tn_consistent}
\end{equation}
Let us denote by $x_n$ the node of $t_n$ such that $t_{n+1}= t_n [\hat{\Delta}(t_n)/x_n]$. 
Let us notice that , as every $x_n$ is unclosed in $t_n$, one can prove by induction that 
every $t_n$ is repetition-free. Hence
\begin{equation}
t_\infty \mbox{ is repetition-free}.
\label{tinfty_without_repetition}
\end{equation}
By Koenig's lemma, $t_\infty$ contains an infinite branch $y_0 y_1 \cdots y_s \cdots$ 
whose (infinite) labelling word is $A_0 A_1 \cdots A_s \cdots$ ( where $A_s = t_\infty(y_s)$).\\ 
Condition (C3) in the definition of $T^{(\OR)}_C$, combined with lemma \ref{rhoe_is_nice},  
shows that every equation $(\pi,T,U)$ produced by $T_C$ has size
\begin{equation}
\ma\{\|T\|,\|U\|\}\leq D_2,
\label{TC_is_bounded}
\end{equation}
hence that the number of possible unweighted equations produced by $T_C$ is {\em finite}. 
Hence $T_C$ occurs only a finite number of times
on this branch ( because $t_\infty$ is repetition-free (\ref{tinfty_without_repetition}) and 
$T_{cut}$ cannot occur on an infinite branch). Let $n_0$ be the last point where $T_C$ occurs 
( or $n_0 =0$ if $T_C$ never occurs on this branch). 
$(y_{{n_0}+i})_{i\geq 0}$ is a branch of a tree $t' \in {\cal T} (\Str_{AB}, A_{n_0})$. 
Let us notice also that
\begin{equation}
\mbox{ every equation produced by $T_C$ is unmarked},
\label{TC_is_unmarked}
\end{equation}
( by condition (C4) in the definition of $T^{(\OR)}_C$, see section \ref{strategies_B0}),
and 
\begin{equation}
\mbox{ every equation produced by $T_C$ has a length } \lambda \leq \lambda_2,
\label{TC_is_short}
\end{equation}
because it has a length $\leq d_0$ and $d_0 \leq \lambda_2$ by definition (\ref{lambda2}) 
in section \ref{constants}.
Moreover, the root $A_0$ of $t_\infty$ is supposed  to have a size $\leq D_2$ 
( by definition (\ref{D2}), in section \ref{constants}), to be unmarked ( by the hypothesis of the lemma),
and to have a length $\lambda_0 \leq \lambda_2$ ( by definition (\ref{lambda2}) in section
\ref{constants}).
Hence, in either case, $t'$ fulfills the hypotheses (\ref{unmarked})(\ref{lambda_small})
 stated in section \ref{stacking_derivations} and assumed in section \ref{trees}.\\
As $\Str_{ABC}$ is a strategy for ${\cal B}_0$and $A_0$ is true, $A_{n_0}$ is also true, hence hypothesis
(\ref{root_is_true}) assumed in section \ref{trees} is fulfilled. 
We may apply now the results obtained in \S \ref{NSS}.\\
 By lemma
\ref{K5K6}, the branch $(y_{n_0+i})_{i \geq 0}$ must contain an infinite B-stacking
sequence. Let us remark that, as $T_{\emptyset}$ does not occur (otherwise the branch would be 
finite) every equation $(\pi,U^{-},U^{+})$ labelling this branch is such that $U^{-} \neq 
\emptyset,U^{+}\neq \emptyset$. By lemma \ref{TC_is_defined} such a B-stacking  sequence 
contains a subsequence
$(A_{\kappa_1},A_{\kappa_2}, \cdots, A_{\kappa_d})$ with $d \leq d_0$, fulfilling hypotheses 
(1,2) of lemma \ref{T3}, and by the above remark it fulfills hypothesis (\ref{non_null_series}) 
of section \ref{triangulations} too. 
Let $n_i \in \bbbn$ such that $x_{n_i} = y_{\kappa_i}$, for $1 \leq i \leq d$.
By (\ref{tn_consistent}), $\Omega(\bar{\Pi}(t_{n_d}),\im(t_{n_d}))\neq \emptyset$.
Let us consider some 
$$\OR \in \Omega(\bar{\Pi}(t_{n_d}),\im(t_{n_d})).$$ 
Let $\Sys_d= (A_{\kappa_i})_{1 \leq i \leq d}$ and $D = \D(\Sys_d)$. By lemma 
\ref{T3},
\begin{equation} 
\INV(\Sys_d) \neq \bot, D \in [0,d-1] \mbox{ and }
\NN{\INV(\Sys_d)} \leq \Sigma_{d_0} + s_{d_0}.
\label{correct_TC}
\end{equation}
Let $\Sys_{D+1}= (A_{\kappa_i})_{1 \leq i \leq D+1}$.
By hypothesis (2) of lemma \ref{T3} (we established that this hypothesis is true), 
$$ \bar{\Pi}(t_{n_{D+1}}) \leq \bar{\Pi}(t_{n_{d}}),$$
and it is straightforward that
$$ \im(t_{n_{D+1}}) \subseteq \im(t_{n_{d}}),$$ 
hence, 
\begin{equation} 
\OR \in \Omega(\bar{\Pi}(t_{n_{D+1}}),\im(t_{n_{D+1}})).
\label{OR_is_correctforD}
\end{equation}
Let $W_{D+1} = A_0 \cdot A_1 \cdots A_{\kappa_1} \cdots A_{\kappa_{D+1}}$ ( the word from the root to $y_{\kappa_{D+1}}$).
Let us notice that 
\begin{equation}
\D(\Sys_{D+1})= \D(\Sys_{d})=D,   \INV(\Sys_{D+1}) = \INV(\Sys_d).
\label{Cut_is_correct}
\end{equation}
By (\ref{correct_TC}),(\ref{Cut_is_correct}),
\begin{equation}
\rho_e(\INV(\Sys_{D+1})) \in T^{(\OR)}_C(W_{D+1}).
\label{T_OR_definedat_WD}
\end{equation}
By (\ref{OR_is_correctforD})(\ref{T_OR_definedat_WD}), the set 
$\{\OR \in \Omega(\bar{\Pi}(t_{n_{D+1}}), \im(t_{n_{D+1}})), T^{(\OR)}_C(W_{D+1}) \neq 
\emptyset\}$ is not empty, so that case $(0)$ of the definition of $\hat{\Delta}(t)$ ( see section 
\ref{strategies_B0}) is fulfilled and 
$$\hat{\Delta}(t_{n_{D+1}})= A_{n_{D+1}}(T^{(\OR_0)}_C(W_{D+1})),$$
i.e. $T_C$ occurs at $y_{\kappa_{D+1}+1}$. This is a contradiction with the minimality of $n_0$.
We have proved that hypothesis (\ref{hypo_to_kill}) is impossible. Hence the lemma is proved.
\end{prooof}

\section{Elimination}
\label{elimination}
\subsection{System ${\cal B}_1$}
We prove here that the new formal system ${\cal B}_1$ obtained by {\em elimination} of  
meta-rule (R5) in ${\cal B}_0$ is recursively enumerable and complete. The decidability of the 
bisimulation problem follows.\\
Let ${\cal B}_1 = < {\cal A}, H, \deducedir_{{\cal B}_1}>$ where ${\cal A}, H,$ are the same as
in ${\cal B}_0$ , but the {\em elementary deduction relation} $\ededucedir_{{\cal B}_1}$ is 
the relation generated by the subset of 
metarules $R0,R1,R2,R3,R'3,R4,R6,R7,R8$, i.e. all the metarules of ${\cal B}_0$ except
$R5$. The deduction relation $\deducedir_{{\cal B}_1}$ is now defined by:
$${\deducedir}_{{\cal B}_1} = \ededuce{*}_{{\cal B}_1} \circ
{\stackrel{[1]}{\ededucedir}}_{R0,R3,R'3,R4} \circ \ededuce{*}_{{\cal B}_1}.$$

\begin{lem} : ${\cal B}_1$ is a deduction system.
\end{lem}
\begin{sketch}
As $\deducedir_{{\cal B}_1} \subseteq \deducedir_{{\cal B}_0}$, property (A1) is fulfilled by
$\deducedir_{{\cal B}_1}$.\\
By the well-known decidability properties for finite-automata, rules $R0,R1,R2,R3,$
$R'3,R4,R6,R7,R8$ are recursively enumerable. 
Hence  property (A2) is fulfilled by 
${{\cal B}_1}$.
\end{sketch}
\subsubsection{Completeness}
\label{B1_is_complete}
\begin{defi} Let $P$ be a finite subset of ${\cal A}$ and let $\bar{\pi} \in \bbbn$.
$P$ is said {\em locally $\bar{\pi}$-consistent} iff, for every $(\pi,S,S') \in P$, if
$$\pi < \bar{\pi},$$
then, there exists ${\cal R}_1 \in \bar{\cal B}_1$
such that
$$[\pi,S,S',{\cal R}_1] \subseteq \CO(P).$$
\label{local_consistency}
\end{defi}
\begin{lem}
Let $P$ be a finite subset of ${\cal A}$ and let $\bar{\pi} \in \bbbn$. If $P$ is 
locally $\bar{\pi}$-consistent, then  $P$ is $\bar{\pi}$-consistent.
\label{local_implies_global}
\end{lem}
\begin{prooof}
Let us consider, for every integers $n \geq 0, p \geq 0$,  the following property ${\cal Q}(n,p)$:
$\forall \pi \in \bbbn, \lambda \in \bbbn-\{0\}, S,S' \in {\sf DRB}_{1,\lambda} \pa{V},$
\begin{eqnarray}
(\pi,S,S')\in \CO_p(P) \mbox{ and } \pi + n -1 < \bar{\pi} 
& \Rightarrow & \nonumber\\
\exists {\cal R}_n \in {\cal B}_n(S,S'), 
 [\pi,S,S',{\cal R}_n] \subseteq \CO(P).&&
\label{Rn_in_congre}
\end{eqnarray}
Let us prove by induction on $(n,p)$ that
\begin{equation} 
\forall (n,p) \in \bbbn \times \bbbn, {\cal Q}(n,p).
\label{Qnp_istrue}
\end{equation}
{\bf\underline{$n=0,p=0$:}}$\;\;$\\
The only possible value of ${\cal R}_0 \in {\cal B}_0(S,S')$ is 
${\cal R}_0= \{(\epsilon,\epsilon)\},$ and $[\pi,S,S',{\cal R}_0]= \{(\pi,S,S')\} \subseteq 
\CO_0(P)$.\\
{\bf\underline{ $p>0$:}}$\;\;$\\
There exists a subset $Q \subseteq {\cal P}_f({\cal A})$, such that
$$ P \ededuce{p-1}_{\cal C} Q \mbox{ and } Q \ededuce{1}_{\cal C} \{(\pi,S,S')\}.$$
As every rule of ${\cal B}_0$ increases the weight, we can suppose that every assertion of $Q$
has a weight $\leq \pi$. Hence,
by induction hypothesis,
\begin{equation} 
\forall (\pi',T,T') \in Q, \exists {\cal R}_n \in {\cal B}_n(T,T'),
[\pi',T,T',{\cal R}_n] \subseteq \CO(P).
\label{P-induction_hypo}
\end{equation}
Let us consider the type of rule used in the last step, $Q \ededuce{1}_{\cal C} \{(\pi,S,S')\}$, 
of the above deduction.
\begin{description}
\item{R0:} $(\pi-1,S,S') \in Q$.\\
By (\ref{P-induction_hypo}), $\exists {\cal R}_n \in {\cal B}_n(S,S')$,
$$[\pi-1,S,S',{\cal R}_n] \subseteq \CO(P).$$
As $[\pi-1,S,S',{\cal R}_n] \ededuce{1}_{\cal C} [\pi,S,S',{\cal R}_n]$, 
$$[\pi,S,S',{\cal R}_n] \subseteq \CO(P).$$
\item{R1:}$(\pi,S',S) \in Q$.\\
(analogous to the above case) 
\item{R2:}$(\pi,S,T) , (\pi,T,S') \in Q$.\\
By (\ref{P-induction_hypo}), $\exists {\cal R}_n \in {\cal B}_n(S,T),
{\cal R}'_n \in {\cal B}_n(T,S') $,
$$[\pi,S,T,{\cal R}_n] \subseteq \CO(P),[\pi,T,S',{\cal R}'_n] \subseteq \CO(P) .$$
Using the properties mentionned in section \ref{congruence_closure}, we get that:
$$[\pi,S,S',{\cal R}_n \circ {\cal R}'_n] \subseteq \CO(P).$$
\item{R3:}$\;\;$\\
In this case, ${\cal R}_n = {\rm Id} \cap X^{\leq n} \times X^{\leq n} \in {\cal B}_n(S,S')$, and
$$ [\pi,S,S',{\cal R}_n] \subseteq \{(\pi,S,S')\} \cup \{ ( \pi + k, T, T), 1 \leq k \leq n, T \in {\sf DRB}_{1,\lambda}\pa{V}\} \subseteq \CO(P).$$
\item{R'3:}$\;\;$\\
In this case, ${\cal R}_n = {\rm Id} \cap X^{\leq n} \times X^{\leq n} \in {\cal B}_n(S,S')$, and
$$ [\pi,S,S',{\cal R}_n]= \{(\pi+k,S \odot u,\rho_e(S)\odot u) \mid 0 \leq k \leq n, u \in X^k\} \subseteq \CO(P),$$ 
(because $\rho_e(S)\odot u= \rho_e (S\odot u)$).
\item{R6:}$(\pi,S_1 \cdot S' +U,S') \in Q, S = S^*_1 \cdot U.$\\
By (\ref{P-induction_hypo}), $\exists {\cal R}_n \in {\cal B}_n(S_1 \cdot S' +U,S'),$
$$[\pi,S_1 \cdot S' +U,S',{\cal R}_n] \subseteq \CO(P).$$
Using the properties mentionned in section \ref{congruence_closure}, we get that:
\begin{eqnarray*}
[\pi, S, S', {\cal R}^{<S_1,*>}_n] &=& [\pi, S_1^* \cdot U, S', {\cal R}_n^{<S_1,*>}]\\
                                 & \subseteq & \CO [\pi,S_1 \cdot S'+ U ,S' ,{\cal R}_n]\\
& \subseteq & \CO (Q) \subseteq \CO(P).
\end{eqnarray*}
\item{R7:}$(\pi,S_1,S'_1) \in Q, S = S_1 \cdot T, S' = S'_1 \cdot T.$\\
By (\ref{P-induction_hypo}), $\exists {\cal R}_n \in {\cal B}_n(S_1,S'_1),$
$$[\pi,S_1,S'_1,{\cal R}_n] \subseteq \CO(P).$$
Using the properties mentionned in section \ref{congruence_closure}, we get that:
\begin{eqnarray*}
[\pi, S, S', <S_1 |{\cal R}_n>] &=& [\pi, S_1 \cdot T, S'_1 \cdot T,<S_1 |{\cal R}_n> ]  \\
& \subseteq & \CO ([\pi,S_1,S'_1,{\cal R}_n])\\
& \subseteq & \CO (Q) \subseteq \CO(P).
\end{eqnarray*}
\item{R8:}$\forall i \in[1,\delta],(\pi,T_{i,*},T'_{i,*}) \in Q, S = S_1 \cdot T, 
S' = S_1 \cdot T'.$\\
By (\ref{P-induction_hypo}), $\exists {\cal R}_{1,n }, \ldots,{\cal R}_{\delta,n }\in 
{\cal B}_n(T_{i,*},T'_{i,*}),$ such that
$$[\pi,T_{i,*},T'_{i,*},{\cal R}_{i,n }] \subseteq \CO(P).$$
Using the properties mentionned in section \ref{congruence_closure}, we get that:
\begin{eqnarray*}
[\pi, S,S', <S ,{\cal R}_{*,n}>] &=& [\pi, S_1 \cdot T, S_1 \cdot T',<S ,{\cal R}_{*,n}>]  \\
& \subseteq & \CO (\bigcup_{1 \leq i \leq \delta}[\pi,T_{i,*},T'_{i,*},{\cal R}_{i,n } ])\\
& \subseteq & \CO (Q) \subseteq \CO(P).
\end{eqnarray*}
\end{description} 
In all cases ${\cal Q}(n,p)$ has been established.\\ 
{\bf \underline{$n>0, p=0$:}}$(\pi,S,S') \in P.$\\
As $P$ is locally $\bar{\pi}$-consistent and $\pi \leq \pi + n-1 < \bar{\pi}$ , there exist 
${\cal R}_1 \in {\cal B}_1(S,S'), q \in \bbbn$ such that:
\begin{equation}
[\pi,S,S',{\cal R}_1] \subseteq \CO_q(P).
\label{R1_in_congre}
\end{equation}
As $(n-1,q)< (n,0)$, by induction hypothesis,$\forall (x,x') \in {\cal R}_1 \cap X \times X, \exists {\cal R}_{x,x',n-1} \in {\cal B}_{n-1}(S \odot x,S' \odot x')$ such that
\begin{equation}
[\pi +1, S \odot x, S' \odot x',{\cal R}_{x,x',n-1}] \subseteq \CO(P).
\label{Rx_in_congre}
\end{equation}
Let us consider ${\cal R}_n = \{(\epsilon,\epsilon)\} \bigcup_{(x,x') \in {\cal R}_1 \cap X \times X} \{(x,x')\}\cdot {\cal R}_{x,x',n-1}$.
One can check that ${\cal R}_n \in {\cal B}_n(S,S')$ and , by (\ref{R1_in_congre}, \ref{Rx_in_congre}) we obtain:
$$[\pi ,S, S',{\cal R}_{n}] = \{(\pi,S,S')\} \bigcup_{(x,x') \in {\cal R}_1 \cap X \times X} 
[S \odot x,S' \odot x',{\cal R}_{x,x',n-1}] \subseteq \CO(P).$$
Let us define now an oracle $\OR \in \Omega$ which is $\bar{\pi}$-consistent with $P$.
For every $(S,S') \in \bigcup_{\lambda \geq 1} {\sf DRB}_{1,\lambda}\pa{V}$ occuring in $\CO(P)$ 
(i.e. as  the projection on $\bigcup_{\lambda \geq 1}{\sf DRB}_{1,\lambda}\pa{V} \times {\sf DRB}_{1,\lambda}\pa{V}$ of an assertion in  $\CO(P)$), let us note
$$ W(S,S')= \mi(\{ \pi \in \bbbn \mid (\pi,S,S') \in \CO(P)\}).$$
$$ D(S,S')= \ma\{\bar{\pi}-W(S,S'), 0\}.$$
$$ C(S,S')= \mi\{ {\cal R} \in {\cal B}_{D(S,S') }(S,S') \mid  [W(S,S'),S,S',{\cal R}] \subseteq
\CO(P) \}.$$
Notice that $C(S,S')$ is well-defined, owing to property (\ref{Qnp_istrue}).
We then define $\OR$ by: for every $(S,S')$ occuring in $\CO(P)$,
\begin{equation}
\OR(S,S') = \mi\{ {\cal R} \in {\cal B}_\infty(S,S') \mid C(S,S')={\cal R} \cap (X^{\leq D(S,S')}
\times X^{\leq D(S,S')}) \},
\label{THE_oracle}
\end{equation}
and for every $(S,S')$ not occuring in $\CO(P)$,
\begin{equation}
\OR(S,S') = \mi\{ {\cal R} \in {\cal B}_\infty (S,S')\} (\mbox{ if } S \sim S'),\;\;
 \OR(S,S') = {\rm Id}_{X^*} (\mbox{ if } S \not\sim S').
\label{trivial_case}
\end{equation} 
One can check that, by the choice of $C(S,S')$, $\OR$ is $\bar{\pi}$-consistent with $P$.
\end{prooof}
\begin{lem}
Let $A_0 \in {\cal A}$ such that $H(A_0) = \infty$. Let us consider the sequence of trees
$t_n=\hat{\Str}^n_{ABC}(A_0)$. For every integer $n \geq 0$, $t_n$ is consistent.
\label{consistency_persists}
\end{lem}
Let us say that the strategy $T$ ``applies on'' node $x$ iff, $x$ has exactly $m$ sons 
$x\cdot 1, x \cdot2, \ldots, x \cdot m$ and 
$$ \tau(x1) \cdot \tau(x \cdot 2) \cdots \tau( x \cdot m) \in T(\tau(x[0]) \cdot \tau(x[1]) \cdots \tau(x[|x|])),$$
i.e. the word consisting of the labels of the sons of $x$ belongs to the image of the path 
from $\epsilon$ (included) to $x$ (included) by the strategy $T$.\\
\begin{prooof}
For every $k \in \bbbn$ we define 
$$\bar{\pi}_k= \bar{\Pi}(t_k).$$
We prove by induction on $(n,\pi)$ the following property ${\cal R}(n,\pi)$:
\begin{eqnarray}
\label{Rnpi_hypo}
\forall x \in \dom(t_n),\mbox{ if } t_n(x) = (\pi,S,S') \mbox{ with } \pi < \bar{\pi}_n,
& \mbox{ then } &\\
\exists {\cal R}_1 \in {\cal B}_1(S,S'), [\pi,S,S',{\cal R}_1] \subseteq \CO(\im(t_n)).&&
\label{Rnpi_conclus}
\end{eqnarray}
At every step of our proof by induction, we consider some node $x$ of $t_n$ fulfilling hypothesis
(\ref{Rnpi_hypo}) and we show that it must fulfill (\ref{Rnpi_conclus}).
Let us notice that , if $x$ is not closed, then hypothesis (\ref{Rnpi_hypo}) cannot be true, 
by minimality of $\bar{\pi}_n$. Let us notice also that, if $x$ is closed , but there is some 
$x' \prec x$ such that $t_n(x') = t_n(x)$, then (\ref{Rnpi_conclus}) on
$x$  is the same property as  (\ref{Rnpi_conclus}) for $x'$. Hence , in the sequel, we can suppose
  that $x$ is closed and that it is minimal (w.r.t. to $\preceq$):
\begin{equation}
 x = \mi_{\preceq}\{ y \in \dom(t_n) \mid t_n(y) = t_n(x) \}.
\label{x_minimal}
\end{equation}
{\bf \underline{$n=0,\pi=0$:}} $\dom(t_0)= \{\epsilon\}$, $t_0(\epsilon)=A_0$. If $\epsilon$ is 
not closed, then $\bar{\pi}_0=\pi=0$, hence there is no node $x$ fulfilling hypothesis (\ref{Rnpi_hypo}). 
Otherwise, $\bar{\pi}_0 = \infty$ and  $x=\epsilon$ is closed: either $T_{\emptyset}(A_0)= \{\epsilon\}$ or $T_{\varepsilon}(A_0)= \{\epsilon\}$. Let us choose 
\begin{equation}
{\cal R}_1= {\rm Id}_{X^*} \cap X^{\leq 1} \times X^{\leq 1}.
\label{choose_identity}
\end{equation}
If we note $A_0 = (\pi,S^-_0,S^+_0)$, then 
$$[\pi,S^-_0,S^+_0,{\cal R}_1]= \{(\pi,S^-_0,S^+_0)\} \cup\{(\pi+1,S^-_0 \odot x,S^+_0 \odot x)
\mid x \in X\},$$
where , $\forall x \in X, S^-_0 \odot x \equiv S^+_0 \odot x \equiv \emptyset$.
Using rule $R'3$, we see that
\begin{equation} 
[\pi,S,S',{\cal R}_1] \subseteq \CO(\emptyset) \subseteq \CO(\im(t_n)).
\label{good_choice}
\end{equation}
{\bf \underline{$n>0,\pi=0$:}} Let $x$ be some node of $t_n$ such that $ \exists S,S', 
t_n(x) = (\pi,S,S')$ and $\pi < \bar{\pi}_n$.
Let us denote by $W_x$ the word labelling the path from the root of $t_n$ ( included) to $x$
 ( included ).\\
{\bf case 1:} $\;\;\exists x' \in \dom(t_n), x' \mbox{ internal node },
 \mbox{ such that } t_n(x') = t_n(x)$.\\
As $\pi = 0$, the sons $x'\cdot 1,x'\cdot 2,\ldots,x' \cdot m$ of $x'$ are such that 
$t_n(x' \cdot 1)\cdot t_n(x' \cdot 2)\cdots t_n(x' \cdot m) \in T_A^{(\OR)}(W_{x'}),$ for 
some oracle $\OR$. Let us choose 
\begin{equation}
{\cal R}_1 = \OR(S,S') \cap X^{\leq 1} \times X^{\leq 1}.
\label{choice_for_TA}
\end{equation} 
Then
\begin{equation}
[\pi,S,S',{\cal R}_1] \subseteq \im(t_n).
\label{TA_is_OK}
\end{equation}
{\bf case 2:} $\;\;T_{\emptyset}(W_x)= \{\epsilon\}$ or $T_{\varepsilon}(W_x)= \{\epsilon\}$.\\
In this case the choice ${\cal R}_1= {\rm Id}_{X^*} \cap X^{\leq 1} \times X^{\leq 1}$ satisfies 
again (\ref{good_choice}).\\
{\bf \underline{$\pi>0$:}}\\
Let $x$ fulfilling hypothesis (\ref{Rnpi_hypo}). As $t_n$ is a proof-tree for ${\Str}_{ABC}$, and
as we suppose $x$ is closed and minimal (\ref{x_minimal}), one of the following cases must occur.\\
{\bf case 1:}$\;\;T_{cut}$ applies on $x$.\\
There exists $x' \in \dom(t_n), \exists \pi' \in \bbbn$, such that
$$ t_n(x') = (\pi', S,S') \mbox{ and }\pi' <\pi.$$
By induction hypothesis 
$$\exists {\cal R}_1 \in {\cal B}_1(S,S'), [\pi',S,S',{\cal R}_1] \subseteq \CO(\im(t_n)),$$
and by means of rule $R0$:
$$[\pi,S,S',{\cal R}_1] \subseteq \CO([\pi',S,S',{\cal R}_1]).$$
Hence (\ref{Rnpi_conclus}) is true.\\ 
{\bf case 2:}$\;\;T_{\emptyset}$ ot $T_{\epsilon}$ applies on $x$.\\
Here again, the choice (\ref{choose_identity}) fufills property (\ref{good_choice}). \\
In the remaining cases we use the following notation: 
for every $k \in \bbbn$ such that $t_k$ is not closed,
$$x_k= \mi\{ x \in \dom(t_k), x \mbox{ is not closed for }\Str_{ABC} \mbox{ and }\exists S,S',
t(x) = (\bar{\pi}_k,S,S') \}.$$
If $ \exists k < n \mid t_k$ is not consistent or is closed, then by (\ref{StrABC_def2}),
$t_k = t_{k+1} = \cdots = t_n$, hence ${\cal R}(n,\pi) \Leftrightarrow {\cal R}(k,\pi)$, and 
this last property is true by induction hypothesis.\\
Let us suppose now that $\forall k < n$, $t_k$ is consistent and unclosed.
According to formula (\ref{StrABC_def1}), 
$$t_{k+1}=t_{k}[e_{k+1}/x_{k}],$$ for some tree of depth one, $e_{k+1}$. \\
Let $k \in [0,n-1], x = x_k, \pi= \bar{\pi}_k$ (such a $k$ must 
exist because $x$ is internal). Let $x \cdot 1,\ldots, x \cdot \mu$ be the sequence of sons of $x$.\\
{\bf case 3:}$\;\;T_{A}$ applies on $x$.\\
Hence there exists some oracle $\OR$ such that $T^{(\OR)}_{A}$ applies on $x$.
The choice (\ref{choice_for_TA}) fufills property (\ref{TA_is_OK}).\\
{\bf case 4:}$\;\;T^{\alpha}_{B}$ applies on $x$ ( for some $\alpha \in \{-,+\}$).\\
Let us suppose $\alpha= +$. Let $x' = x( |x| - k_1)$ ( the prefix of $x$ having length $|x|-k_1$),
$t_n(x') = (\pi',\bar{U},U').$
By definition of $\hat{\Str}_{ABC}$, there exists some oracle $\OR$ which is 
$\bar{\pi}_k$-consistent with $\im(t_k)$ and such that:
$$\mu =1 \mbox{ and }t_n(x \cdot 1)= T^{(\OR),+}_{B}(W_x).$$
Let us look at the proof of lemma \ref{TB_are_strategies} in the particular case  of this oracle 
$\OR$: as the pairs $(u_\ell,u'_\ell)$ belong to $\OR(\bar{U},U')$ (for every $\ell \in [1,q]$) and
$\pi'+ |u_\ell| -1 < \pi' + k_0 \leq \pi' + 2 \cdot k_0 < \bar{\pi}_k$,
 deduction (\ref{num1}) can be obtained just by using rules in ${\cal C}$. As deduction (\ref{num1}) is 
the only one 
( in the proof of lemma \ref{TB_are_strategies}) using rules in ${\cal B}_0 - {\cal C}$ we conclude that deduction (\ref{num0}) can be replaced by:
\begin{equation}
\{ t_n(x'),t_n(x \cdot 1)\}\cup \im(t_k) \ededuce{*}_{\cal C} \tau_{-1}(t_n(x)).
\label{TB_is_congre}
\end{equation}
(We recall $\tau_{-1}$ consists in replacing the weight of a given weighted equation into its
predecessor).
Deduction (\ref{TB_is_congre}) implies that
\begin{equation}
\exists p \in \bbbn, (\pi-1,S,S') \in \CO_p(\im(t_n)).
\label{x_in_congre_p}
\end{equation}
By induction hypothesis, as $\pi-1 < \bar{\pi}_n$, $\im(t_n)$ is locally $\pi-1$-consistent, hence, by lemma \ref{local_implies_global}, $\im(t_n)$ is $\pi-1$-consistent. Hypothesis (\ref{x_in_congre_p}) implies that
$$\exists {\cal R}_1 \in {\cal B}_1(S,S'), [\pi-1,S,S',{\cal R}_1] \subseteq \CO(\im(t_n)),$$
hence, using R0,  that
$$\exists {\cal R}_1 \in {\cal B}_1(S,S'), [\pi,S,S',{\cal R}_1] \subseteq \CO(\im(t_n)).$$
{\bf case 5:}$\;\;T_{C}$ applies on $x$.\\
By definition of $\hat{\Str}_{ABC}$, there exists some oracle $\OR$ which is 
$\bar{\pi}_k$-consistent with $\im(t_k)$ and such that:
$$\mu=1 \mbox{ and }t_n(x \cdot 1)= T^{(\OR)}_{C}(W_x).$$
Let $W_x= A_1\cdots A_\ell \cdots A_{|x|+1}$, $ \kappa_1< \cdots<\kappa_i<\kappa_{i+1}<  \cdots 
\kappa_{D+1}=|x|+1$,
 ${\Sys}= ({\cal E}_{i})_{1 \leq i \leq D+1}$, where , for every $1 \leq i \leq d$,
 $${\cal E}_{i}=A_{\kappa_i}= (\pi_i, \sum^{d}_{j=1} \alpha_{i,j} S_j \;,\;
\sum^{d}_{j=1}\beta_{i,j} S_j)$$ and
$$T^{(\OR)}_{C}(W_x)= \rho_e(\INV(\Sys)), \W(\Sys) \neq \bot,\D({\Sys}) = D \leq d-1.$$ 
Let us look at the proof of lemma \ref{T1} in the particular case  of this oracle 
$\OR$: the only place where a rule in ${\cal B}_0 - {\cal C}$ is used, is in deduction (\ref{EmE'm}), when case 2, subcase1 (or case 2, subcase 2), of the 
recursive definition of $\INV(\Sys)$ occurs .
Let us recall that the pair $(u,u')$ chosen by the oracle  $\OR$ is such that:
$${\cal R} = \OR (\sum^{d}_{j=1} \alpha_{1,j} S_j,\sum^{d}_{j=1}\beta_{1,j} S_j),$$
$$\nu = \Div(\alpha_{1,*},\beta_{1,*}),\;\;
{\cal R}_\nu = {\cal R} \cap X^{\leq \nu} \times X^{\leq \nu},\;\;
(u,u') \in {\cal R}_\nu.$$ 
Let us notice that $\pi_1 + \nu -1 < \pi_1 + 2 \cdot \nu < \pi_2 \leq \W(\Sys) +1 = \pi =\bar{\pi}_k$.
As $\OR$ is $\bar{\pi}_k$-consistent with $\im(t_k)$, we conclude that
\begin{eqnarray*}
(\pi_1 + |u|, (\sum^{d}_{j=1} \alpha_{i,j} S_j)\odot u, (\sum^{d}_{j=1}\beta_{i,j} S_j) 
\odot u') &\in &[\pi_1 , \sum^{d}_{j=1} \alpha_{i,j} S_j, \sum^{d}_{j=1}\beta_{i,j} S_j,
{\cal R}_\nu] \\
& \subseteq & \CO (\im(t_k)).
\end{eqnarray*}
Hence deduction (\ref{EmE'm}) can be replaced by
\begin{equation}
{\cal E'}_1 \in \CO(\im(t_k)).
\label{E1E'1_congre}
\end{equation}
Similarly, for every $i \in [2,D]$, as $\pi_i + 2 \cdot \Div(\alpha^{(i-1)}_{i,*},
\beta^{(i-1)}_{i,*}) < \pi_{i+1} \leq \W(\Sys) +1 = \pi =\bar{\pi}_k$, and ${\cal E}^{(i-1)}_{i} \in \CO(\im(t_k))$,
\begin{equation}
({\cal E}^{(i-1)}_{i})' \in \CO(\im(t_k)).
\label{EiE'i_congre}
\end{equation}
It follows that  deduction (\ref{one_strengthened}) can be replaced by
\begin{equation}
\{\INV(\Sys)\} \cup 
\im(t_k) \ededuce{*}_{\cal C} \tau_{-1}(t_n(x)).
\label{one_strengthened_congre}
\end{equation}
using the facts that $ \rho_e(\INV(\Sys)) \ededuce{*}_{\cal C} \INV(\Sys)$ and $\im(t_k) \subseteq \im(t_n)$
we may conclude that:
\begin{equation}
\{t_n(x \cdot 1)\} \cup 
\im(t_n) \ededuce{*}_{\cal C} \tau_{-1}(t_n(x))= (\pi-1,S,S').
\label{one_strengthened_congre}
\end{equation}
From (\ref{one_strengthened_congre}) and the induction hypothesis, we can conclude, as in case 4,
that
$$\exists {\cal R}_1 \in {\cal B}_1(S,S'), [\pi,S,S',{\cal R}_1] \subseteq \CO(\im(t_n)).$$
(End of the induction).\\
By the above induction, for every $n \in \bbbn$, $\im(t_n)$ is $\bar{\pi}_n$-consistent i.e. 
$t_n$ is consistent.
\end{prooof}
\begin{lem}$\hat{\Str}_{ABC}$ is closed. 
\label{hatSABC_closed}
\end{lem}
\begin{prooof}
Let $A_0 \in {\cal A}$. By lemma \ref{consistency_persists}, 
$\forall n \in \bbbn,\hat{\Str}_{ABC}^{n}(A_0)$ is consistent.\\
If $\hat{\Str}_{ABC}^{n}(A_0)$ is consistent and is not closed, then , by definition (\ref{StrABC_def1}),
$$\hat{\Str}_{ABC}^{n}(A_0) \neq \hat{\Str}_{ABC}^{n+1}(A_0);$$
if $\hat{\Str}_{ABC}^{n}(A_0)$ is consistent and is closed, then , by definition (\ref{StrABC_def2}),
$$\hat{\Str}_{ABC}^{n}(A_0) = \hat{\Str}_{ABC}^{n+1}(A_0).$$
Hence the equivalence (\ref{closed_global_strategy}), which defines the notion of closed global 
strategy, is fulfilled by $\hat{\Str}_{ABC}$.
\end{prooof}
\begin{theo} : ${\cal B}_0, {\cal B}_1$ are complete formal systems.
\label{B1_complete}
\end{theo}
\begin{prooof}
By lemma \ref{SABC_terminating} $\hat{\Str}_{ABC}$ is terminating on every unmarked assertion 
and by
lemma \ref{hatSABC_closed} $\hat{\Str}_{ABC}$ is closed. Let $A_0$ be some unmarked true assertion.
According to the proof of lemma \ref{term_closed_implies_complete}, $\exists n_0 \in \bbbn$ such 
that $t_{\infty} = \hat{\Str}^{n_0}(A_0)$ is a proof-tree  which is closed, hence such that
$\bar{\Pi}(t_\infty) = \infty$. By lemma \ref{hatSABC_closed}, $t_\infty$ is consistent, i.e.
$\im(t_\infty)$ is $\infty$-consistent: $\forall (\pi,S,S') \in \im(t_\infty)$,
$$\exists {\cal R}_1 \in {\cal B}_1(S,S'), [\pi,S,S',{\cal R}_1] \subseteq \CO(\im(t_\infty)),$$
hence,
\begin{equation}
\im(t_\infty) \ededuce{*}_{\cal C} [\pi,S,S',{\cal R}_1] \deducedir_{R4} (\pi,S,S').
\label{THE_B1_deduction}
\end{equation}
As the rules of ${\cal C}$ and $R4$ are rules of ${\cal B}_1$, deduction (\ref{THE_B1_deduction})
shows that
\begin{equation}
\im(t_\infty) \deducedir_{{\cal B}_1} (\pi,S,S').
\label{final_deduction}
\end{equation}
i.e. $\im(t_\infty)$ is a ${\cal B}_1$-proof.\\
In the general case where $A_0=(\pi_0,U^-_0,U^+_0)$ might be marked, 
we observe that, owing to rules (R1)(R2)(R'3):

$$ \{\rho_e(A_0)\} \ededuce{*}_{\cal C} \{A_0\} .$$
This deduction combined with some ${\cal B}_1$-proof of $\rho_e(A_0)$ gives a ${\cal B}_1$-proof
of $A_0$.
\end{prooof}

\begin{theo}
The bisimulation  problem for rooted equational 1-graphs of finite out-degree is decidable.
\label{decidable}
\end{theo}
\begin{prooof}
Let us consider the sequence of statements: lemma \ref{reduction1}, lemma \ref{reduction2}, 
corollary \ref{2graph_and_pda} and lemma \ref{gamma_reducesto_sigma}.
By means of the above statements, the bisimulation  problem for rooted equational 1-graphs of 
finite out-degree reduces to the following decision problem ( we call it the bisimulation 
problem for deterministic vectors):
\begin{description}
\item{INSTANCE}: a bi-rooted, normalized  dpda ${\cal M}$, its terminal alphabet $X$, a
surjective litteral morphism ${\psi}:X^* \rightarrow Y^*$ (we denote its kernel by $\bar{\psi}$), 
and $\lambda \in \bbbn-\{0\},S,S' \in 
\bbbD \bbbR \bbbB_{1,\lambda}\pa{V}$ 
(where $V$ is the structured alphabet associated with ${\cal M}$).
\item{QUESTION}: $S \sim S'$?
(where $\sim$ is the $\bar{\psi}$-bisimulation relation). 
\end{description}
Let us consider ${\cal M},X,V,\bar{\psi}$ given by some instance. \\
The equivalence relation $\sim$ on $\bbbD \bbbR \bbbB_{1,\lambda}\pa{V}$ has a recursively 
enumerable complement (this is
well-known). By theorem \ref{B1_complete} and lemma \ref{complete_implies_pre}, relation  $\sim$ is 
recursively enumerable too. Hence $\sim$ is recursive.\\
But the function associating to every ${\cal M},X,V,\bar{\psi}$ the corresponding deduction sytem 
${\cal B}_1$ is recursive. Hence the bisimulation problem for deterministic 
vectors is decidable.
\end{prooof}

\subsection{System ${\cal B}_2$}
We exhibit here a deduction system ${\cal B}_2$ which is simpler than ${\cal B}_1$
and is still complete. 
\subsubsection{Elementary rules}
Let us {\em eliminate} the weights in the rules of ${\cal B}_1$: we define a new set of assertions,
${\cal A}_2$ by
$${\cal A}_2= \bigcup_{\lambda \in \bbbn - \{0\}} \bbbD \bbbR \bbbB_{1,\lambda}\pa{V} \times 
\bbbD \bbbR \bbbB_{1,\lambda}\pa{V}.$$
We define a binary relation ${\ededucedir} \subseteq {\cal P}_f({\cal A}_2) \times
{\cal A}_2$, the {\em elementary deduction relation},  as the set of all the 
pairs having one of the following forms:
\begin{description}

\item[(R21)]$$ \; \; \{(S, T)\}  {\ededucedir} (T, S)$$
for $\lambda  \in \bbbn-\{0\}, S, T \in \bbbD \bbbR \bbbB_{1,\lambda}\pa{V}$,
\item[(R22)]$$ \; \; \{(S, S'), (S', S'')\}  {\ededucedir}  (S, S'')$$
for $\lambda  \in \bbbn-\{0\},  S, T \in \bbbD \bbbR \bbbB_{1,\lambda}\pa{V}$,
\item[(R23)]$$ \; \; \emptyset   {\ededucedir} (S, S)$$
for $ S \in \bbbD \bbbR \bbbB_{1,\lambda}\pa{V}$,
\item[(R'23)]$$ \; \; \emptyset   {\ededucedir} (S, \rho_e(S))$$
for $ S \in \bbbD \bbbR \bbbB_{1,\lambda}\pa{V}$,
\item[(R24)]$$ \; \; \{(S \odot x, T \odot x') \mid (x,x') \in {\cal R}_1 \}  {\ededucedir}(S, T)$$
for $\lambda  \in \bbbn-\{0\},  S, T \in \bbbD \bbbR \bbbB_{1,\lambda}\pa{V},
(S  \not\equiv \epsilon \wedge T \not\equiv \epsilon)$ and ${\cal R}_1 \in \bar{{\cal B}}_1$,
\item[(R25)]$$ \; \{(S_1 \cdot T+ S, T)\} {\ededucedir}  (S_1^* \cdot S , T)$$
for $\lambda \in \bbbn-\{0\}, S_1 \in \bbbD \bbbR \bbbB_{1,1}\pa{V}, S_1 \not\equiv \epsilon, 
(S_1,S) \in \bbbD \bbbR \bbbB_{1,\lambda+1}\pa{V}, T \in \bbbD \bbbR \bbbB_{1,\lambda}\pa{V}$,
\item[(R26)]$$ \; \; \{(S, S')\} {\ededucedir} (S \cdot T, S' \cdot T)$$
for $\delta, \lambda \in \bbbn-\{0\},  S, S' \in \bbbD \bbbR \bbbB_{1,\delta}\pa{V},
T \in \bbbD \bbbR \bbbB_{\delta,\lambda}\pa{V}$,
\item[(R27)]$$ \; \; \{(T_{i,*}, T'_{i,*}) \mid 1 \leq i \leq \delta\} {\ededucedir} 
(S \cdot T, S \cdot T')$$
for $\delta, \lambda \in \bbbn-\{0\},  S \in \bbbD \bbbR \bbbB_{1,\delta}\pa{V},
T, T' \in \bbbD \bbbR \bbbB_{\delta,\lambda}\pa{V}$,
\end{description}
We define ${\deducedir}_{{\cal B}_2}$ by : for every $ P \in {\cal P}_f({\cal A}_2), 
A \in {\cal A}_2$,\\
$$P {\deducedir} A  \Longleftrightarrow P \ededuce{*} \circ
{\stackrel{[1]}{\ededucedir}}_{23,24} \circ \ededuce{*} \{A\}.$$\\
where $\ededucedir_{23,24}$ is the relation defined by
$R23,R'23,R24$ only. \\
We define a simpler cost function $H_2: {\cal A}_2 \rightarrow \bbbn \cup \{\infty\}$ by :
$$ \forall (S,S') \in {\cal A}_2, H_2(S,S') = \Div(S,S').$$ 
We let 
$${\cal B}_2 = <{\cal A}_2, H_2, \deducedir_{{\cal B}_2}>.$$
\begin{lem} : ${\cal B}_2$ is a deduction system.
\end{lem}
\subsubsection{Completeness}$\;\;$\\
Let us denote by ${\cal C}_2$ the subset of rules of ${\cal B}_2$ obtained by removing the weights
in the rules of ${\cal C}$. 
\begin{defi}
Let $P \in {\cal P}_f({\cal A}_2)$. $P$ is said to be {\em self-generating }iff, for every 
$(S,S') \in P$,
\begin{enumerate}
\item either $S=S'=\epsilon$
\item or $\exists {\cal R}_1 \in \bar{\cal B}_1(S,S'), \forall (x,x') \in {\cal R}_1,
P \ededuce{*}_{{\cal C}_2}(S \odot x, S' \odot x') $.
\end{enumerate}
\label{self-proof}
\end{defi}
(See in remark \ref{chapeau_caucal} below, the origins of this notion).
\begin{lem}
Let $A \in {\cal A}_2$ such that $A$ is unmarked. Then $H(A) = \infty$ iff there exists a finite self-generating set 
$P \subseteq {\cal A}_2$ such that $A \in P$. 
\label{true_iff_self_proof} 
\end{lem}
\begin{prooof}
Owing to metarules $R23, R24$ it is clear that every self-generating set $P \in 
{\cal P}_f({\cal A}_2)$ is a ${\cal B}_2$-proof. Hence , if $A$ belongs to some self-generating set, then
$H(A) = \infty$.\\
Let us suppose now that $H_2(A) = \infty$.
Let us consider the closed proof-tree $t_\infty$ obtained by applying 
the global strategy $\hat{S}_{ABC}$ on the assertion $(0,A)$. By lemma 
\ref{SABC_terminating} $t_\infty$ is finite and by lemma 
\ref{hatSABC_closed}, $t_\infty$ is consistent, which means that $\im(t_\infty)$ is 
$\infty$-consistent. Let
$$P = pr_{2,3}(\im(t_\infty)),$$
(where
$pr_{2,3}: {\cal A} \rightarrow {\cal A}_2$ is the map erasing the weights).\\
As $\im(t_\infty)$ is $\infty$-consistent, $P$ is self-generating and $A \in P$.
\end{prooof}
\begin{theo} : ${\cal B}_2$ is a complete deduction system.
\label{B2_is_complete}
\end{theo}
\begin{prooof}
We already noticed that every self-generating set is a ${\cal B}_2$-proof. Hence lemma 
\ref{true_iff_self_proof} proves that every true, unmarked assertion 
possesses some finite  ${\cal B}_2$-proof.\\
Let $A$ be any true assertion. $\rho_e(A)$ has a finite proof $P$. Owing to rules
(R1)(R2)(R'3), $Q=P \cup \{A\}$ is a ${\cal B}_2$-proof of $A$.
\end{prooof}
\subsection{System ${\cal B}_3$}
We exhibit here a deduction system ${\cal B}_3$ which is even simpler than ${\cal B}_2$
and is still complete.
Let us consider ${\cal B}_3 = <{\cal A}_3, H_3, \deducedir_{{\cal B}_3}>,$
where 
$${\cal A}_3= \bigcup_{\lambda \in \bbbn - \{0\}} \bbbD \bbbR \bbbB_{1,\lambda}\pa{V_0} \times 
\bbbD \bbbR \bbbB_{1,\lambda}\pa{V_0}.$$
, $H_3 = {H_2}_|{\cal A}_3$ and $\deducedir_{{\cal B}_3}$ is defined below:
the metarules of ${\cal B}_3$ are essentially those of ${\cal B}_2$, but 
restricted to the unmarked vectors. 
\begin{description}

\item[(R31)]$$ \; \; \{(S, T)\}  {\ededucedir} (T, S)$$
for $\lambda  \in \bbbn-\{0\}, S, T \in \bbbD \bbbR \bbbB_{1,\lambda}\pa{V_0}$,
\item[(R32)]$$ \; \; \{(S, S'), (S', S'')\}  {\ededucedir}  (S, S'')$$
for $\lambda  \in \bbbn-\{0\},  S, T \in \bbbD \bbbR \bbbB_{1,\lambda}\pa{V_0}$,
\item[(R33)]$$ \; \; \emptyset   {\ededucedir} (S, S)$$
for $ S \in \bbbD \bbbR \bbbB_{1,\lambda}\pa{V_0}$,
\item[(R34)]$$ \; \; \{(S \odot x, T \odot x') \mid (x,x') \in {\cal R}_1 \}  {\ededucedir}(S, T)$$
for $\lambda \in \bbbn-\{0\}, S, T \in \bbbD \bbbR \bbbB_{1,\lambda}\pa{V_0}, 
(S  \not\equiv \epsilon \wedge T \not\equiv \epsilon)$ and ${\cal R}_1 \in \bar{{\cal B}}_1$,
\item[(R35)]$$ \; \{(S_1 \cdot T+ S, T)\} {\ededucedir}  (S_1^* \cdot S , T)$$
for $\lambda \in \bbbn-\{0\}, S_1 \in \bbbD \bbbR \bbbB_{1,1}\pa{V_0}, S_1 \not\equiv \epsilon, 
(S_1,S) \in \bbbD \bbbR \bbbB_{1,\lambda+1}\pa{V_0}, T \in \bbbD \bbbR \bbbB_{1,\lambda}\pa{V_0}$,
\item[(R36)]$$ \; \; \{(S, S')\} {\ededucedir} (S \cdot T, S' \cdot T)$$
for $\delta, \lambda \in \bbbn-\{0\},  S, S' \in \bbbD \bbbR \bbbB_{1,\delta}\pa{V_0},
T \in \bbbD \bbbR \bbbB_{\delta,\lambda}\pa{V_0}$,
\item[(R37)]$$ \; \; \{(T_{i,*}, T'_{i,*}) \mid 1 \leq i \leq \delta\} {\ededucedir} 
(S \cdot T, S \cdot T')$$
for $\delta, \lambda \in \bbbn-\{0\},  S \in \bbbD \bbbR \bbbB_{1,\delta}\pa{V_0},
T, T' \in \bbbD \bbbR \bbbB_{\delta,\lambda}\pa{V_0}$,
\end{description}

We then define ${\deducedir}_{{\cal B}_3}$ by : for every $ P \in {\cal P}_f({\cal A}_3), 
A \in {\cal A}_3$,\\
$$P {\deducedir}_{{\cal B}_3} A  \Longleftrightarrow P \ededuce{*}_{{\cal B}_3} \circ
{\stackrel{[1]}{\ededucedir}}_{33,34} \circ \ededuce{*}_{{\cal B}_3} \{A\}.$$\\
where $\ededucedir_{33,34}$ is now the relation defined by
$R33,R34$ only. \\

As $ {\deducedir}_{{\cal B}_3} \subseteq {\deducedir}_{{\cal B}_2}$ , $H_3 = H_2$, it is clear that
${\cal B}_3$ is a deduction system. 
\subsubsection{Completeness}$\;\;$\\
Let us call ${\cal C}_3$ the intersection of set of the rules of ${\cal C}$ whith the set of rules 
of ${\cal B}_3$ (it is also equal to the set of instances of $R31,R32,R33,R35,R36,R37$).
Let us call now $P \in {\cal P}_f({\cal A}_3)$ a ${\cal C}_2$-{\em self-generating } set iff it fulfills
definition \ref{self-proof} and a  {\em self-generating } set iff it fulfills
definition \ref{self-proof} but where ${\cal C}_2$ is replaced by ${\cal C}_3$.
\begin{rem}$\;\;$\\
1-This notion of ``self-generating set ( of pairs)`` is a straightforward adaptation to our 
d-space of vectors of the notion of
``self-proving set of pairs`` defined in \cite[p.162]{Cou83t} for the magma $M(F \cup \Phi,V)$.\\
2-The notion of ``self-bisimulation'' (introduced in \cite{Cau90canon} and also used in \cite{Sti91,Hir-Jer-Mol94})
was also such an adaptation, but  in the context of a monoid-structure. The notion we use in this work can 
be seen, as well, as a generalisation of this notion of self-bisimulation:
when every class in $V_0/\smile$ has just one element, the only ``rational deterministic boolean series''
over $V_0$ are the words; in this case the self-bisimulations are exactly the self-generating sets.
\label{chapeau_caucal}
\end{rem}
\begin{lem}
Let $A \in {\cal A}_3$. Then $H_3(A) = \infty$ iff there exists a finite self-generating set 
$P \subseteq {\cal A}_3$ such that $A \in P$. 
\label{true_iff_self_proof3} 
\end{lem}
\begin{prooof}
Owing to metarules R33  and R34, every self-generating set is a 
${\cal B}_3$-proof.\\
Let $A \in {\cal A}_3$ such that $H_3(A) = \infty$.
By lemma \ref{true_iff_self_proof}, there exists some ${\cal C}_2$-self-generating set $P$
such that $A \in P$.\\
Let us consider $Q=\{ \rho_e(B)\mid B \in P\}$.\\
One can check that, $\rho_e$ maps the set of rules of ${\cal C}_2$ is into the set of rules of 
${\cal C}_3$. One can also check that $\rho_e$ and $\odot$ are commuting (i.e. $\rho_e(S \odot u)=
\rho_e(S) \odot u$).
Hence $Q$ is such that, 
for every 
$(S,S') \in Q$,
\begin{enumerate}
\item either $S=S'=\epsilon$
\item or $\exists {\cal R}_1 \in \bar{\cal B}_1(S,S'), \forall (x,x') \in {\cal R}_1,
Q \ededuce{*}_{{\cal C}_3}(S \odot x, S' \odot x') $.

\end{enumerate}
i.e. $Q$ is self-generating.
\end{prooof}

\begin{theo} : ${\cal B}_3$ is a complete deduction system.
\label{B3_complete}
\end{theo}
\begin{prooof}
Lemma \ref{true_iff_self_proof3}
implies the completeness property.
\end{prooof}

\subsubsection{Aknowledgement}
I am indebted to the CNRS ( especially to my collegues in the CNRS comitee) and to 
Bordeaux 1 university ( i.e. to my collegues in Bordeaux 1), who allowed 
me to be {\em free of teaching} during the academic years 1996/1998.
I thank O. Burkart, D. Caucal, P. Jancar, F. Moller and C. Stirling for useful discussion or
information about the subject treated in this work. This work has also benefited from the
incisive questions, comments and criticism of O. Burkart and C. Stirling about \cite{Sen97}.

\section*{ANNEX}
Let us sketch here a proof of theorem \ref{1graph_and_pda}.
\begin{lem}
Let $\Gamma= (\Gamma_0,v_0)$ be 
the computation 1-graph $({\cal C}({\cal M}), v_{{\cal M}})$ of some normalized
pushdown automaton ${\cal M}$. Then $\Gamma$ is equational and has finite 
out-degree.
\label{computgraphs_are_equa}
\end{lem}
\begin{prooof}
Let ${\cal M} =<X, Z, Q,\delta, q_0, z_0,F>$ be a normalized pda.
Let us consider a new letter $e \notin X$ and build the real-time pda
${\cal M}_e =<X \cup\{e\}, Z, Q,\delta_e, q_0, z_0,F>$ obtained
by setting that, for every $x \in X$ and $q\in Q, z \in Z$:
$$ \delta_e (qz, x) = \delta(qz,x);\;\;
 \delta_e (qz, e) = \delta(qz,\epsilon).
$$
By \cite[theorem 2.6 p.62]{Mul-Sch85}, the computation-graph 
${\cal C}({\cal M}_e)$
is context-free  and by \cite[theorem 6.3 p. 187]{Bau92} 
every context-free graph is equational.
Hence ${\cal C}({\cal M}_e)$ is equational. Let us remark that ${\cal C}({\cal M})$ is obtained from this graph just by contracting all the edges labelled by $e$. Let us contract the edges labelled by $e$ in some system of equations $S_e$
defining ${\cal C}({\cal M}_e)$: we  obtain
 a system of equations $S$ defining ${\cal C}({\cal M})$.
\end{prooof}
We use now the notation of \cite{Cou90c}. Given a system of graph equations
$S= <u_i=H_i; i \in [1,n]>$,  by ${\cal G}(S,u_i)$ we denote the $i$-th 
component of the canonical solution of $S$.
\begin{defi}
\label{standard_equations}
Let $S= <u_i=H_i; i \in [1,n]>$ be a system of graph equations.
It is said {\em standard} iff it fulfills the conditions\\
(1) for every $i\in [1,n]$ and every distinct integers  $k,\ell \in [1,\tau(H_i)]$, the sources 
$src(H_i,k)$, $src(H_i,\ell)$ are distinct vertices of $H_i$,\\
(2)for every $i \in [1,n]$ and every hyperedge $h$ of $H_i$ 
which is labelled by some unknown, all the vertices of $h$ are distinct,\\
(3)for every $i \in [1,n],k \in [1, \tau(u_i)], \lambda \in \bbbn$,
if there exist $\lambda$ edges going out of $src({\cal G}(S,u_i),k)$,inside the
graph ${\cal G}(S,u_i)$
then there exists also $\lambda$ edges going out of $src(H_i,k)$, inside
the graph $H_i$.
\end{defi}
\begin{lem}
Let $S= <u_i=H_i; i \in [1,n]>$ be a system of graph equations where the 
unknown
$u_1$ has type 1.
One can compute from $S$ a {\em standard} system of graph equations 
$S'= <u'_i=H'_i; i \in [1,n']>$ such that the canonical solution 
of $S'$ has a first component  ${\cal G}(S',u'_1)= {\cal G}(S,u_1)$.
\label{standardisation}
\end{lem}
\begin{prooof}
From $S$ one can construct a first system $S_1$ which generates the same first component ${\cal G}(S_1,u_1) = {\cal G}(S,u_1)$ and such that restrictions (1)(2)
of the lemma are fulfilled: this  follows from 
\cite[proposition 2.10 p.209]{Cou90c},(notice that the condition ``separated''
in this reference is exactly the conjunction (1)$\wedge$ (2)).\\
Let $S_1= <v_i=K_i; i \in [1,m]>$.
Let us replace every right-hand side $K_i$ by a finite hypergraph $L_i$
obtained by unfolding the graph $K_i$,  according to the rules $v_j
\rightarrow K_j$, as many times as necessary in order that every source 
$src(K_i,k)$ gets as many 
outgoing edges in $L_i$ as in the ``complete unfolded graph'' 
${\cal G}(S_1, v_i)$. The new system
$S'= <v_i=L_i; i \in [1,m]>$ still fulfills conditions (1)(2), it 
fulfills also condition (3) and for every $i \in [1,m]$,
${\cal G}(S_1, v_i)= {\cal G}(S', v_i)$. Hence $S'$ satisfies the
conclusion of the lemma.
 \end{prooof}
\begin{lem}
Let $\Gamma= (\Gamma_0,v_0)$ be a rooted 1-graph over $X$ which
is the first component of the canonical solution of some standard 
system of graph equations
.
Then, $\Gamma$ is isomorphic to the computation 1-graph 
$({\cal C}({\cal M}), v_{{\cal M}})$ of some normalized
pushdown automaton ${\cal M}$.
\label{standardgraphs_are_comput}
\end{lem}
\begin{sketch}
Let $S= <u_i=H_i; i \in [1,n]>$ be a satandard system of graph equations 
such that
$\Gamma= {\cal G}(S,u_1)$.\\
Let us define ${\cal M} =<X, Z, Q,\delta, q_0, z_0,F>$ as follows.
In every right-hand side $H_i$ we number bijectively all the
unknown hyperedges: $\{h_{1,i},\ldots,h_{j,i},\ldots,h_{n_i,i}\}$ and
all the vertices :$\{v_{1,i},\ldots,v_{q,i},\ldots,v_{N_i,i}\}$.
We note $\beta(j,i)= label(h_{i,j})$.
$$Z = \{ [j,i] \mid 1 \leq i \leq n , 1 \leq j \leq n_i\} 
\cup \{[1,0]\}.$$
(We extend $\beta$ by defining $\beta(1,0)=1$).\\  
Intuitively every symbol $[j,i]$ describes the situation of a vertex which
belongs to a component which has been glued on the $j$-th unknown hyperedge
of $H_i$.\\ 
Let $Q=[1,N]$ where $N$ is the maximum number of vertices in
the graphs $H_i$.
Intuitively, the transitions of ${\cal M}$ starting from a mode $q[j,i]$ 
describe
the edges starting from the $q$-th vertex of 
$H_{\beta(j,i)}$. Let us define precisely the transitions starting from a mode
$q[j,i]$:\\
{\bf case 1}: $q$ is strictly larger than the number of vertices of 
$H_{\beta(j,i)}$.\\
Then there is no transition starting from $q[j,i]$.\\
{\bf case 2}: vertex number $q$ of 
$H_{\beta(j,i)}$ is a source of $H_{\beta(j,i)}$ and $i \neq 0$.\\
Then
$$q[j,i] \deriv{\varepsilon}q',$$
where $q'$ is the number of the vertex of $H_i$ on which it is glued
(it is some vertex of $h_{j,i}$).\\
{\bf case 3}: vertex number $q$ of 
$H_{\beta(j,i)}$ is not a source of $H_{\beta(j,i)}$ or $i=0$.\\
{\bf internal edges}:\\
For every edge $(v_{q,\beta(j,i)},x,v_{q',\beta(j,i)})$, we add the transition
$$q[j,i] \deriv{x}q'[j,i].$$
{\bf external edges}:\\
Let $k=\beta(j,i)$.For every $\ell$ such that $v_{q,\beta(j,i)}$ is a vertex 
of $h_{\ell,k}$ and every edge $(v_{r,\beta(\ell,k)},x,v_{q',\beta(\ell,k)})$
where the vertex $v_{r,\beta(\ell,k)}$ of $H_{\beta(\ell,k)}$ is glued on
the vertex $v_{q,\beta(j,i)}$ by the rewriting rule
$u_{\beta(\ell,k)} \rightarrow H_{\beta(\ell,k)}$, we add the transition:
$$ q[j,i] \deriv{x} q'[\ell,k][j,i].$$
The starting configuration is $1[1,0]$ (i.e $q_0=1$,$z_0= [1,0]$).\\
This pda is normalized (this is easy to check) and has a computation graph
whose isomorphism-class is exactly ${\cal G}(S,u_1)$ 
(this would be much more tedious to prove formally).
\end{sketch}
Theorem \ref{1graph_and_pda} clearly follows from these three lemmas.
\bibliographystyle{alpha}
\bibliography{gagb}

\end{document}